\documentclass[aps, showpacs, floatfix, twocolumn, reprint, 
twoside, prc, preprintnumbers,superscriptaddress]{revtex4-2}
\usepackage{array}  
\usepackage{graphicx}  
\usepackage[export]{adjustbox}  
\usepackage{float}

\usepackage{epstopdf}
\usepackage{amsmath}
\usepackage{amssymb}
\usepackage{bm}
\usepackage{xcolor,xspace}
\usepackage{multirow}
\usepackage{mhchem}  
\usepackage{hyperref}
\hypersetup{colorlinks=True,urlcolor=blue,linkcolor=blue,
citecolor=blue,filecolor=black}
\usepackage{braket}
\usepackage{wasysym}  
\usepackage{scrextend}  
\usepackage[percent]{overpic}  
\usepackage[capitalise]{cleveref}  

\newcommand{\eq}{\!=\!}
\newcommand{\tpi}{\mbox{Type~I}}
\newcommand{\tpii}{\mbox{Type~II}}

\begin{document}
\title{Configuration mixing and intertwined quantum phase 
transitions in odd-mass niobium isotopes}
\author{N.~Gavrielov}\email{noam.gavrielov@yale.edu}
\affiliation{Center for Theoretical Physics, Sloane Physics 
Laboratory, Yale University, New Haven, Connecticut 
06520-8120, USA}
\affiliation{Racah Institute of Physics, The Hebrew 
University, Jerusalem 91904, Israel}

\date{\today}  
\begin{abstract}
Nuclei in the $Z\!\approx\!40,N\!\approx\!60$ region have 
one of the most complicated structural evolution across the 
nuclear chart, with coexisting shapes arising from 
different mixed configurations. In such a region, it is 
difficult to investigate odd-mass nuclei. In this paper a 
new algebraic framework is introduced, the interacting 
boson-fermion model with configuration mixing. 
Using this framework, with a boson core and a proton in 
the $1f_{5/2},2p_{3/2},2p_{1/2},1g_{9/2}$ orbits, a 
calculation is carried out to understand the structural 
evolution of the odd-mass niobium isotopes ($Z\eq41$) 
with neutron number 52--62. 
The calculated results are compared to energy levels, two 
neutron separation energies, $E2$ and $M1$ transition 
rates, and to quadrupole and magnetic moments. 
The detailed analysis discloses the effects of an abrupt 
crossing of states between normal and intruder 
configurations (Type~II QPT), which is accompanied by a 
gradual evolution from spherical- to deformed-core shapes
within the intruder configuration (Type~I QPT), where both 
types of QPTs occur around the critical point of neutron 
number 60. The identification of both types of QPTs in the 
same chain of isotopes provides an empirical manifestation 
of intertwined quantum phase transitions (IQPTs) in 
odd-mass nuclei and the relevance of IQPTs to the niobium 
chain. 
\end{abstract}

\maketitle
\section{Introduction}\label{sec:intro}
\subsection{Intertwined quantum phase transitions (IQPTs) 
in odd-mass nuclei}\label{sec:iqpt}
Quantum phase transitions (QPTs) \cite{Gilmore1978b, 
Gilmore1979} are structural changes induced by variation of 
parameters in the Hamiltonian, and are considered pivotal 
for understanding the dynamics of atomic nuclei 
\cite{Cejnar2010} and other systems \cite{carr2010QPT}. 

In nuclear structure, most of the attention has 
been devoted to the evolution of structure exhibiting two 
types of phase transitions. 
The first type of QPT, denoted as 
\tpi~\cite{Dieperink1980}, is a shape-phase transition in a 
single configuration. One common approach for investigating 
\tpi~QPTs is by using 
Hamiltonians composed of two (or more) different parts 
\cite{Iachello2011}
\begin{equation}\label{eq:type-i}
\hat H = (1-\xi)\hat H_1 + \xi\hat H_2~.
\end{equation}
In \cref{eq:type-i} one examines the equilibrium shape and 
symmetry of the Hamiltonian, which vary from those of $\hat 
H_1$ to those of $\hat H_2$ as the control parameter $\xi$ 
is varied from 0 to 1.
\tpi~QPTs have been established in the neutron number 90 
region for Nd-Sm-Gd-Dy isotopes, where the shape of the 
nuclei evolves from spherical to deformed 
\cite{Cejnar2010}. 

The second type of QPT, denoted as \tpii, is a transition 
in two (or more) configurations that coexist 
\cite{Heyde2011} and cross. One common approach for 
investigating \tpii~QPTs is by using Hamiltonians composed 
of a matrix form \cite{Frank2006}. For two 
configurations, this reads
\begin{equation}\label{eq:type-ii}
\hat H =
\begin{bmatrix}
\hat H_A(\xi_A) & \hat W(\omega) \\ 
\hat W(\omega)  & \hat H_B(\xi_B)
\end{bmatrix}~,
\end{equation}
where $\hat H_A$ and $\hat H_B$ denote the $A$ 
configuration (normal) and $B$ configuration (intruder) 
Hamiltonians and $\hat W$ their coupling.
In \cref{eq:type-ii}, one examines the evolution of 
structure from $A$ to $B$ by varying the control parameters 
$\xi_A$, $\xi_B$ and $\omega$.
\tpii~QPTs have been established in nuclei near shell 
closure, e.g., in the light Pb-Hg isotopes, with strong 
mixing between the configurations.

A \tpii~occurs when protons and neutrons that occupy
spin-orbit partner orbitals, 
$\pi(n\ell_{\ell\pm1/2})$--$\nu(n\ell_{\ell\mp1/2})$,
interact via the residual isoscalar proton-neutron 
interaction, $V_{pn}$~\cite{Federman1979}. 
The resulting gain in $n-p$ energy compensates the loss in 
single-particle and pairing energy. As a consequence, a 
mutual polarization effect occurs, which lowers 
single-particle orbitals of higher configurations to near 
(and effectively below) the ground state configuration. 
If the mixing is small, the \tpii~QPT can be accompanied by 
a distinguished \tpi~QPT within each configuration 
separately. Such a scenario, referred to as intertwined 
QPTs (IQPTs), was recently shown to occur in the even-even 
zirconium (Zr) isotopes \cite{Gavrielov2019, Gavrielov2020, 
Gavrielov2022}.

Most studies of QPTs in nuclei have focused on systems
with even numbers of protons and neutrons
~\cite{Cejnar2010,Heyde2011,Casten2009,Iachello2011,
Fortunato2021}. The structure of odd-mass nuclei is more 
complex due to the simultaneous presence of both collective 
and single-particle degrees of freedom. Consequently, QPTs 
in such nuclei have been far less studied. 
Fully microscopic approaches to QPTs in 
medium-heavy odd-mass nuclei, such as large-scale shell 
model~\cite{Caurier2005} and beyond-mean-field 
methods~\cite{Bally2014}, where suggested. However they are 
computationally demanding.
Other approaches have also been proposed including 
algebraic frameworks (shell-model inspired 
\cite{Scholten1982, IachelloVanIsackerBook} and
symmetry-based \cite{Jolie2004, Alonso2005, Alonso2007, 
Alonso2009, Boyukata2010,Petrellis2011a, Iachello2011b, 
Boyukata2021})
and density functionals-based mean-field methods 
\cite{Nomura2016a,Nomura2016b,Nomura2020,Quan2018}, 
involving particle-core coupling schemes with
boson-fermion or collective Hamiltonians.
So far, these approaches were restricted to \tpi~QPTs in 
odd-mass nuclei without configuration mixing.

As mentioned in previous works \cite{Brant1998, 
Rodriguez-Guzman2011, Nomura2020, Spagnoletti2019a, 
Garrett2021}, there is a growing need to develop a 
tractable framework that incorporates mixing of multiple 
configurations in odd-mass nuclei. This task is carried our 
in the present paper.

\subsection{The niobium isotopes}\label{sec:nb}
One region in the nuclear chart that is considered to 
accommodate mixed configurations and undergo a \tpii~QPT is 
the $Z\!\approx\!40$ region near neutron number 60. In this 
region, the ground state wave function seems to be 
dominated by a spherical configuration for neutron number 
50--58 and by a deformed configuration for neutron number 
60 and above \cite{Cheifetz1970, Federman1979, Heyde1985, 
Heyde1987}.

The sudden onset of deformation has been ascribed in the 
shell-model to $V_{pn}$ between nucleons that occupy the 
$\pi(1g_{9/2})$--$\nu(1g_{7/2})$ spin-orbit 
partners~\cite{Federman1977, Federman1979, Heyde1985, 
Heyde1987}, which results in a crossing between the normal 
and intruder configurations. 
The crossing arises as the $\nu(2d_{5/2}, 3s_{1/2}, 
2d_{3/2}, 1g_{7/2}, 1h_{11/2})$ orbits are filled, which 
induces a promotion of the protons across the $Z\eq40$ 
sub-shell gap. This promotion creates $2p\text{--}2h$ 
intruder excitations \cite{Federman1979, Federman1979b} and 
the so called configuration mixing scenario in this region. 
The protons promotion also generates a quenching of the 
$\pi(2p_{1/2})$--$\pi(2p_{3/2})$ orbits 
\cite{Federman1984}. Subsequently, it was also found 
\cite{Mach1990} that alongside the $\pi(2p_{1/2}), 
\pi(2p_{3/2})$ orbits the $\pi(1f_{5/2})$ orbit contributes 
significantly to the intruder excitations of the 
$\pi(1g_{9/2})$ orbit in the lighter $^{92,94,96}$Zr 
isotopes. This contribution was also demonstrated in the 
recent Monte-Carlo shell model \cite{Togashi2016} 
calculation for the chain of the even-even Zr isotopes with 
neutron number 50--70.

These dramatic structural changes have attracted 
considerable theoretical and experimental interest (for 
reviews see \cite{Heyde2011,Garrett2022}). For odd-$A$ 
nuclei, different theoretical approaches have studied this 
region, including non-relativistic mean-field based methods 
\cite{Lhersonneau1995, Lhersonneau1997a, 
Rodriguez-Guzman2011, Esmaylzadeh2019, Nomura2020}, 
shell model approaches \cite{Gloeckner1975, Mach1990, 
Bucurescu2005, Orce2006, Orce2010} and algebraic approaches 
\cite{Brant1988, Lhersonneau1990, Lhersonneau1998, 
Brant1998, Spagnoletti2019a, Boulay2020}, where large-scale 
shell model approaches are scarce 
\cite{Sieja2009,Sieja2021}. 

The structure of niobium (Nb) isotopes ($Z\eq41$) with 
neutron number 52--62 was recently investigated for first 
time in within the new framework of the interacting 
boson-fermion model with configuration mixing 
\cite{Gavrielov2022c}. The positive-parity states were 
analyzed to exemplify the occurrence of IQPTs, similarly to 
the adjacent even-even Zr isotopes \cite{Gavrielov2019, 
Gavrielov2022}.
In this work, we extend our analysis to the negative parity 
states, along with more observables that are compared to 
experimental data. This comparison is further supported by 
analyzing the configuration and single-particle content of 
the wave functions for the entire chain.

\subsection{Layout}\label{sec:layout}
The paper is divided into the following sections. 
\cref{sec:theo-fram} presents the theoretical framework, 
which includes the boson Hamiltonian 
(\cref{sec:boson-ham}), fermion Hamiltonian 
(\cref{sec:fermion-ham}), boson-fermion interaction 
(\cref{sec:boson-fermion-int}), electromagnetic transitions 
operators (\cref{sec:em-oper}) and wave functions 
(\cref{sec:wf}). 
In \cref{sec:qpts-nb} QPTs in the Nb chain are discussed, 
presenting Type~I and Type~II QPTs in odd-mass nuclei 
(\cref{sec:type-i-ii}) and the Nb model space for IBFM-CM 
(\cref{sec:model-space}).

The results are divided into two main sections.
In \cref{sec:results-indi} the results for the individual 
isotopes are presented, which include spectrum analysis. 
This section is further partitioned into positive-parity 
states (\cref{sec:positive-states}), in the 
$^{93\text{--}97}$Nb region (\cref{sec:93-97nb_p}) and the 
$^{99\text{--}103}$Nb region (\cref{sec:99-103nb_p}), and 
negative-parity states (\cref{sec:negative-states}), 
in the $^{93\text{--}97}$Nb region (\cref{sec:93-97nb_m}) 
and the $^{99\text{--}103}$Nb region 
(\cref{sec:99-103nb_m}).
\cref{sec:res_evo_obs} presents results for the evolution 
of configuration and single-particle content 
(\cref{sec:evo-conf}), energy levels 
(\cref{sec:evo-energy}), two-neutron separation energies 
(\cref{sec:evo-s2n}), $E2$ transition rates and quadrupole 
moments (\cref{sec:evo-e2-q}), $M1$ and magnetic moments 
(\cref{sec:evo-mag-m1}). The conclusions and outlook are in 
\cref{sec:conc}.

\section{Theoretical framework}\label{sec:theo-fram}
For the study of QPTs in the Nb isotopes we use the 
algebraic framework of the interacting boson-fermion model 
(IBFM)~\cite{IachelloVanIsackerBook}. The IBFM treats 
odd-$A$ nuclei as a system of monopole ($s$) and quadrupole 
($d$) bosons, representing valence nucleon pairs, and a 
single (unpaired) nucleon. In a previous paper 
\cite{Gavrielov2022c}, the IBFM was extended to include 
core excitations and obtain a boson-fermion model with 
configuration mixing (IBFM-CM). In such a model, the 
Hamiltonian has the form
\begin{equation}\label{eq:ham}
\hat H = \hat H_{\rm b} + \hat H_{\rm f} + \hat V_{\rm bf}~,
\end{equation}
where $\hat H_{\rm b}$ is the boson core Hamiltonian, $\hat 
H_{\rm f}$ is fermion single-particle Hamiltonian and $\hat 
V_{\rm bf}$ is the boson-fermion interaction.

\subsection{The boson Hamiltonian}\label{sec:boson-ham}
For a single configuration, the interacting boson model 
(IBM) Hamiltonian consists of Hermitian and 
rotational-scalar interactions that conserve the total 
number of $s$ and $d$ bosons,
\begin{equation}\label{eq:boson-number}
\hat N=\hat n_s+\hat n_d \eq s^\dagger s+\sum_\mu 
d^\dagger_\mu d_\mu~.
\end{equation}
The latter is fixed by the microscopic interpretation of 
the IBM \cite{IachelloTalmi1987} to be 
$N\eq N_{\pi}+N_{\nu}$, where $N_{\pi}$ ($N_{\nu}$) is the 
number of proton (neutron) particle or hole pairs counted 
from the nearest closed shell. 
For multiple shell model configurations, different shell 
model spaces of 0p-0h, 2p-2h, 4p-4h, $\ldots$ particle-hole 
excitations are associated with the corresponding boson 
spaces of $N,N+2,N+4,\ldots$ bosons, respectively, which 
are subsequently mixed.
The boson Hamiltonian ($\hat H_{\rm b}$) is that of the 
configuration mixing model (IBM-CM) of \cite{Duval1981, 
Duval1982}, and has the form 
\cite{Frank2006}
\begin{equation}\label{eq:H_b}
\hat H_{\rm b} = 
\begin{bmatrix}
\hat H_{\rm b}^{\rm A}(\xi^{(\rm A)}) &
\hat{W}_{\rm b}(\omega)\\
\hat{W}_{\rm b}(\omega) 
& \hat H_{\rm b}^{\rm B}(\xi^{(\rm B)})
\end{bmatrix} ~.
\end{equation}
Here $\hat H_{\rm b}^{\rm A}(\xi^{(\rm A)})$
represents the normal A configuration 
($N$ boson space) and $\hat H_{\rm b}^{\rm B}(\xi^{(\rm 
B)})$
represents the intruder B configuration
($N\!+\!2$ boson space), corresponding to 2p-2h excitations
across the (sub-) shell closure.
Standard forms of $\hat H_{\rm b}^i(\xi^i)$ 
with $i = {\rm A,B}$
include pairing, quadrupole, and rotational terms, in the 
following form \cite{Gavrielov2022}
\begin{equation}\label{eq:H_b_consis}
\hat H_{\rm b}^i = \epsilon^{(i)}_d \hat n_d + 
\kappa^{(i)} \hat Q_\chi \cdot \hat Q_\chi + \kappa^{\prime 
(i)} \hat L\cdot \hat L + \delta_{i,{\rm B}}\Delta_p,
\end{equation}
where $\Delta_p$ is the off-set energy between 
configurations A and B, the quadrupole operator is 
\begin{equation}\label{eq:quad_op}
\hat Q_\chi = d^\dag s+s^\dag \tilde d\!+\! \chi (d^\dag 
\tilde d)^{(2)}~,
\end{equation}
and the mixing term is
\begin{equation}\label{eq:mixing_int}
\hat W_{\rm b} \eq \omega 
[(d^\dag d^\dag)^{(0)} \!+\! (s^\dag)^2] +\text{H.c.}~,
\end{equation} 
where H.c. stands for Hermitian conjugate. In 
\cref{eq:quad_op,eq:mixing_int} $\tilde d_\mu \eq
(-)^\mu d_{-\mu}$.
Such IBM-CM Hamiltonians have been used extensively for the
study of shape-coexistence, configuration-mixed and QPTs
in even-even nuclei~\cite{Duval1981,Duval1982,Sambataro1982,
GarciaRamos2014a,GarciaRamos2014b,GarciaRamos2015a,
Nomura2016c,Leviatan2018a,MayaBarbecho2022,
Gavrielov2019,Gavrielov2020,Gavrielov2022}.
\subsection{The fermion Hamiltonian}\label{sec:fermion-ham}
The fermion Hamiltonian ($\hat H_{\rm f}$) of
Eq.~(\ref{eq:ham}) has the form
\begin{equation}\label{eq:H_F}
\hat H_{\rm f} = \begin{bmatrix}
\sum_j\epsilon^{(\rm A)}_j \hat n_j & 0 \\
0 & \sum_j\epsilon^{(\rm B)}_j \hat n_j
\end{bmatrix} ~,
\end{equation}
where $j$ is the angular momentum of the occupied
orbit, $\hat n_j = \sum_\mu a^\dagger_{j\mu}a_{j\mu}$ the 
corresponding number operator and $\epsilon^{(\rm 
i)}_j\,({i=\rm A,B)}$ are the single-particle energies for 
each configuration, A or B.
In this work, the single-particle energies are determined 
using the microscopic interpretation of the 
IBFM~\cite{IachelloVanIsackerBook} (see \cref{app:bcs}  
for more details).
\subsection{The boson-fermion 
interaction}\label{sec:boson-fermion-int}
The boson-fermion interaction has the form
\begin{equation}\label{eq:V_BF}
\hat V_{\rm bf} = \begin{bmatrix}
  \hat V^{\rm A}_{\rm bf}(\zeta^{(\rm A)}) &
 \hat{W}_{\rm bf}(\omega_j)\\
\hat{W}_{\rm bf}(\omega_j) &
\hat V^{\rm B}_{\rm bf}(\zeta^{(\rm B)})
\end{bmatrix} ~.
\end{equation}
Here, $\hat V^{(i)}_{\rm bf}\,({i=\rm A,B})$ is the general 
boson-fermion interaction \cite{IachelloVanIsackerBook} for 
each configuration. In this work it involves monopole, 
quadrupole and exchange terms 
\begin{equation}\label{eq:V_BF_i}
\hat V^{(i)}_{\rm bf} = V^{{\rm MON}(i)}_{\rm bf} + V^{{\rm 
QUAD}(i)}_{\rm bf} + V^{{\rm EXC}(i)}_{\rm bf},
\end{equation}
which read
\begin{subequations}
\begin{align}\label{eq:v_bf_expl}
V^{{\rm MON}(i)}_{\rm bf} & = \sum_j A^{(i)}_j [[d^\dagger 
\times 
\tilde d]^{(0)} \times [a^\dagger_j \times \tilde 
a_j]^{(0)}]^{(0)}_0~,\\
V^{{\rm QUAD}(i)}_{\rm bf} & = \sum_{jj^\prime} 
\Gamma^{(i)}_{jj^\prime} [\hat Q_\chi \cdot [a^\dagger_j 
\times 
\tilde a_{j^\prime}]^{(2)}]^{(0)}_0~,\\
V^{{\rm EXC}(i)}_{\rm bf} & = \nonumber\\
 & \mkern-45mu \sum_{jj^\prime j^{\prime\prime}} 
\Lambda^{j^{\prime\prime}(i)}_{jj^\prime} :[[d^\dagger 
\times 
\tilde a_j]^{(j^{\prime\prime})} \times [\tilde d \times 
a^\dagger_{j^\prime}]^{(j^{\prime\prime})}]^{(0)}_0:~,
\end{align}
\end{subequations}
where $\tilde a_{j\mu} = (-)^{j+\mu}a_{j-\mu}$.
Using the microscopic interpretation of the 
IBFM~\cite{IachelloVanIsackerBook}, these couplings can be 
expressed in terms of strengths 
($A^{(i)}_{0},\Gamma^{(i)}_{0},\Lambda^{(i)}_{0}$) and
occupation probabilities $(u_j,v_j)$ (see \cref{app:bcs}  
for more details).
The new off-diagonal term contributes to the $j$-dependent 
mixing
\begin{equation}
\hat{W}_{\rm bf}(\omega_j) = \sum_j\omega_j \hat n_j 
[(d^{\dag}d^{\dag})^{(0)} \!+\! (s^{\dag})^2 + 
\text{H.c.}]~.	
\end{equation}
\subsection{Electromagnetic transitions 
operators}\label{sec:em-oper}
Operators inducing electromagnetic transitions of type
$\sigma$ and multipolarity $L$, contain boson and fermion
parts,
\begin{equation}\label{eq:TsigL}
\hat{T}(\sigma L) =
\hat{T}_{\rm b}(\sigma L) + \hat T_{\rm f}(\sigma L) ~.
\end{equation}

For $\sigma L \eq E2$ transitions, the boson and fermion 
parts of \cref{eq:TsigL} are
\begin{subequations}\label{eq:Te2}
\begin{align}
\hat{T}_{\rm b}(E2) & =
e^{(\rm A)}\hat Q^{(N)}_{\chi} + e^{(\rm B)}\hat 
Q^{(N+2)}_{\chi}~,\label{eq:te2_b}\\
\hat T_{\rm f}(E2) & = 
\sum_{jj^\prime}f^{(2)}_{jj^\prime}[a^\dagger_j \times 
\tilde a_{j^\prime}]^{(2)},\label{eq:te2_f}
\end{align}
\end{subequations}
In \cref{eq:te2_b}, $e^{(\rm A)},e^{(\rm B)}$ are the 
boson effective charges for configuration A and B, 
respectively. The superscript $(N)$ denotes a projection 
onto the $[N]$ boson space and in \cref{eq:te2_f} we have
\begin{equation}\label{eq:f2}
f^{(2)}_{jj^\prime} = 
-\frac{e_f}{\sqrt{5}}\braket{j||Y^{(2)}_{lm}||j^\prime}~,
\end{equation}
where $e_f$ is the effective charge for $E2$ transitions. 

For $\sigma L \eq M1$ transitions, the boson and fermion 
parts of \cref{eq:TsigL} are
\begin{subequations}\label{eq:Tm1}
\begin{align}
\hat T_{\rm b}(M1) & =
\sum_i \sqrt{\frac{3}{4\pi}}g^{(i)} \hat L^{(N_i)} 
\nonumber\\
& \qquad\qquad + \tilde g^{(i)}[\hat Q^{(N_i)}_{\chi} 
\times \hat 
L^{(N_i)}]^{(1)}~.\label{eq:tm1_b}\\
\hat T_{\rm f}(M1) & = 
\sum_{jj^\prime}f^{(1)}_{jj^\prime}[a^\dagger_j \times 
\tilde a_{j^\prime}]^{(1)}~,\label{eq:tm1_f}
\end{align}
\end{subequations}
with
\begin{equation}
f^{(1)}_{jj^\prime} = -\frac{f_1}{\sqrt{3}}\braket{j||g_l 
\hat l + g_s \hat s||j^\prime}~.
\end{equation}
Here, $i \eq ({\rm A,B})$ and $N_{\rm A} \eq N,\, N_{\rm 
B} \eq N\!+\!2$.
For a proton, the free value for the spin $g$-factor is  
$g_s \eq 5.5857~\mu_N$, and for the angular g-factor is
$g_l \eq 1~\mu_N$ (see \cref{app:bcs} for more details 
about the quenching of $g_s$).

\subsection{Wave functions}\label{sec:wf}
The Hamiltonian of \cref{eq:ham} is diagonalized 
numerically. The resulting eigenstates, $\ket{\Psi;J}$, 
are linear combinations of wave functions $\Psi_{\rm A}$ and
$\Psi_{\rm B}$, involving bosonic basis states in the two 
spaces $\ket{[N],\alpha,L}$ and $\ket{[N+2],\alpha,L}$, 
where $\alpha$ denoted additional quantum numbers 
characterizing the boson basis used.
The boson ($L$) and fermion ($j$) angular momenta
are coupled to $J$ and the combined wave function has the 
form
\begin{multline}\label{eq:wf}
\ket{\Psi;J} \eq
\sum_{\alpha,L,j}C^{(N,J)}_{\alpha,L,j}
\ket{\Psi_{\rm A};[N],\alpha,L,j;J} \\
+ \sum_{\alpha,L_,j}C^{(N+2,J)}_{\alpha,L,j}
\ket{\Psi_{\rm B};[N+2],\alpha,L,j;J}~,
\end{multline}
For such a wave function, it is possible to examine the 
probability of normal-intruder mixing 
\begin{equation}\label{eq:prob_norm_int}
P^{(N_i,J)} = \sum_{j}P^{(N_i,J)}_{j}~,
\end{equation}
with $P^{(N_\text{A},J)} + P^{(N_\text{B},J)} \eq 
1$. Here, $P^{(N_i,J)}_{j}$ is the probability 
of a single-particle orbit $j$ within a certain 
configuration $i$
\begin{equation}\label{eq:prob_spe}
P^{(N_i,J)}_{j}
 \eq \sum_{\alpha,L}|C^{(N_i,J)}_{\alpha,L,j}|^2~.
\end{equation}

\section{QPTs in the niobium chain}\label{sec:qpts-nb}
\subsection{Type~I and Type~II QPTs} 
\label{sec:type-i-ii}
The occurrence of QPTs in Bose-Fermi systems for a 
single-configuration in the framework of the IBFM is more 
complicated than in the case of boson systems (IBM) 
\cite{Petrellis2011a}.
QPTs in odd-mass nuclei consider the effect of the odd 
nucleon on the phase transitions of the boson core.
Considering the  U(5)-SU(3) boson QPT (Type I), in the 
adjacent odd-mass system the transition is from a weak 
coupling (spherical U(5) boson core) to a strong coupling 
(axially deformed SU(3) boson core) spectrum.

A U(5) spherical spectrum is typically identified with 
couplings of the fermion orbits $j_{\rm f}$ with states 
$L_{\rm b}$ of the adjacent even-even system to give a 
total angular momentum $J \eq L \otimes j_{\rm f}$,
\begin{equation}\label{eq:l_j_mult}
|L-j_{\rm f}| \le J \le 
|L+j_{\rm f}|.
\end{equation}
For the a $0^+$ state of the adjacent even-even isotope, 
this results in states with a total $J \eq j_{\rm 
f}$, while for a first-excited $2^+$ it results 
with a multiplet of states, for each of the $j$-orbits. 
Their respective irreducible representations (irreps) in 
the boson U(5) limit are $n_d \eq 0$ and 1. 
For a single-$j$ coupling, $L\otimes j_{\rm f}$, one can 
compare the energy of the state $L$ of the adjacent 
even-even nuclei to the ``center of gravity'' 
(CoG)~\cite{Lawson1957} of a 
multiplet of states \cref{eq:l_j_mult} by calculating
\begin{equation}\label{eq:cog}
\Delta E_{\rm CoG} = \frac{\sum_J(2J+1)E_J}{(2L+1)(2j_{\rm 
f}+1)},
\end{equation}
where $E_J$ are the excitation energies of the states with 
total $J$ that belong to the multiple.
The $E2$ [$M1$] transitions between members of the 
weakly-deformed multiplet (originating from the $2^+$ with 
$n_d \eq 1$) and the single state (originating from the 
$0^+$ with $n_d \eq 0$) are comparable to [weaker 
than] those of the adjacent even-even isotope while $E2$ 
[$M1$] transitions between the multiplet members are weak 
[strong].

An SU(3) deformed spectrum is typically identified with 
rotational bands starting at some value $K$ of angular 
momentum. For a single-$j$ scenario (ignoring $K$ bands 
mixing), the energy typically behaves as
\begin{equation}\label{eq:moment_iner}
E_K(J) = BJ(J+1),
\end{equation}
where $B$ is the moment of inertia. The magnetic and 
quadrupole moments for each of the states in the band and 
the $M1$ and $E2$ transitions among them can be compared to 
the geometric collective model expressions of Bohr and 
Mottelson (BM) \cite{BohrMott-II} for $K\not=1/2$
\begin{subequations}\label{eq:GCM-qm}
\begin{align}
Q(J,K) & = Q_0 
\braket{J,K,2,0|J,K}\braket{J,J,2,0|J,J}\label{eq:GCM-q},\\
\mu(J,K) & = g_R J + 
(g_K-g_R)\frac{K^2}{J+1},\label{eq:GCM-mu}
\end{align}
\end{subequations}
and
\begin{subequations}\label{eq:GCM-t}
\begin{align}
B(E2;J^\prime,K\to J,K) & = 
Q_0^2\Big(\frac{5}{16\pi}\Big)  \nonumber\\
& \qquad \times
\braket{J^\prime,K,2,0|J,K}^2,\label{eq:GCM-e2}\\
B(M1;J^\prime,K \to J,K) & = 
\frac{3}{4\pi}(g_K-g_R)^2 K^2 \nonumber\\
& \qquad \times
\braket{J^\prime,K,1,0|J,K}^2,\label{eq:GCM-m1}
\end{align}
\end{subequations}
where $Q_0$, $g_R$ and $g_K$ are fitted to the data, and 
$g_R$ is expected to behave as $g_R\simeq Z/A$.

For multiple configurations, one would expect, as in the 
adjacent even-even case, to observe a crossing of states 
that are associated with the different configurations that 
are mixed (Type II QPT), where in a weak mixing scenario 
the above considerations for transitions would apply for 
each configuration separately.
\subsection{Niobium model space for IBFM-CM} 
\label{sec:model-space}
\begin{figure}[t]
\centering
\includegraphics[width=1\linewidth]{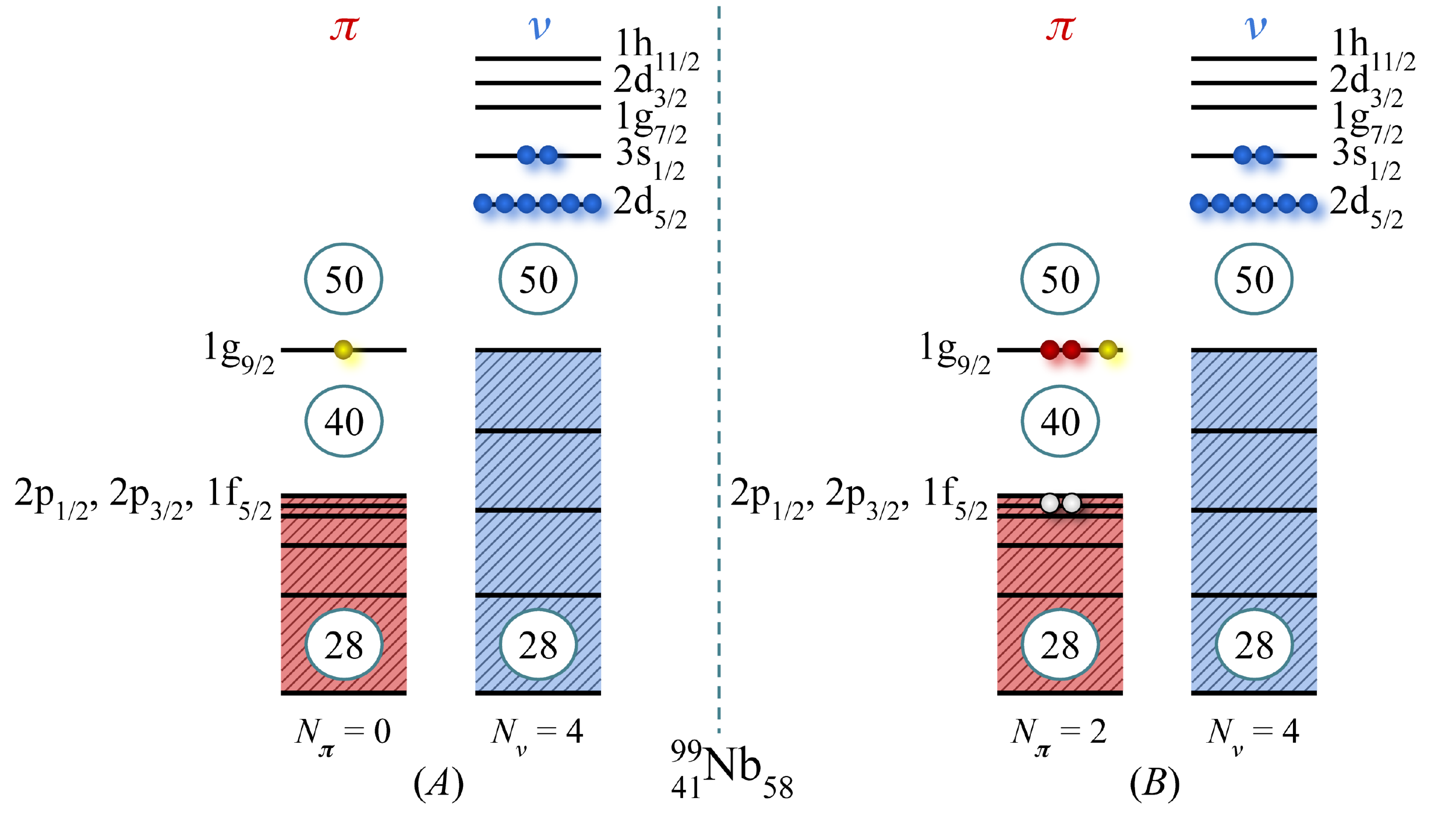}
\caption{Schematic representation of the two coexisting 
shell-model configurations ($A$ and $B$) for 
$^{99}_{41}$Nb$_{58}$. The corresponding numbers of proton 
bosons ($N_{\pi}$) and neutron bosons ($N_{\nu}$), relevant 
to the IBM-CM, are listed for each configuration and are 
depicted by a pair of particles (in red or blue) and a pair 
of holes (in white). There are no active proton bosons for 
configuration A. Alongside them, the extra proton (in 
yellow) is shown at the proton $\pi(1g_{9/2})$ orbit. 
\label{fig:nb-99-shell}}
\end{figure}
The $\ce{_{41}^ANb}$ isotopes with mass number 
\mbox{$A \eq \text{93--103}$} are described by coupling a 
proton to their respective $\ce{_{40}Zr}$ cores with 
neutron number 52--62. The latter isotopes have been 
suggested to have 
\cite{Gavrielov2019,Gavrielov2020,Gavrielov2022} a 
normal A configuration that corresponds to having no active 
protons above the $Z \eq 40$ sub-shell gap, and an intruder 
B configuration that corresponds to two-proton excitation 
from below to above this gap, creating 2p-2h states. 
The parameters of $\hat H_{\rm b}$~\eqref{eq:H_b} and boson 
numbers are taken to be the same as in a previous 
calculation of these Zr isotopes.
According to the usual boson-counting, the corresponding 
bosonic configurations have proton bosons $N_{\pi} \eq 0$ 
for configuration~$A$ and $N_{\pi} \eq 2$ for 
configuration~$B$. Both configurations have neutron bosons 
$N_{\nu} \eq 1,2, \ldots,6$ for neutron number 52--62, 
which sums to a total of $N \eq 1,2,\ldots6$ for 
$^{92-102}$Zr, respectively (see Table~V 
of~\cite{Gavrielov2022} for more details). 
For the odd particle, the valence protons are assumed to 
reside in the $Z \eq \text{28--50}$ shell with the 
$\pi(1f_{5/2}), \pi(2p_{3/2}), \pi(2p_{1/2}), 
\pi(1g_{9/2})$ orbits.

\section{Results: Detailed Quantum Analysis of Individual 
Isotopes}\label{sec:results-indi} 
\begin{figure}[t]
\centering
\includegraphics[width=1\linewidth]{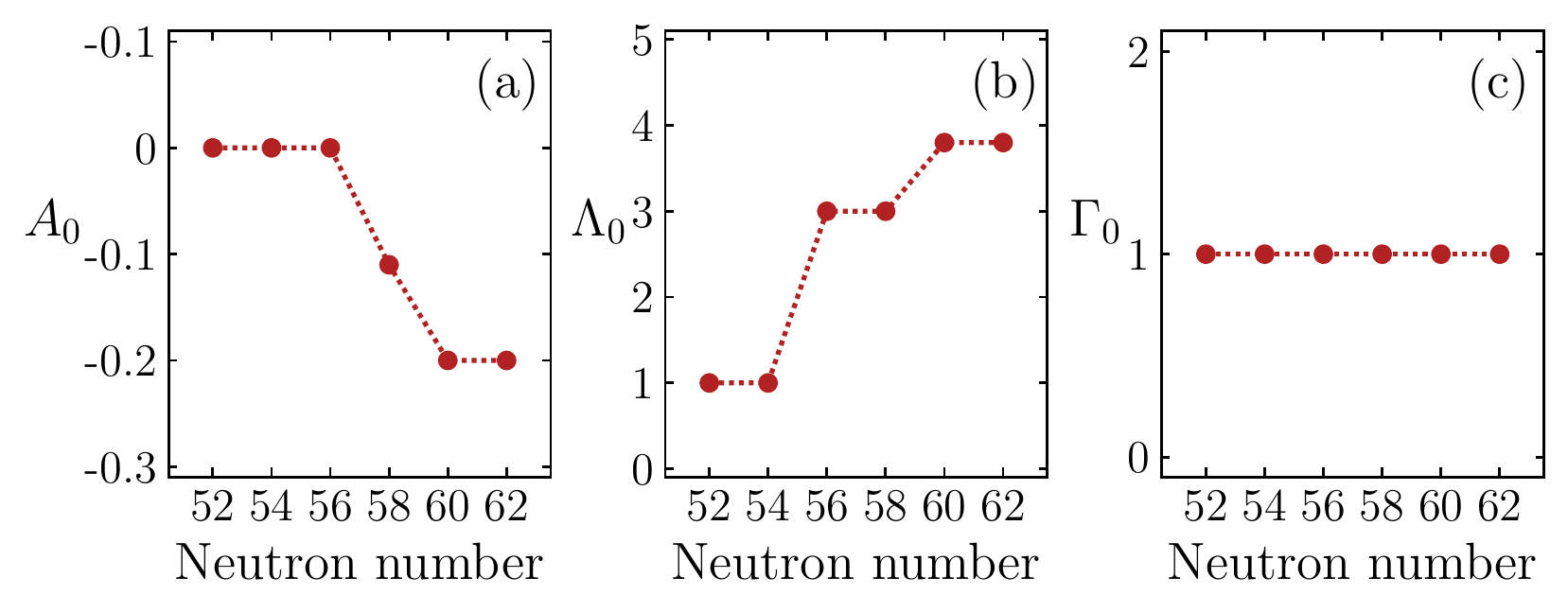}
\caption{Parameters of the IBFM-CM boson-fermion 
interaction, \cref{eq:V_BF_i}, in MeV. For more details see 
\cref{app:bcs}.\label{fig:params}}
\end{figure}
The quantum analysis for $^{93-103}$Nb entails a detailed 
comparison of the experimental energies and $E2$ and $M1$ 
transition rates with the results of the calculation for 
the positive- and negative-parity states. The strengths of 
the boson-fermion interaction  and single quasi-particle 
energies take the same value for both configurations, i.e. 
$A^{(i)}_{0}, \Gamma^{(i)}_{0}, \Lambda^{(i)}_{0} = A_0, 
\Gamma_0, \Lambda_0$ and $\epsilon^{(i)}_j \eq 
\epsilon_j$ for $i \eq \text{A,B}$, and are shown in 
\cref{fig:params} and \cref{tab:bcs,tab:parameters}. The 
BCS calculation and fitting procedure employed to obtain 
them, are discussed in \cref{app:bcs}.

The wavefunctions obtained are of the form as \cref{eq:wf} 
with $j\eq\pi(1g_{9/2})$ for the positive-parity sector and 
$j\eq\pi(2p_{1/2}), \pi(2p_{3/2}), \pi(1f_{5/2})$ for the 
negative-parity sector. The negative- and positive-parity 
calculations are done independently, where the ground state 
is always positive-parity. Therefore, a shift in energy is 
added to the excitations of the negative-parity energies 
that places the lowest calculated energy at the 
experimental value. In 
\cref{fig:93Nb-p,fig:95Nb-p,fig:97Nb-p,fig:99Nb-p,fig:93Nb-m,fig:95Nb-m,fig:97Nb-m,fig:99Nb-m}
states in black (blue) belong to the normal (intruder) A 
(B) configuration.
\subsection{Positive-parity 
states}\label{sec:positive-states}
For the positive parity states only the $\pi(1g_{9/2})$ 
orbit plays a role, which reduces the calculation to a 
single-$j$ one. 
The individual isotopes are divided into two regions: a 
weak coupling region for $^{93-97}$Nb and the IQPT
region for $^{99-103}$Nb, which also incorporates 
strong coupling.

For the region of $^{93-97}$Nb, the calculation is compared 
to the experimental levels in 
\cref{fig:93Nb-p,fig:95Nb-p,fig:97Nb-p}, including $E2$ and 
$M1$ transitions among them. For each isotope, the spectrum 
exhibits coexistence of two spherical configurations with 
weak mixing between them. The corresponding spectra of 
$^{92,94,96}$Zr, the even-even core, are also shown with an 
assignment of selected levels $L$ to the normal A or 
intruder B configurations (in subscript), based on the 
analysis in Ref.~\cite{Gavrielov2022}, which also showed 
that the two configurations in $^{92,94,96}$Zr are 
spherical and weakly deformed, respectively.

For the $^{99-103}$Nb region, the calculation is compared 
to the experimental levels in 
\cref{fig:99Nb-p,fig:101Nb-p,fig:103Nb-p}, including $E2$ 
and $M1$ transitions among them. For $^{99}$Nb, the 
spectrum exhibits coexistence of two configurations, one 
spherical and one weakly deformed, where only the ground 
state seems to belong to the normal A configuration. For 
$^{101,103}$Nb, the spectrum exhibits a rotational pattern 
that resembles a strong coupling scenario within 
the intruder B configuration.

\subsubsection{The $^\text{93--97}$Nb region: weak 
coupling}\label{sec:93-97nb_p}
\begin{figure*}[th!]
\centering
\includegraphics[width=0.533076923\linewidth]
{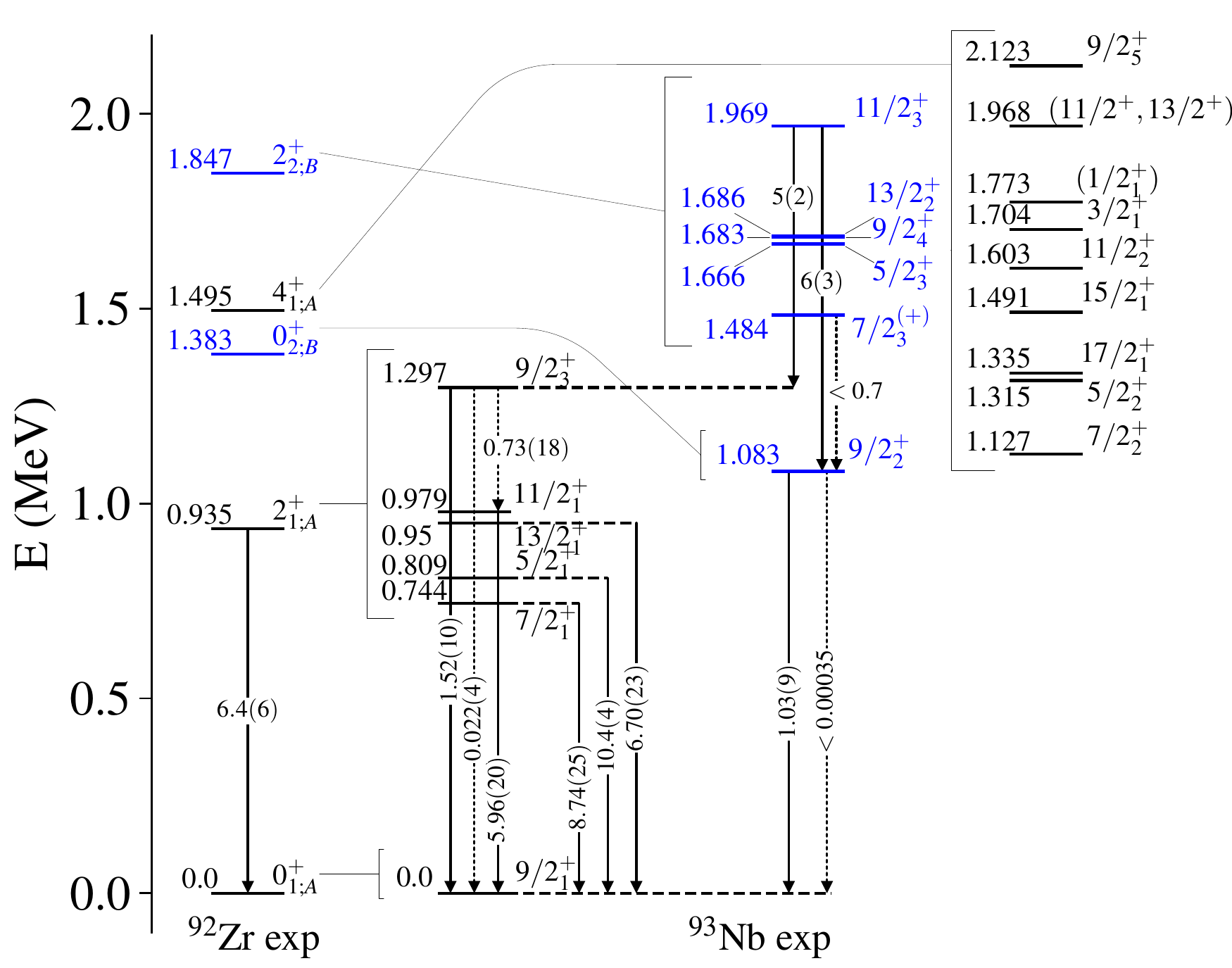}
\includegraphics[width=0.456923077\linewidth]
{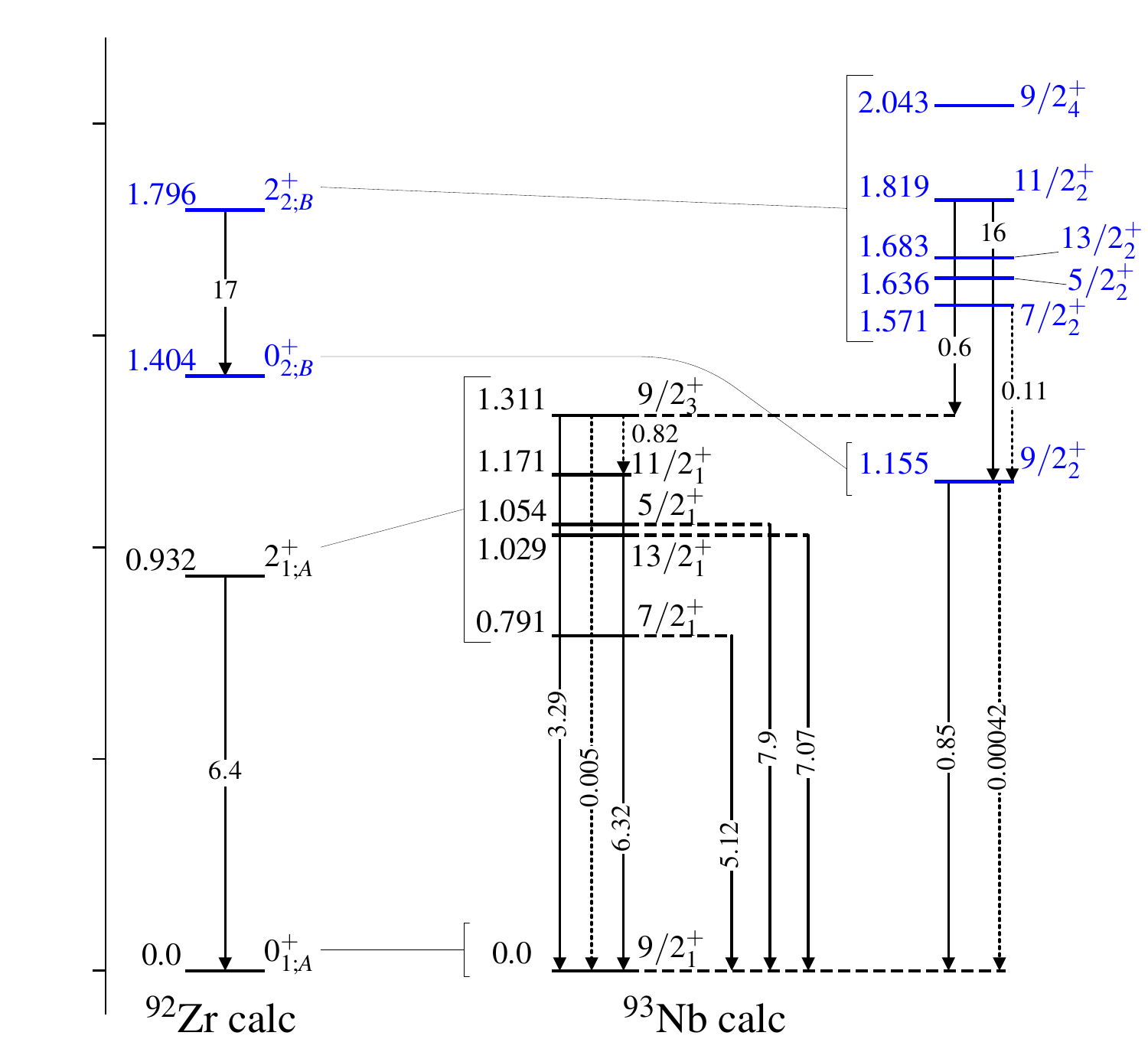}
\caption{Experimental (left) and calculated (right) energy
levels in MeV, and $E2$ (solid arrows) and $M1$
(dashed arrows) transition rates in W.u., for $^{93}$Nb
and $^{92}$Zr. Normal (intruder) states are depicted in 
black (blue). Lines connecting $L$-levels in $^{92}$Zr 
to sets of $J$-levels in $^{93}$Nb indicate the weak 
coupling $(L\otimes \tfrac{9}{2})J$. 
Data taken from \cite{Orce2010,NDS.112.1163.2011}.
Note that the observed $4^+_{\rm 1;A}$ state in $^{92}$Zr
is outside the boson $N \eq 1$ model 
space.\label{fig:93Nb-p}}
\end{figure*}

\begin{figure*}[th!]
\centering
\includegraphics[width=0.609230769\linewidth]
{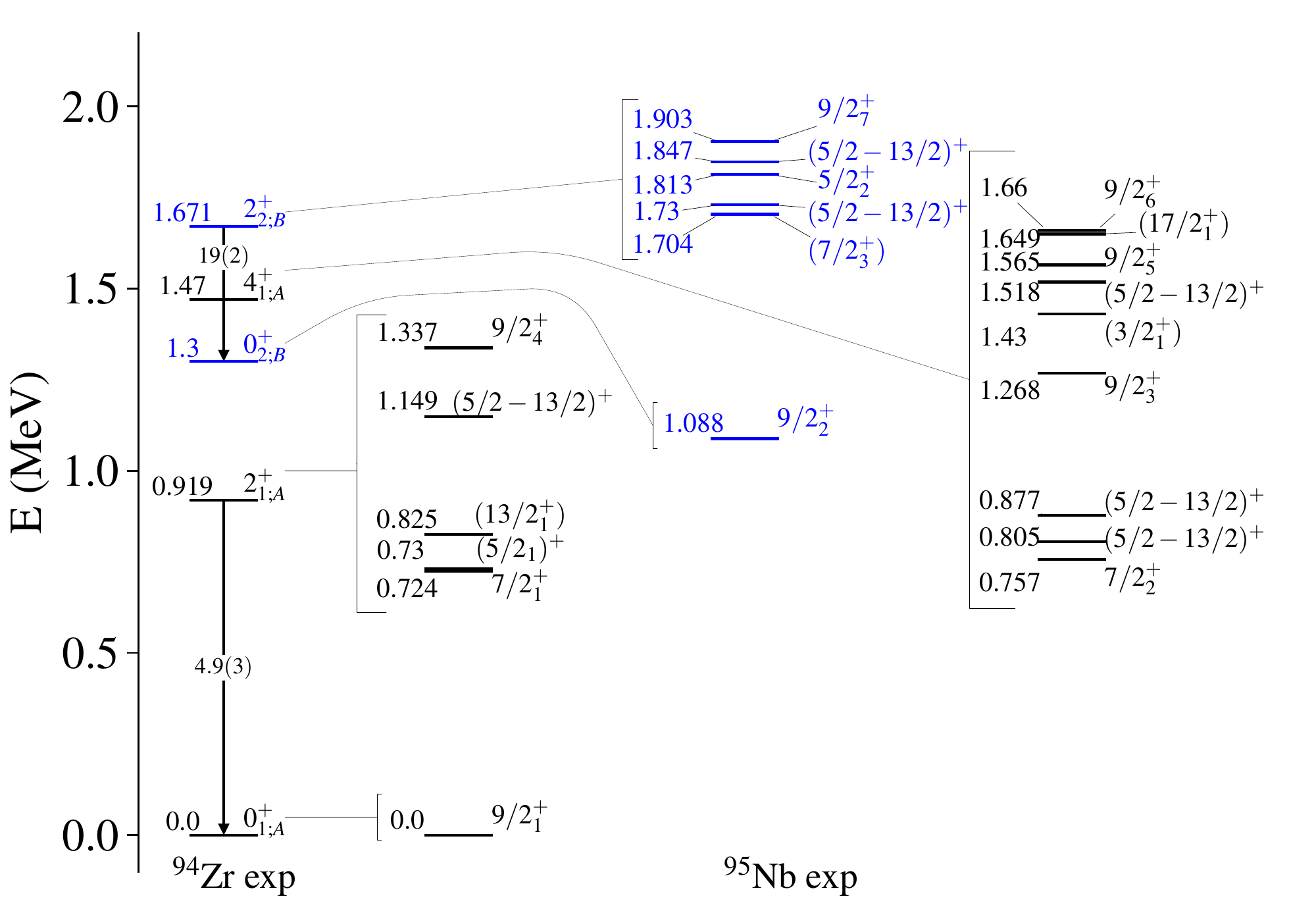}
\includegraphics[width=0.380769231\linewidth]
{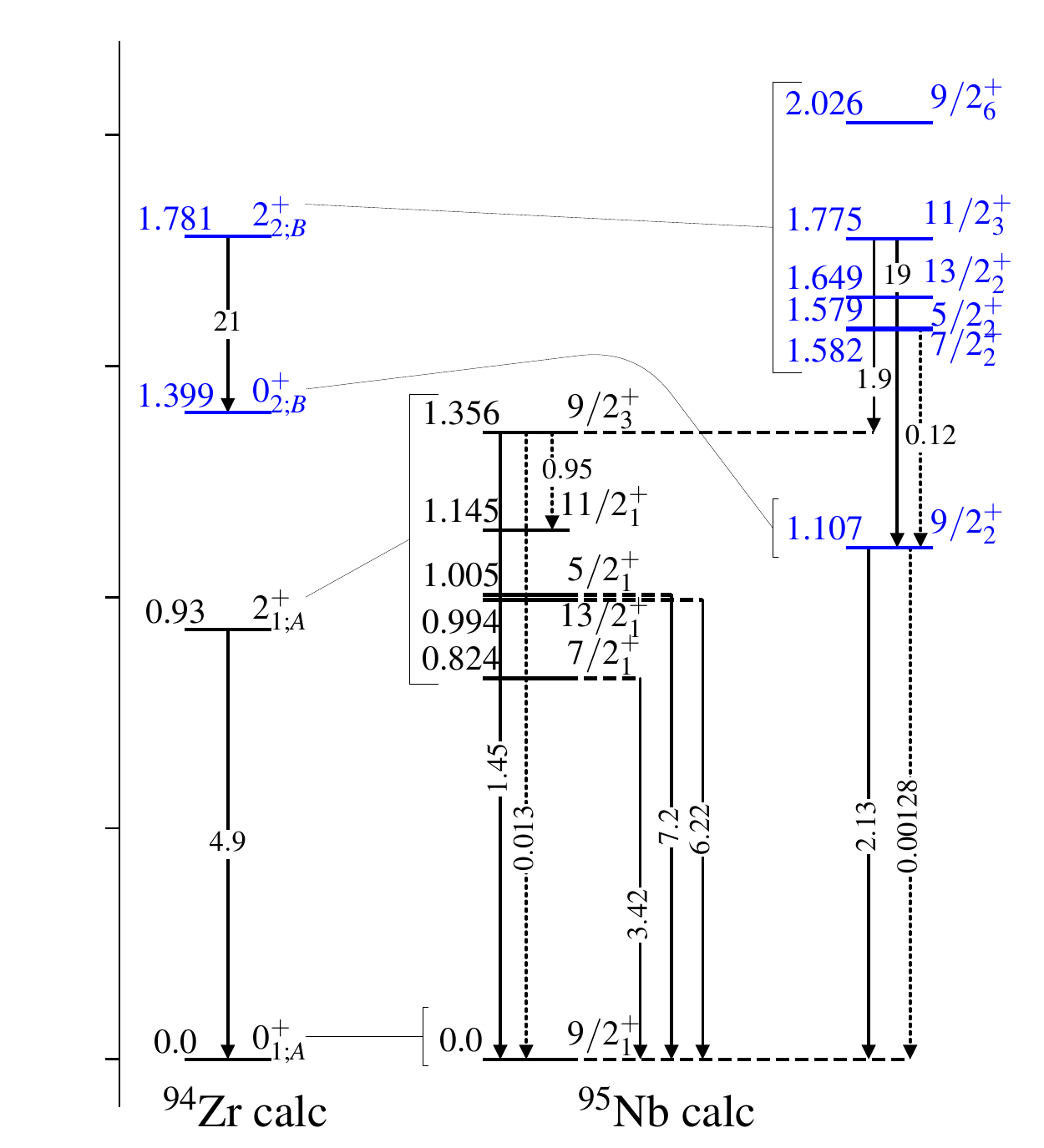}
\caption{Experimental (left) and calculated (right) energy
levels in MeV, and $E2$ (solid arrows) and $M1$
(dashed arrows) transition rates in W.u., for $^{95}$Nb
and $^{94}$Zr. Normal (intruder) states are depicted in 
black (blue).
Lines connecting $L$-levels in $^{94}$Zr 
to sets of $J$-levels in $^{95}$Nb indicate the weak 
coupling $(L\otimes \tfrac{9}{2})J$. 
Data taken from \cite{NDS.111.2555.2010}.
Note that the observed $4^+_{\rm 1;A}$ state in $^{94}$Zr
is considered outside the boson $N \eq 2$ model 
space.\label{fig:95Nb-p}}
\end{figure*}

\begin{figure*}[th!]
\centering
\includegraphics[width=0.49\linewidth]
{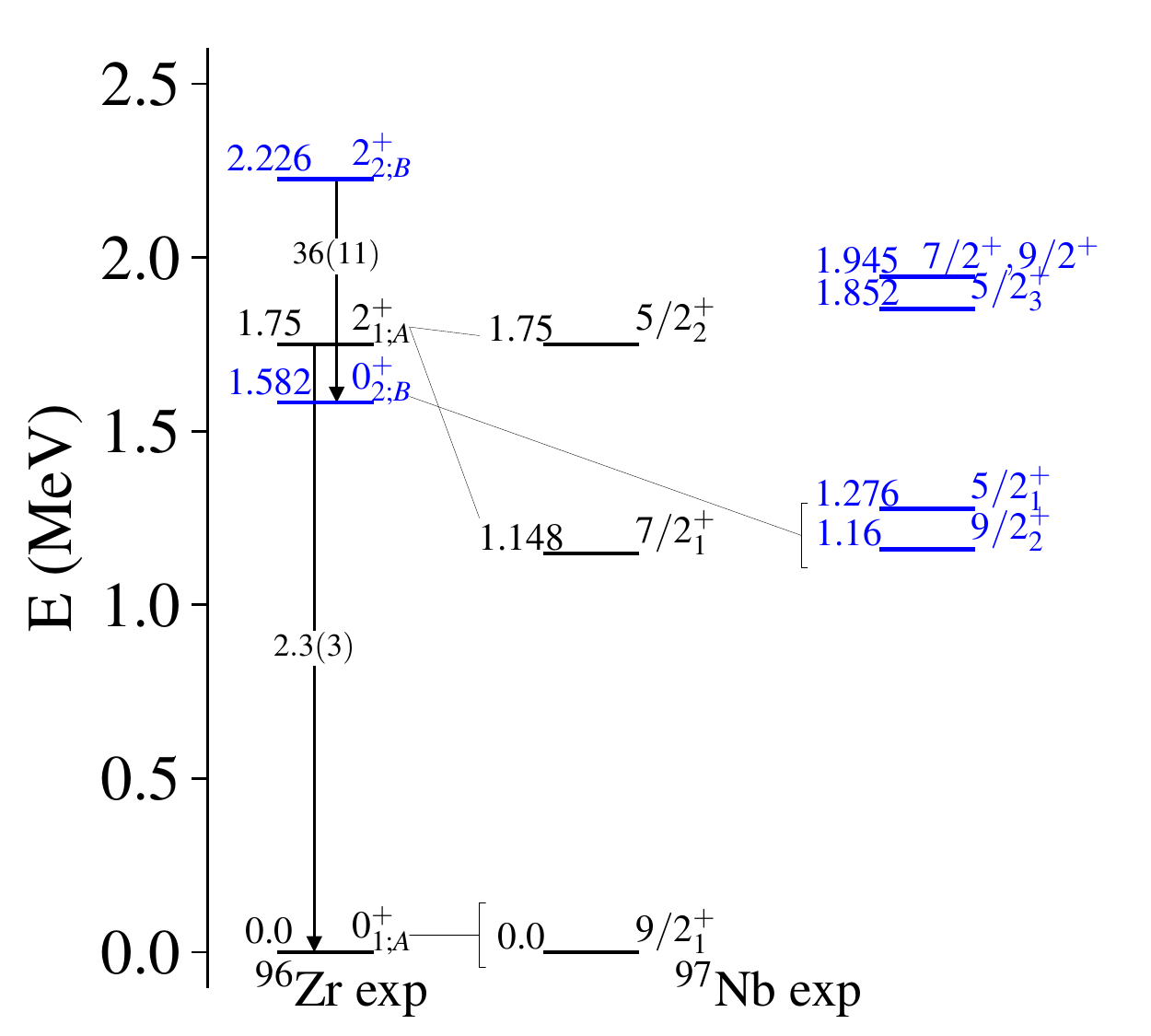}
\includegraphics[width=0.49\linewidth]
{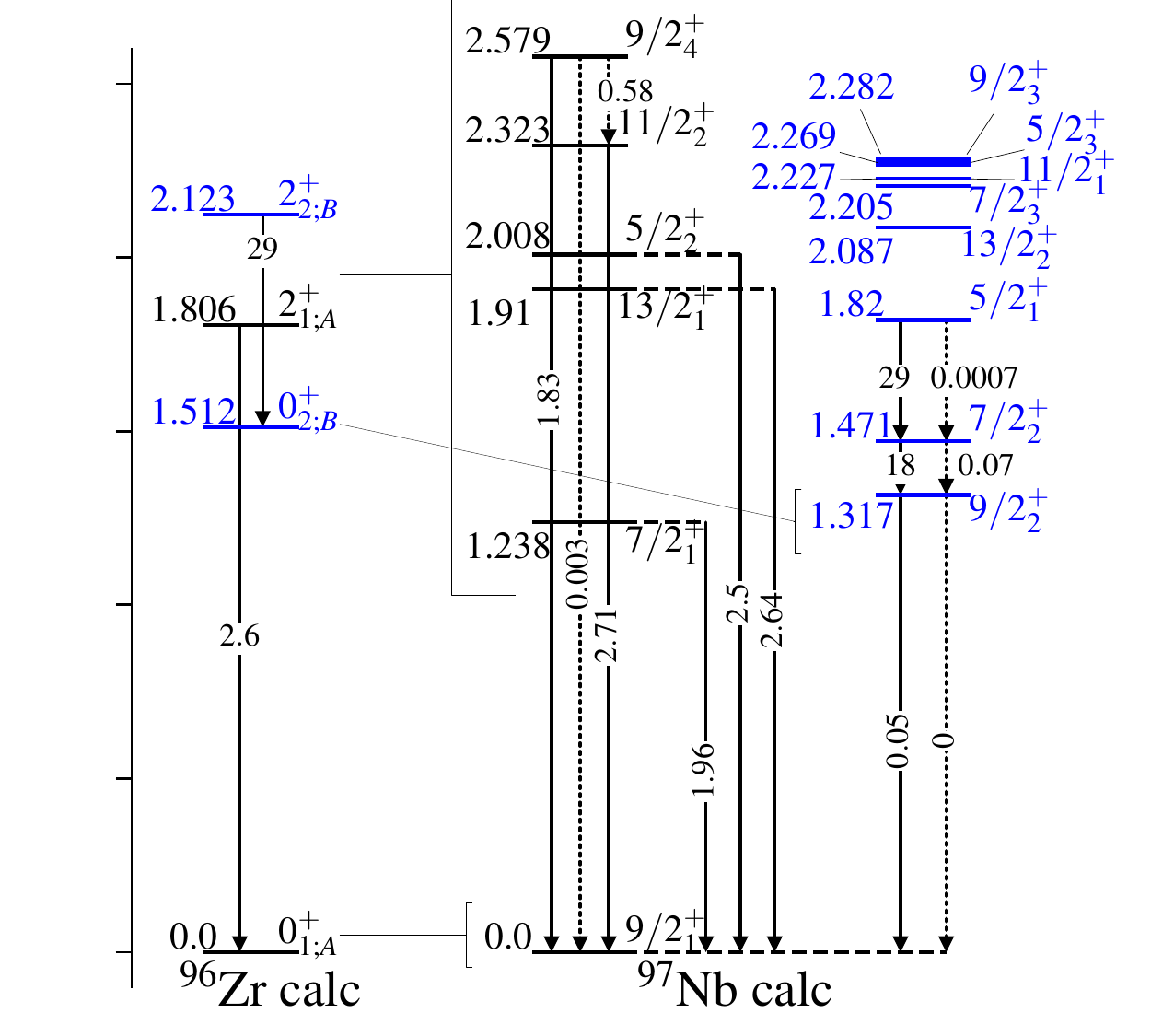}
\caption{Experimental (left) and calculated (right) energy 
levels in MeV, and $E2$ (solid arrows) and $M1$
(dashed arrows) transition rates in W.u., for $^{97}$Nb
and $^{96}$Zr. Normal (intruder) states are depicted in 
black (blue). Lines connecting $L$-levels in $^{96}$Zr 
to sets of $J$-levels in $^{97}$Nb indicate the weak 
coupling $(L\otimes \tfrac{9}{2})J$. 
Data taken from \cite{NDS.111.525.2010}.\label{fig:97Nb-p}}
\end{figure*}

\begin{figure*}[tbh!]
\centering
\includegraphics[width=0.49\linewidth]
{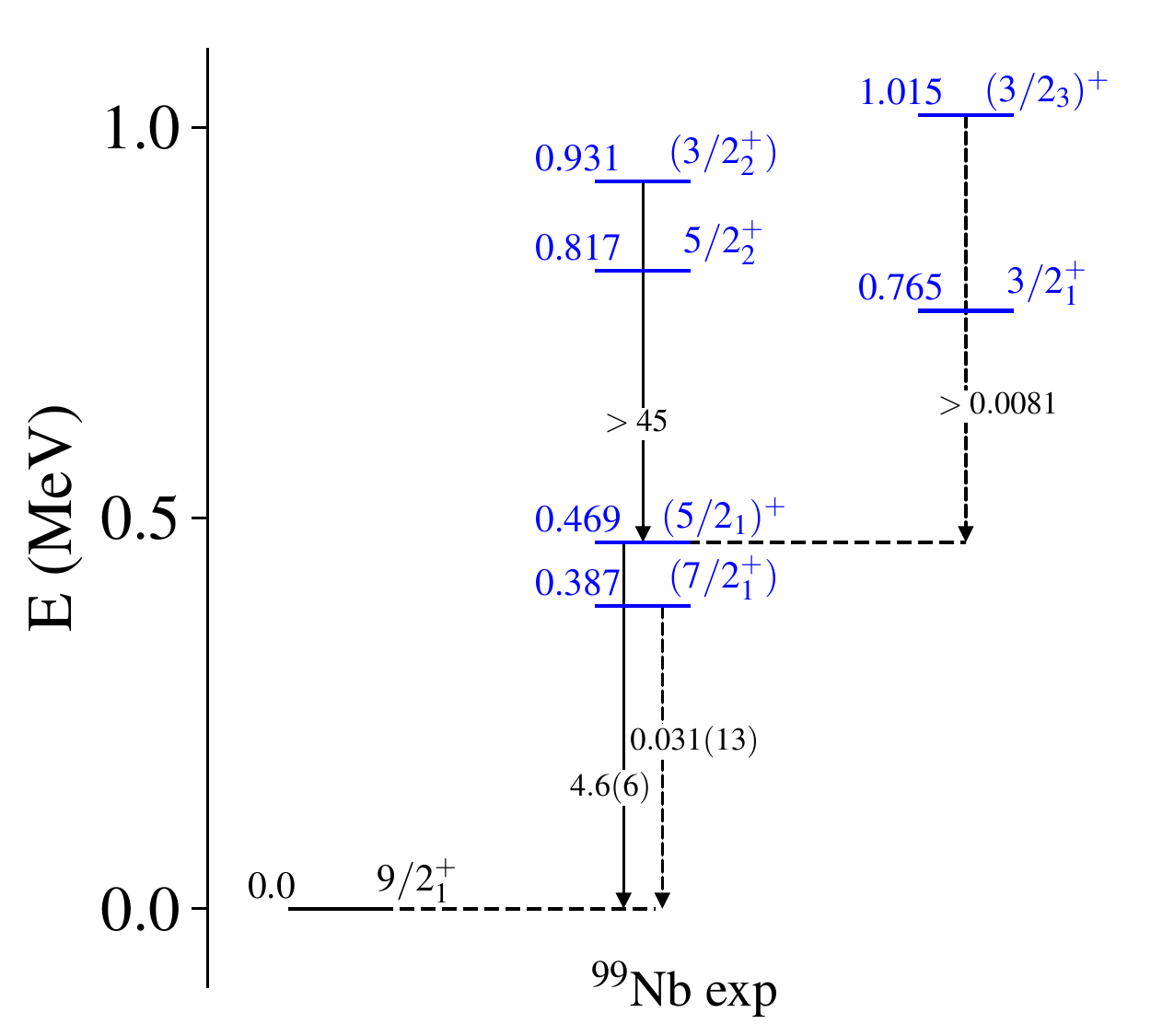}
\includegraphics[width=0.49\linewidth]
{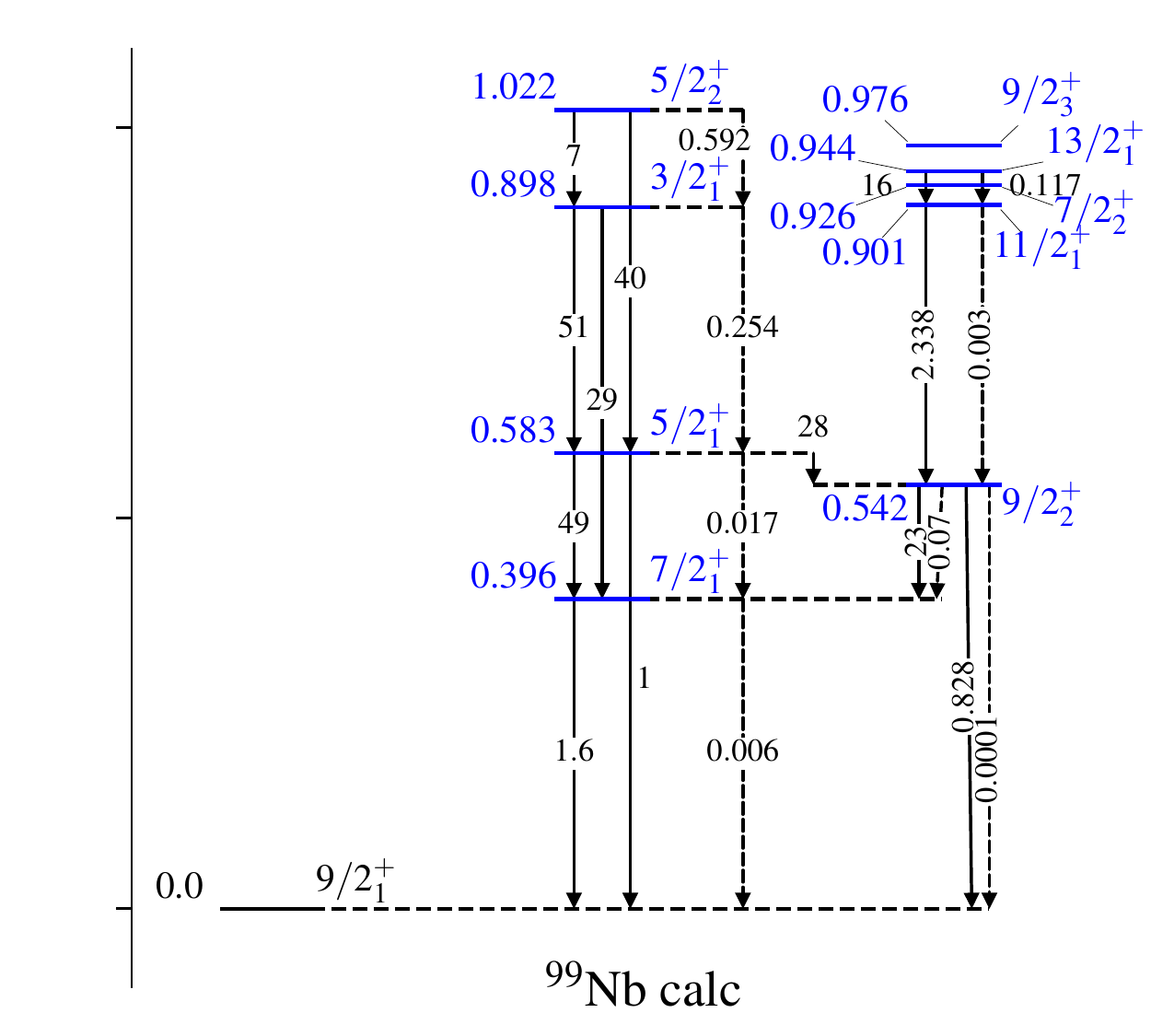}
\caption{Experimental (left) and calculated (right) energy 
levels in MeV, and $E2$ (solid arrows) and $M1$ (dashed 
arrows) transition rates in W.u., for $^{99}$Nb. The 
$9/2^+_1$ state is assigned to the normal A~configuration 
(depicted in black) and the rest of the states to the 
intruder B~configuration (depicted in blue). 
Data taken from \cite{NDS.145.25.2017}. \label{fig:99Nb-p}}
\end{figure*}

\begin{figure*}[th!]
\centering
\includegraphics[width=0.256666667\linewidth]
{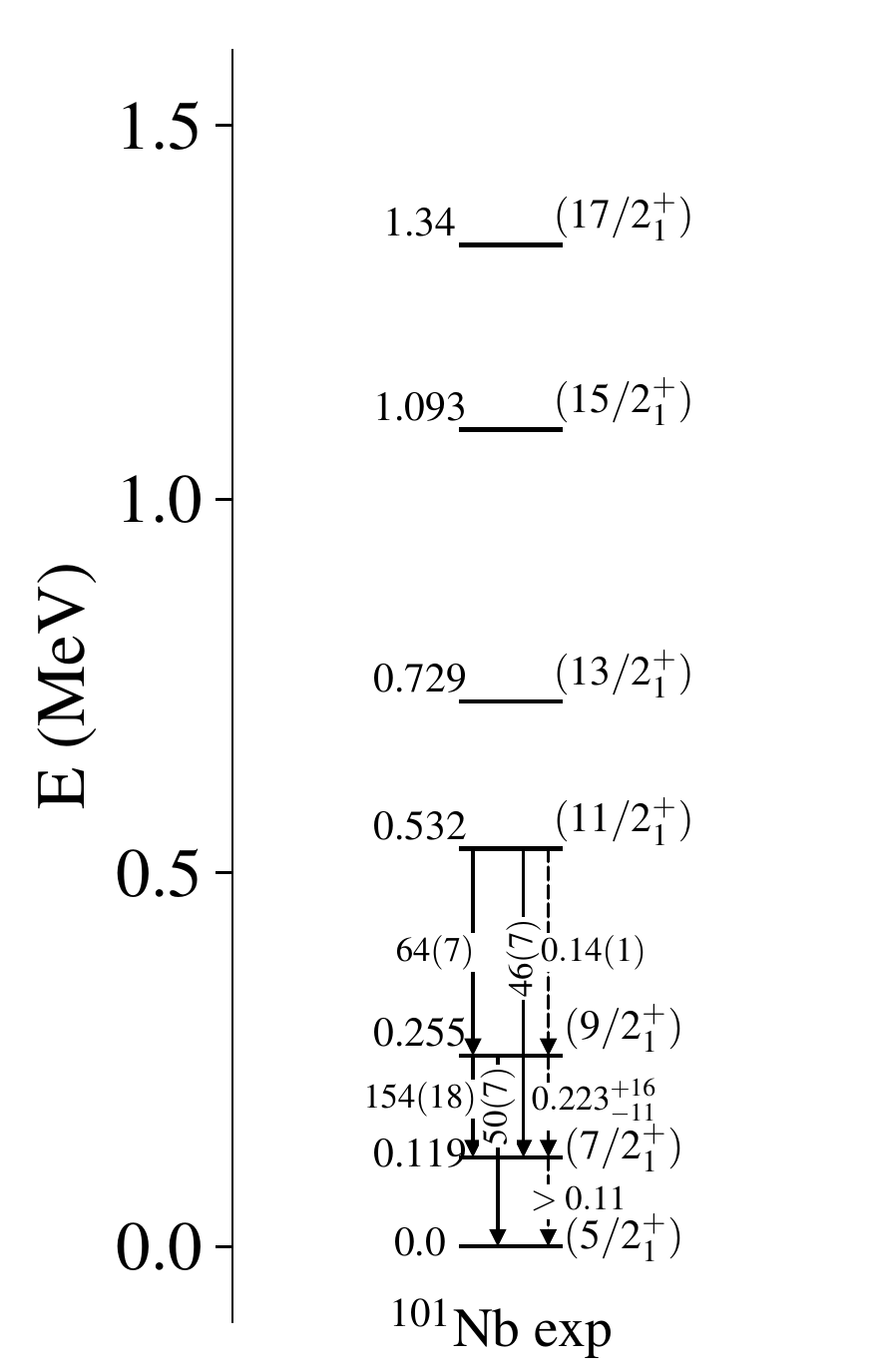}
\includegraphics[width=0.733333333\linewidth]
{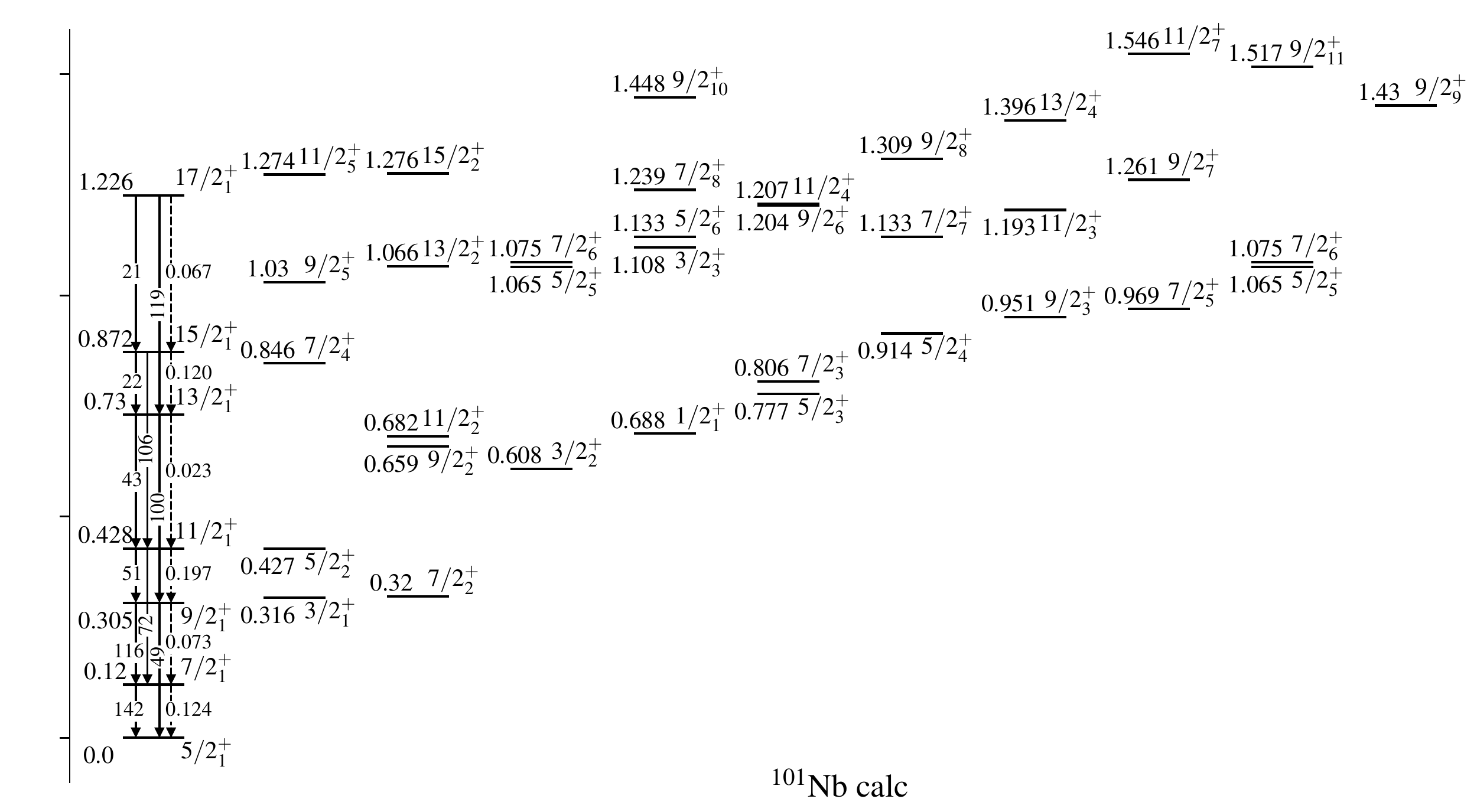}
\caption{Experimental and calculated energy levels in MeV, 
and $E2$ (solid arrows) and $M1$ (dashed arrows) transition 
rates in W.u. for $^{101}$Nb. Shown are states that were 
assigned to different bands in the intruder configuration 
up to $\sim1.5$~MeV (few other states not shown could not 
be associated to a certain band), except the $9/2^+_9$, 
which is normal. Data taken from 
\cite{ensdf,Hagen2017}.\label{fig:101Nb-p}}
\end{figure*}

\begin{figure*}[th!]
\centering
\includegraphics[width=0.2744\linewidth]
{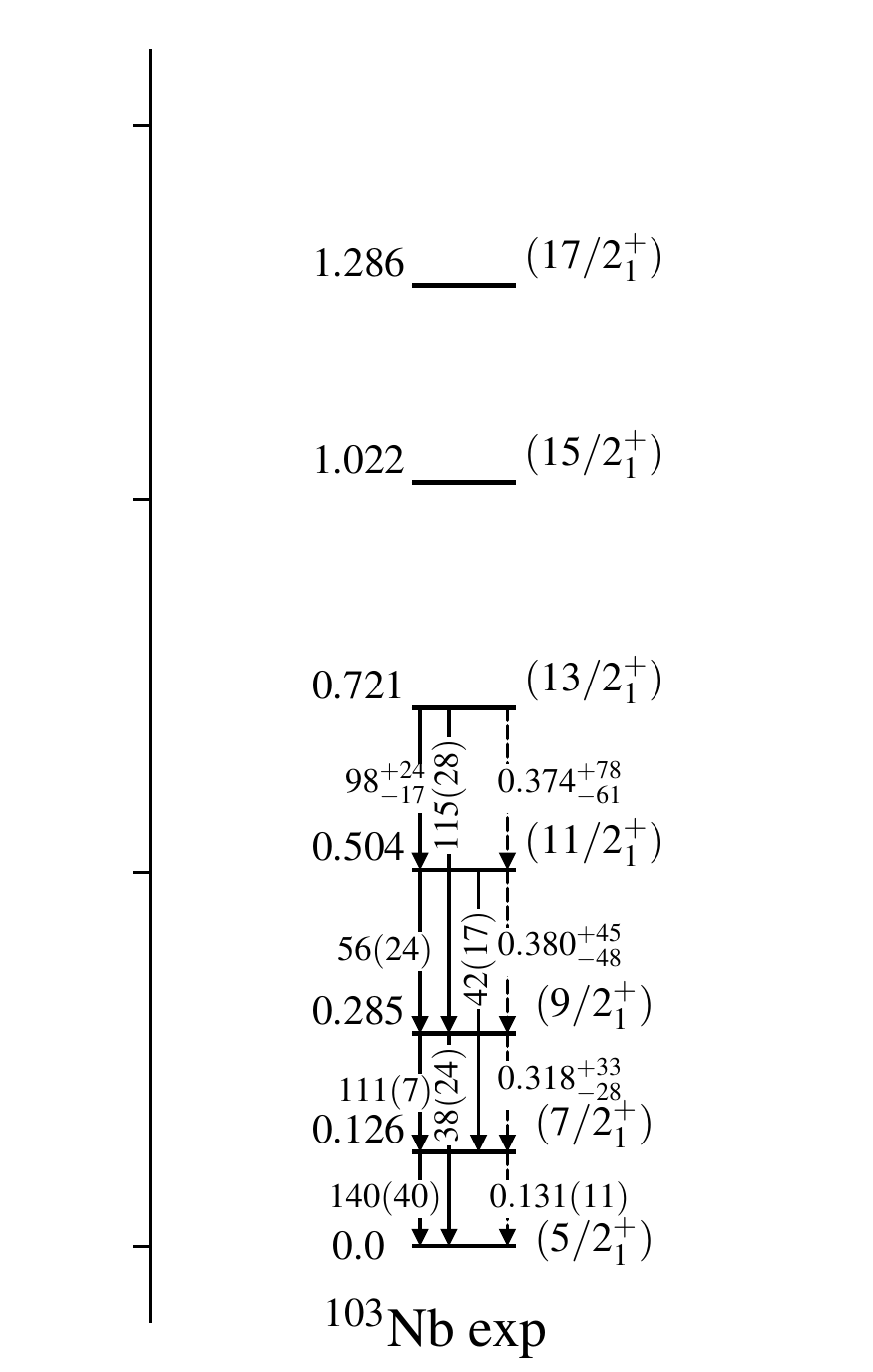}
\includegraphics[width=0.7056\linewidth]
{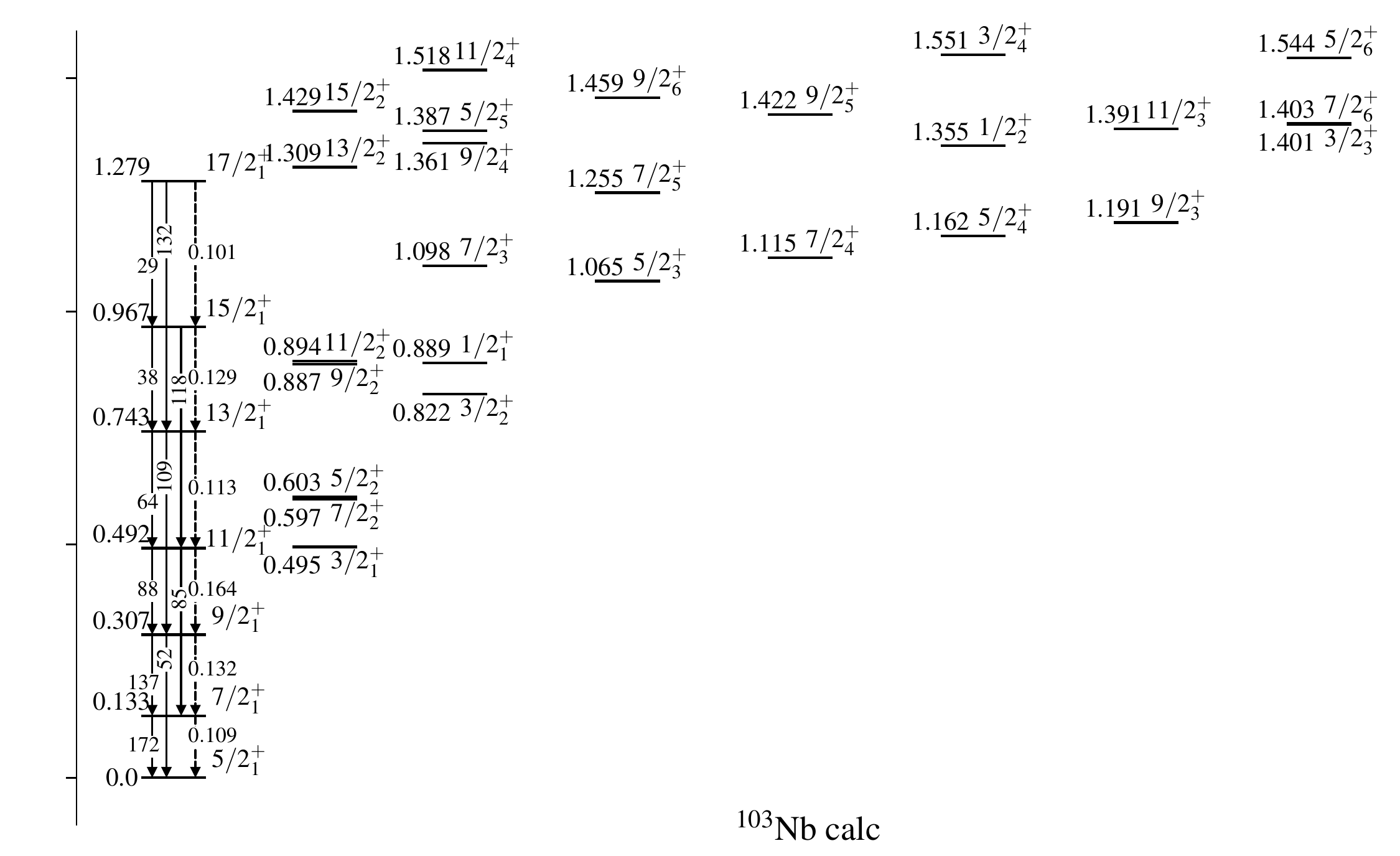}
\caption{Experimental and calculated energy levels in MeV, 
and $E2$ (solid arrows) and $M1$ (dashed arrows) transition 
rates in W.u. for $^{103}$Nb. All states are assigned to 
the intruder configuration. Data taken from 
\cite{NDS.110.2081.2009,Hagen2017}.\label{fig:103Nb-p}}
\end{figure*}

For $^{\text{93--97}}$Nb, shown in 
\cref{fig:93Nb-p,fig:95Nb-p,fig:97Nb-p}, the weak coupling 
between the ground state, $0^+_{1; \rm A}$, of 
$^{\text{92--96}}$Zr and the $\pi(1g_{9/2})$ yields the 
ground state $9/2^+_1$ of $^{\text{93--97}}$Nb.
For the $2^+_{1;\rm A}$, the coupling to the 
$\pi(1g_{9/2})$ yields a quintuplet of states. For 
$^{93}$Nb, the experimental states $7/2^+_1(0.744)$, 
$5/2^+_1(0.809)$, $11/2^+_1(0.979)$, $13/2^+_1(0.950)$, 
$9/2^+_3(1.297)$ (in parentheses are energies in MeV), are 
the members of this quintuplet. They have a CoG 
\eqref{eq:cog} of 0.976~MeV, which is close to the observed 
energy 0.935~MeV of the $2^+_1$ in $^{92}$Zr. 
For $^{95}$Nb, the experimental states $7/2^+_1(0.724)$, 
$5/2^+_1(0.73)$, $13/2^+_1(0.825)$, 
$5/2^+\text{--}13/2^+(1.149)$, $9/2^+_4(1.337)$, are 
members of this quintuplet. They have a CoG \eqref{eq:cog} 
of 0.9776~MeV, which is close to the observed energy 
0.919~MeV of the $2^+_1$ in $^{94}$Zr. 
For both $^{93,95}$Nb the calculation reproduces the 
energies of the quintuplet to a good degree. For $^{97}$Nb 
there is not enough data to clearly assign the existing 
states to a given configuration and this remains to 
be explored.

The $E2$ transitions from the quintuplet states to the 
ground state are comparable in magnitude to the $2^+_{1;\rm 
A}\to0^+_{1;\rm A}$ transition in $\ce{^{\text{92--96}}Zr}$ 
(6.4(6), 4.9(3), 2.3(3)~W.u., respectively). For $^{93}$Nb, 
the calculation reproduces the data to a good degree, 
except for $9/2^+_3$, whose experimental decay 
(1.52(10)~W.u.) is weaker than the others and calculation 
(3.29). For $^{\text{95--97}}$Nb there are no measured $E2$ 
or $M1$ transitions.
The $M1$ transitions of $^{93}$Nb from the quintuplet to 
the ground state are weak, the $7/2^+_1 \to 9/2^+_1$, 
$11/2^+_1 \to 9/2^+_1$, have $B(M1)$ of 0.099(8), 
0.085(6)~W.u., while $M1$ transitions within states of the 
quintuplet are strong, the $5/2^+_1 \to 7/2^+_1$, $9/2^+_3 
\to 11/2^+_1$, $9/2^+_3 \to 7/2^+_1$ have $B(M1)$ of 
0.160(12), 0.73(18), 0.16(3)~W.u., as expected for 
weak coupling to a spherical 
vibrator~\cite{IachelloVanIsackerBook} and which the 
calculation suggests. The situation for calculated $E2$ and 
$M1$ transitions is similar in $^\text{95--97}$Nb.

In \cref{fig:93Nb-p}, one can also identify a 
nontuplet of states, from $(1/2^+_1)$ to $17/2^+_1$, built 
on the $4^+_{1;\rm A}$ state of $^{92}$Zr in the empirical 
spectrum of $^{93}$Nb, with a CoG of 1.591~MeV, close to the
1.495~MeV of $4^+_{1;\rm A}$. This $4^+_{1;\rm A}$ is 
outside the calculated $^{92}$Zr model space (with boson 
number $N_b \eq 1$ for the normal A configuration, see 
\cite{Gavrielov2022} for more details) and as a consequence 
so are the resulting states of $^{93}$Nb. Nevertheless, it 
supports the weak coupling scenario. In \cref{fig:95Nb-p}, 
one can also identify a nontuplet of states, 
built on the $4^+_{1;\rm A}$ state of $^{94}$Zr in 
the experimental spectrum of $^{95}$Nb, however data is 
lacking in order to identify all of them and calculate 
their CoG. 

The IBFM-CM also allows to identify and analyze the 
\textit{intruder B configuration} of $^\text{93--97}$Nb, 
where the weak coupling scenario is also valid. As shown in 
\cref{fig:93Nb-p,fig:95Nb-p,fig:97Nb-p}, the coupling of 
$\pi(1g_{9/2})$ to the $0^+_{2;\rm B}$ state in 
$^\text{92--96}$Zr, yields the excited $9/2^+_2$ state in 
$^\text{93--97}$Nb. For the $2^+_{2;\rm B}$ state 
the coupling yields another quintuplet of states. 
For $^{93}$Nb, it is the experimental $7/2^+_3(1.484)$, 
$5/2^+_3(1.666)$, $9/2^+_4(1.683)$, $13/2^+_2(1.686)$, 
$11/2^+_3(1.969)$, that are reproduced to a good degree by 
the calculation 
and whose experimental CoG is 1.719~MeV, a bit lower than 
the energy 1.847~MeV of the $2^+_{2;\rm B}$ of $^{92}$Zr. 
For $^{95}$Nb, it is the $7/2^+_3(1.704)$, 
$(5/2\text{--}13/2)^+(1.686)$, $5/2^+_2(1.813)$, 
$(5/2^+\text{--}13/2)^+(1.969)$, $9/2^+_4(1.903)$, 
that are reproduced to a good degree by the calculation 
and whose experimental CoG is 1.803~MeV, a bit higher than 
the energy 1.671~MeV of the $2^+_{2;\rm B}$ of $^{94}$Zr.

It is interesting to note that the energy difference from 
the $0^+_{2;\rm B}$ and the $9/2^+_2$ that is associated to 
it, $E(0^+_{2;\rm B})-E(9/2^+_2)$, becomes larger when 
going from $^{93,95}$Nb (where the difference is 0.3, 
0.212~MeV, respectively) to $^{97}$Nb (where the difference 
is 0.422~MeV), suggesting the additional fermion increases 
collectivity, which reduces the energy of the $9/2^+_2$ 
state compared to the $0^+_{2;\rm B}$ state of 
$^\text{92--96}$Zr.

For $^{93}$Nb, the observed $B(E2;9/2^+_2 \to 9/2^+_1) \eq 
1.03(9)$~W.u, is close to the calculated value 0.85~W.u., 
but is smaller than the observed value 
$B(E2;9/2^+_3\to9/2^+_1) \eq 1.52(10)$~W.u, suggesting that 
the $9/2^+_2$ is associated with the B~configuration, but 
that the mixing between these states is possibly stronger 
than predicted.
This is contrary to previous works 
\cite{VanHeerden1973,Orce2010} that assigned the $9/2^+_2$ 
as part of the configuration A quintuplet. 
A similar situation occurs with $11/2^+_3$ state. The 
observed $B(E2;11/2^+_3 \to 9/2^+_2) \eq 6(3)$ and 
$B(E2;11/2^+_3 \to 9/2^+_3) \eq 5(2)$~W.u. suggest a 
fragmentation of the $11/2^+_3$ compared to the calculated 
values of 16 and 0.6~W.u., respectively. The observed value 
of $B(E2;11/2^+_3 \to 11/2^+_1) \eq 21(7)$~W.u., which is 
calculated to be 0.1~W.u., suggests this fragmentation is 
possibly due to stronger mixing between the $11/2^+$ 
states, also due to the stronger $B(E2;11/2^+_2 \to 
7/2^+_1) \eq 17(7)$. The strong $B(E2;5/2^+_3 \to 7/2^+_1) 
\eq 90(35)$ \cite{Orce2010}, calculated to be weak (0.1) 
might suggest stronger mixing for either the $5/2^+_3$ or 
$7/2^+_1$ states.
\begin{figure}[th!]
\centering
\begin{overpic}[width=1\linewidth]{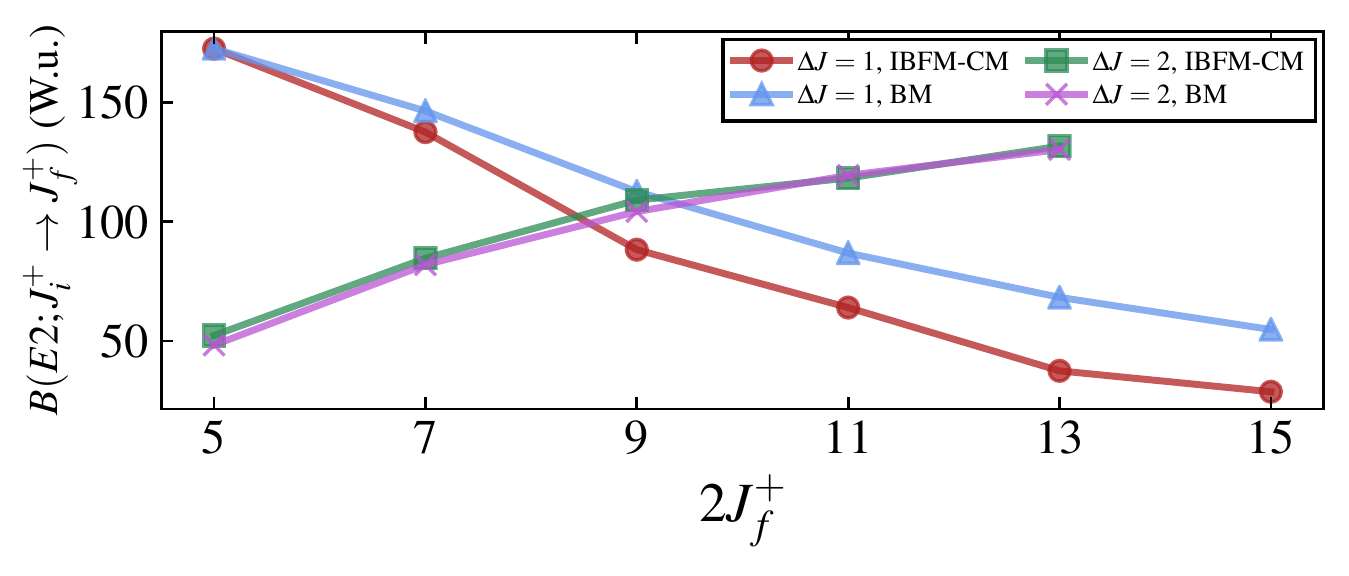}
\put (70,25) {\normalsize $K^\pi \eq 5/2^+$~(a)}
\end{overpic}\\
\begin{overpic}[width=1\linewidth]
{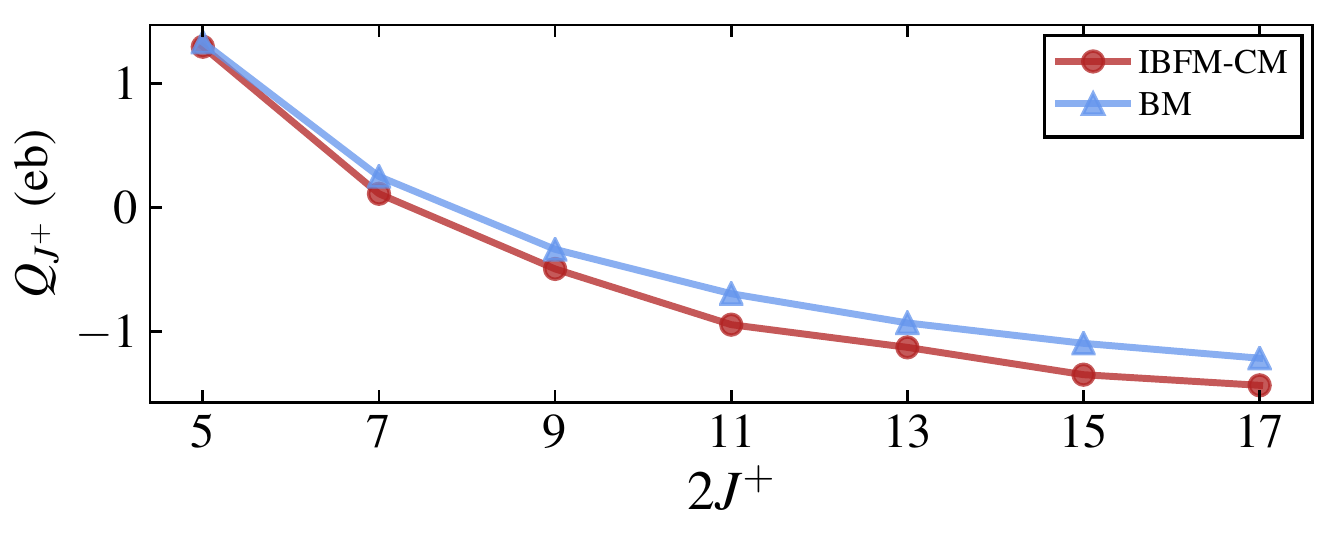}
\put (70,25) {\normalsize $K^\pi \eq 5/2^+$~(b)}
\end{overpic}\\
\caption{Comparison between the present calculation and the 
Bohr and Mottelson model (BM). (a) $E2$ transition rates in 
W.u. between members of the $K^\pi \eq 5/2^+$ band of 
$^{103}$Nb calculated in this work and using the collective 
model (BM), \cref{eq:GCM-e2}, with $\Delta J \eq J_f - 
J_i$. (b) Quadrupole moments in eb for members of the 
$K^\pi \eq 5/2^+$ band in $^{103}$Nb calculated in this 
work and using the collective model, \cref{eq:GCM-q}. 
\label{fig:BM-E2-quad-p}}
\end{figure}
\begin{figure*}[h!]
\centering
\includegraphics[width=0.49\linewidth]
{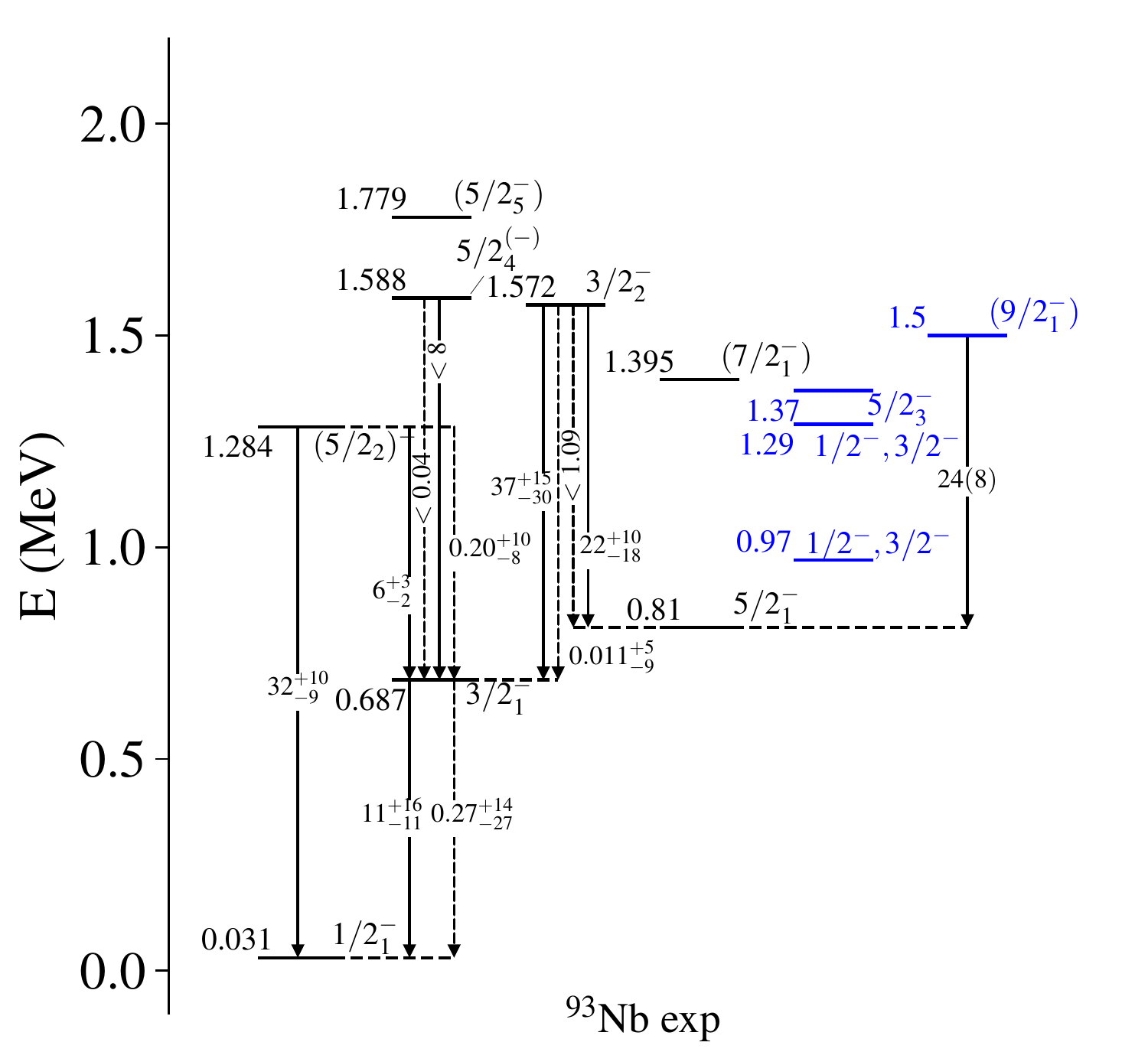}
\includegraphics[width=0.49\linewidth]
{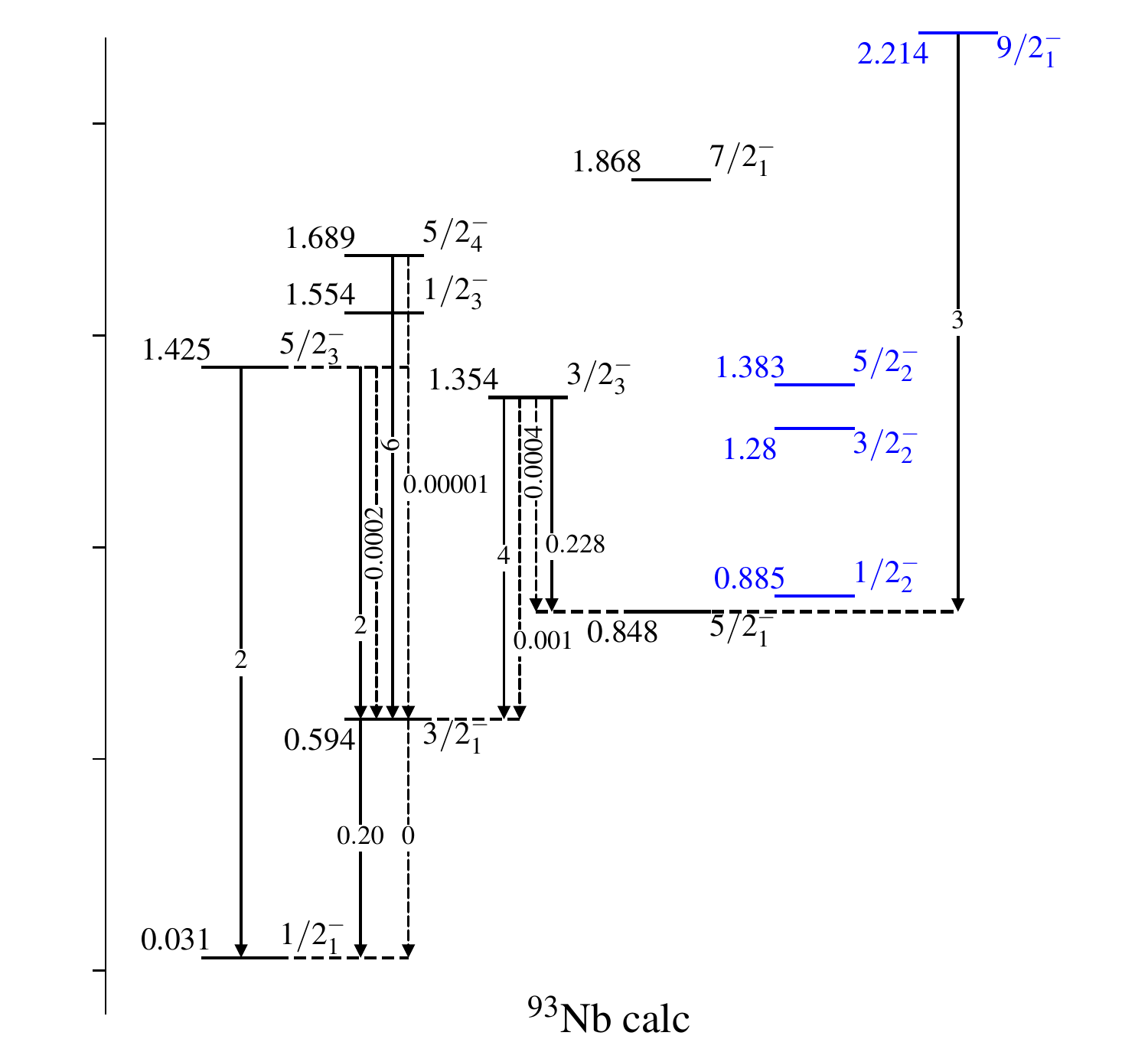}
\caption{Experimental (left) and calculated (right) energy 
levels in MeV, and $E2$ (solid arrows) and $M1$ (dashed 
arrows) transition rates in W.u., for $^{93}$Nb. Data taken 
from \cite{NDS.112.1163.2011,Orce2010}. \label{fig:93Nb-m}}
\end{figure*}
\subsubsection{The $^\text{99-103}$Nb region: 
IQPT and strong coupling}\label{sec:99-103nb_p}
\begin{figure*}
\centering
\includegraphics[width=0.49\linewidth]
{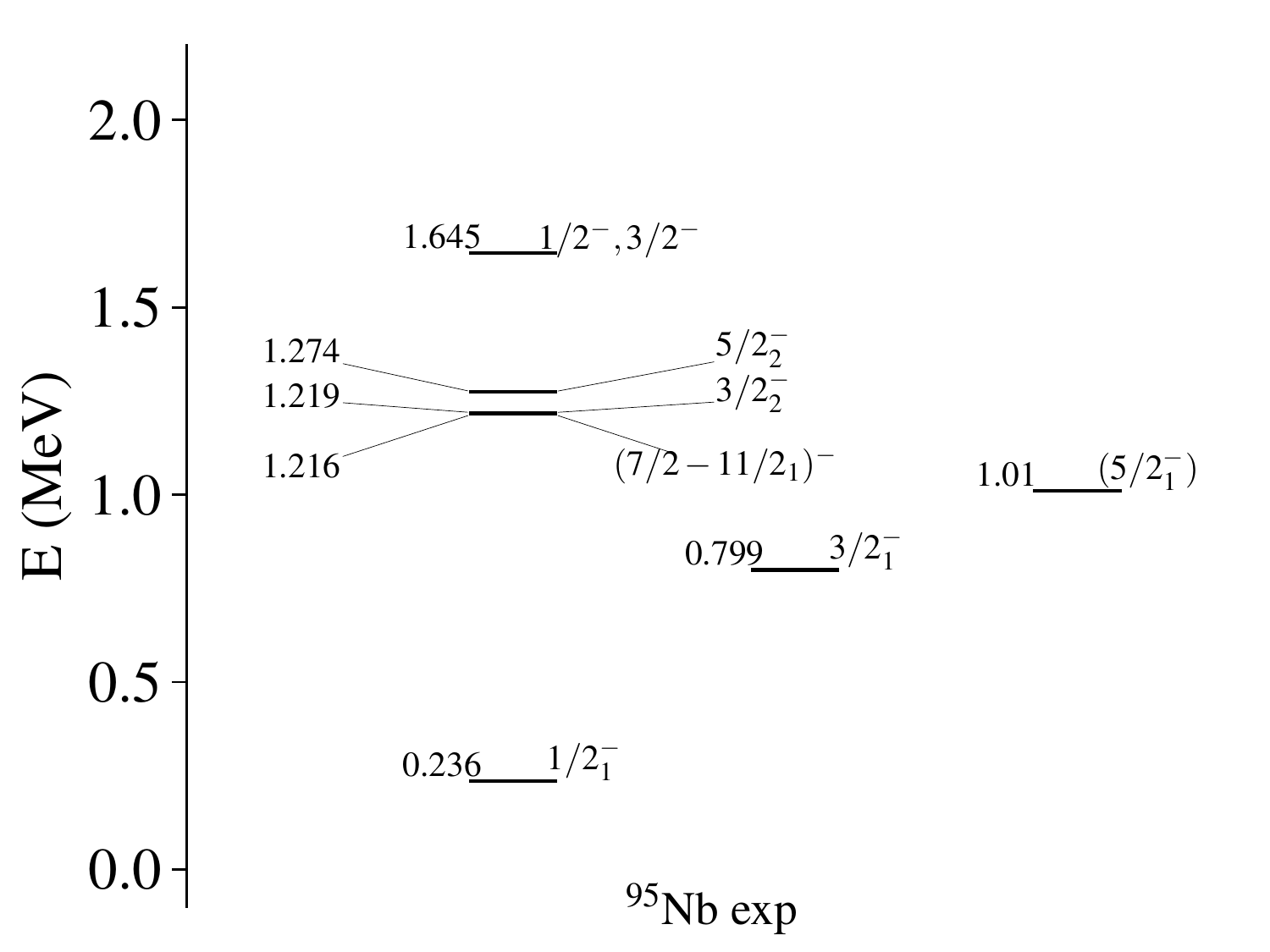}
\includegraphics[width=0.49\linewidth]
{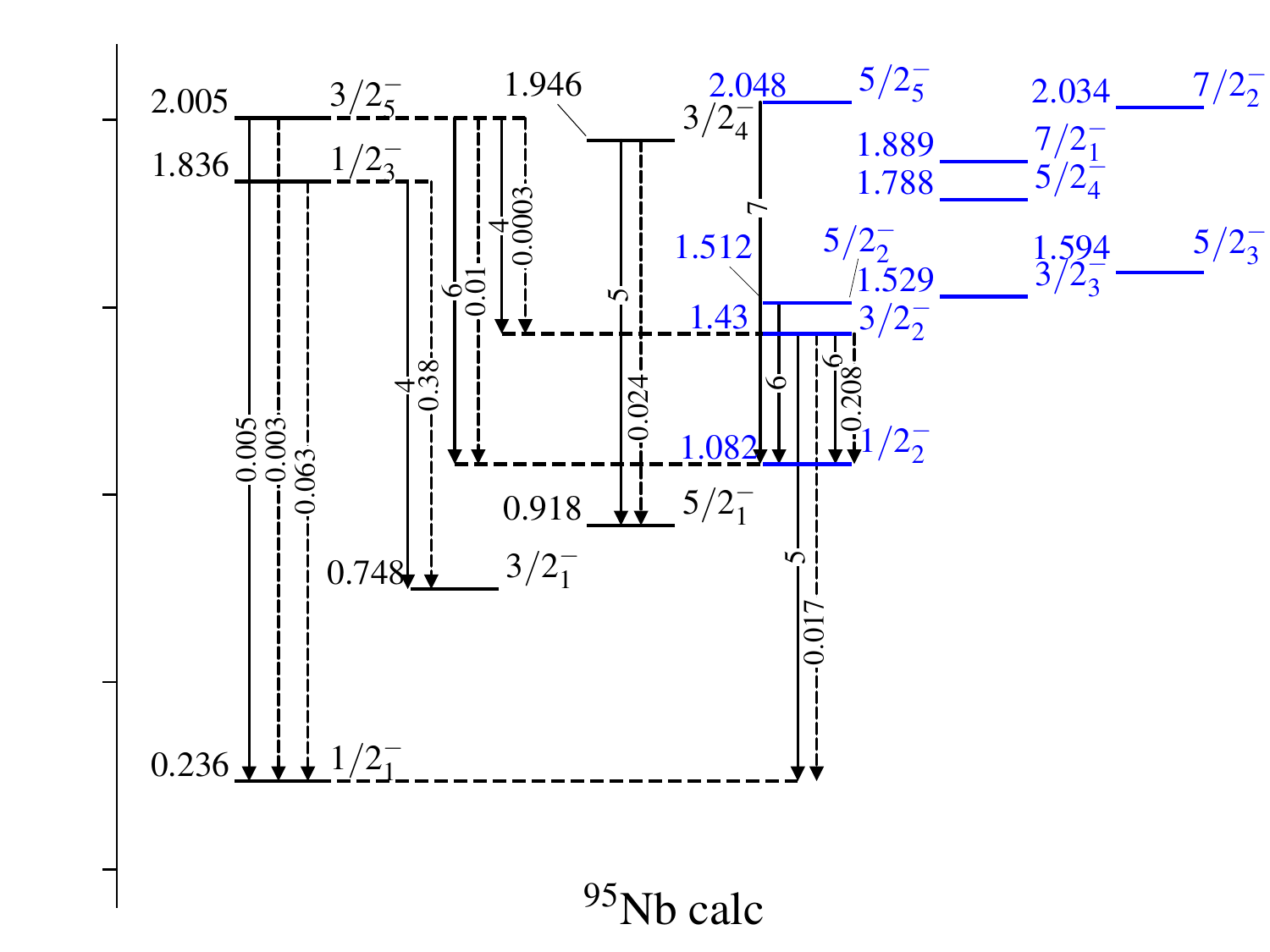}
\caption{Experimental (left) and calculated (right) energy 
levels in MeV, and $E2$ (solid arrows) and $M1$ (dashed 
arrows) transition rates in W.u., for $^{95}$Nb.  Normal 
(intruder) states are depicted in black (blue). Data taken 
from \cite{NDS.111.2555.2010}. \label{fig:95Nb-m}}
\end{figure*}
\begin{figure*}
\centering
\includegraphics[width=0.49\linewidth]
{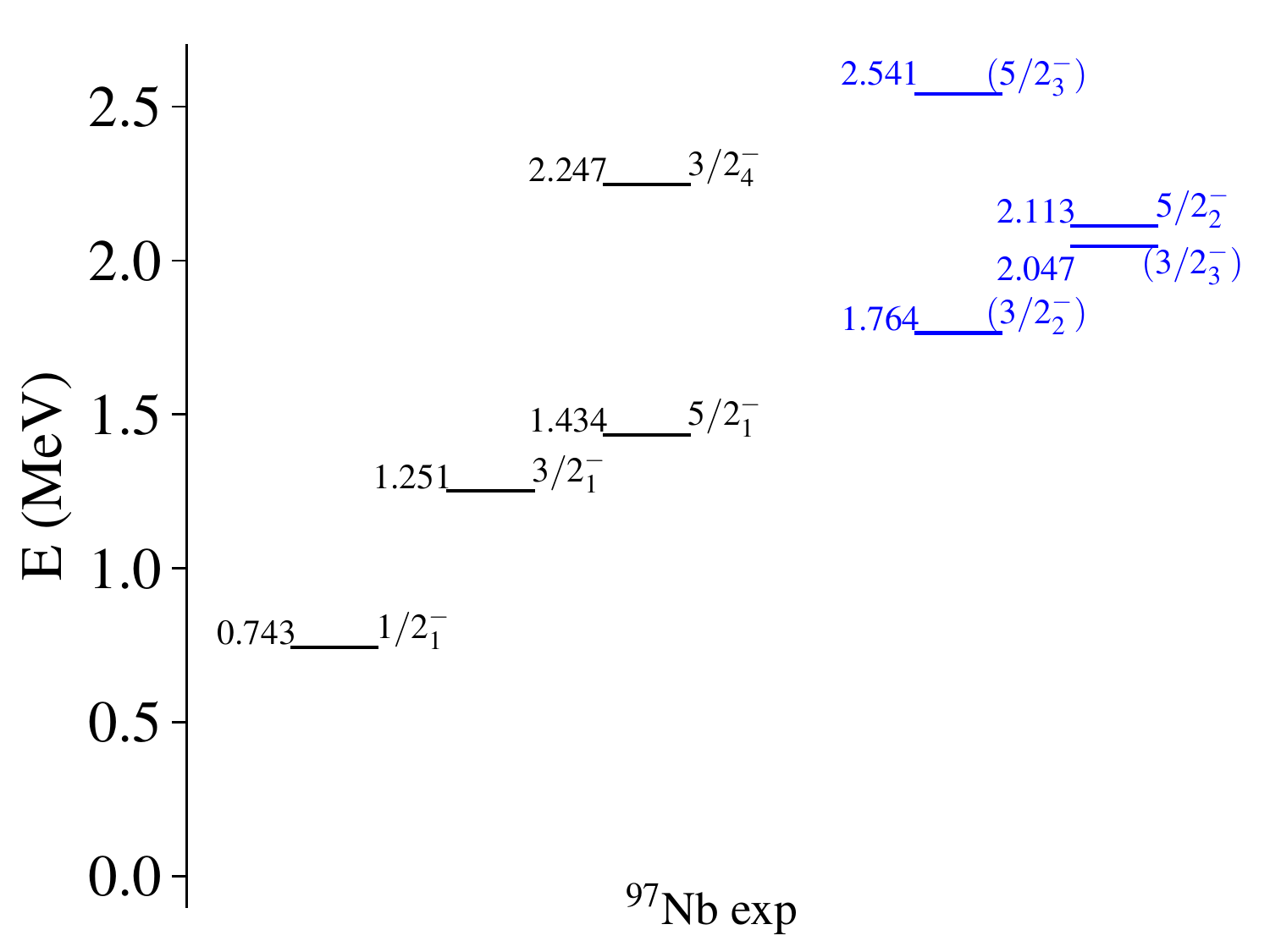}
\includegraphics[width=0.49\linewidth]
{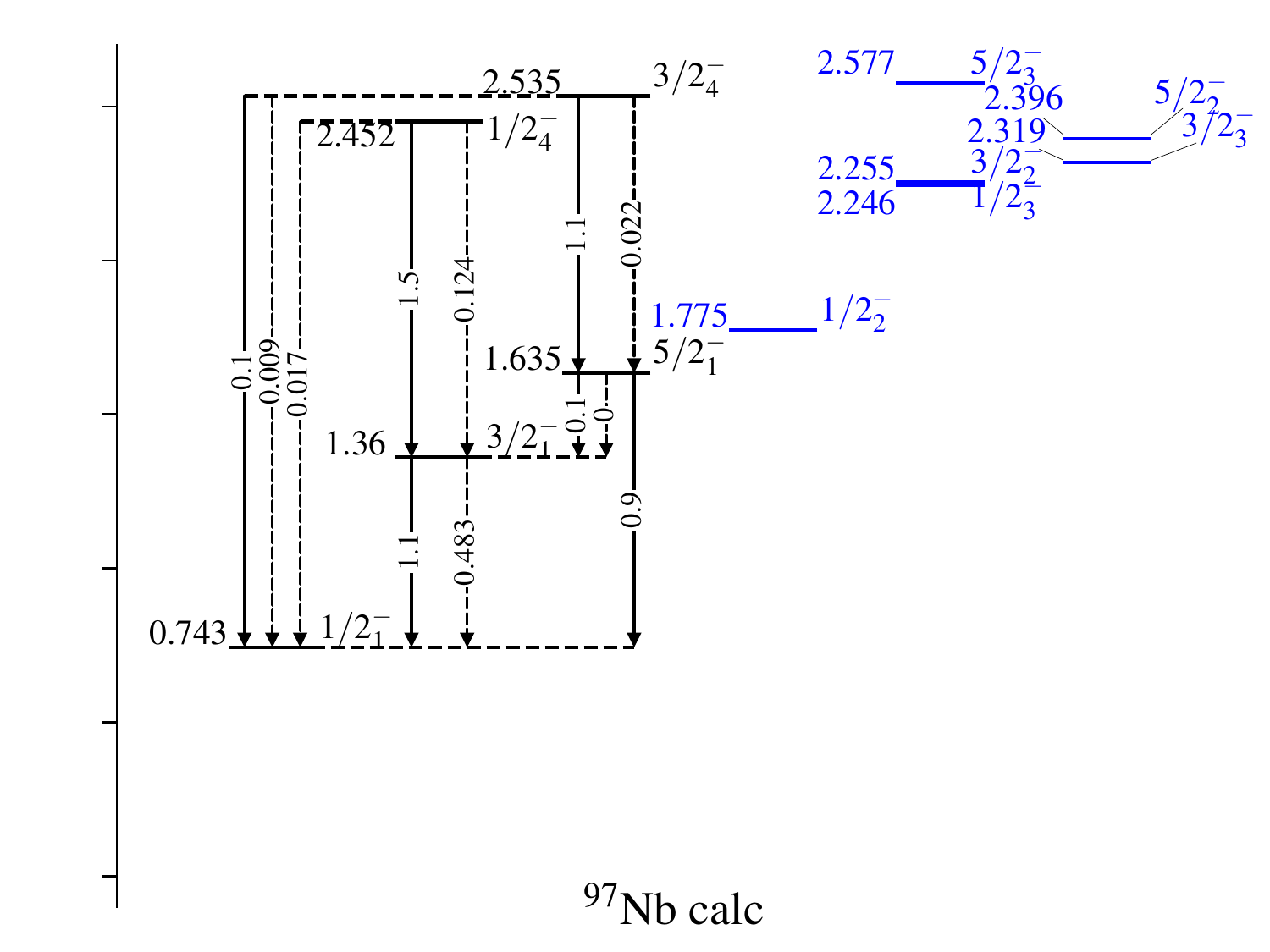}
\caption{Experimental (left) and calculated (right) energy 
levels in MeV, and $E2$ (solid arrows) and $M1$ (dashed 
arrows) transition rates in W.u., for $^{97}$Nb. Normal 
(intruder) states are depicted in black (blue). Data taken 
from \cite{NDS.111.525.2010}. \label{fig:97Nb-m}}
\end{figure*}
For $^{99}$Nb, shown in \cref{fig:99Nb-p}, the ground state 
is a result of the weak coupling between the $0^+_{1; \rm 
A}$ of the $^{98}$Zr-core and the $\pi(1g_{9/2})$ orbit. 
The higher lying states, however, are all intruder. This is 
inline with the case of $^{98}$Zr, where some of the 
configuration B states lie below the first excited $2^+$ of 
configuration A. For example, although mixing is stronger, 
the calculated $7/2^+_1$ has a large $n_d \eq 1$ component 
($\gtrsim60\%$), which associates it as part of the 
quintuplet that originates from the coupling of the 
$\pi(1g_{9/2})$ with the $2^+_{1; \rm B}$ of $^{98}$Zr. The 
higher lying calculated states have larger $n_d$ mixing. 
The calculated $7/2^+_1$ is lower in energy than the 
$9/2^+_2$, which is mainly composed of the coupling between 
the $\pi(1g_{9/2})$ and the $0^+_{2; \rm B}$ of $^{98}$Zr. 
This is an example for the onset of deformation that has 
been identified in $^{98}$Zr 
\cite{Gavrielov2019,Gavrielov2022}.
The few measured $E2$ and $M1$ transitions are reproduced 
qualitatively for the $B(E2;5/2^+_1 \to 9/2^+_1) \eq 
4.6(6)$~W.u. [1], $B(M1;7/2^+_1 \to 9/2^+_1) \eq 
0.031(13)$~W.u. [0.006]  and $B(E2;3/2^+_2 \to 5/2^+_1) > 
45$~W.u. [51], where in square brackets are the calculated 
values.

For $^\text{101--103}$Nb, shown in 
\cref{fig:101Nb-p,fig:103Nb-p}, the yrast states belong to 
the intruder B configuration and are arranged in a 
$K^P\eq5/2^+$ rotational band, with an 
established Nilsson model assignment 
$5/2^+[422]$~\cite{Hotchkis1991}.
The band members can be interpreted in the
strong coupling scheme, where a particle is coupled to an
axially-deformed core. The indicated states are obtained
by coupling the $\pi(1g_{9/2})$ state to the ground band
($L=0^+_1,2^+_1,4^+_1,\ldots$) of $^\text{100--102}$Zr, 
which are all part of the intruder B configuration.
For $^{103}$Nb, the calculation reproduces well the 
observed particle-rotor splitting, with a moment 
of inertia \cref{eq:moment_iner} $B \eq 0.018$~MeV. For 
$^{101}$Nb, the experimental levels follow a less-rotational
pattern. The experimental $E2$ and $M1$ transitions within 
the band of both $^\text{101--103}$Nb are reproduced well 
by the calculation. 
In \cref{fig:BM-E2-quad-p}, the trend in $E2$ transitions 
and quadrupole moments as a function 
of angular momentum $J$, seems to be very similar with 
that of the geometric collective model, 
\cref{eq:GCM-e2,eq:GCM-q}. The the trend of the $M1$ 
transitions and magnetic moments is less similar as these 
observables are less collective in nature and are strongly 
affected by the single-particle character of the wave 
function.

Besides the calculated ground state band, there are 
different $K^\pi$ bands for which states are grouped 
together according to strong $E2$ transitions between them. 
The right-most-one of them in \cref{fig:101Nb-p} of 
$^{101}$Nb is the $9/2^+_9$ state, which is spherical with 
about 76\% for the $n_d \eq 0$ component. Therefore, one 
can observe the change of configuration in the ground 
state, from A to B (Type~II QPT), and also a change in the 
B configuration from spherical spectrum, \mbox{beginning at 
$^{99}$Nb, to deformed in $^{101}$Nb (Type~I QPT)}.

Altogether, we see an evolution of structure
from weak coupling of a spherical shape in $^{93}$Nb,
to strong coupling of a deformed shape in $^{103}$Nb.
Such shape-changes within the B configuration (Type~I QPT),
superimposed on an abrupt configuration crossing (Type-II 
QPT), are the key defining feature of intertwined QPTs 
(IQPTs). Interestingly, the intricate IQPTs scenario, 
originally observed in the even-even Zr 
isotopes~\cite{Gavrielov2019,Gavrielov2022},
persists in the adjacent odd-even Nb isotopes.

\subsection{Negative-parity 
states}\label{sec:negative-states}
For the negative parity states, the individual isotopes 
are divided to two regions: a weak coupling region for 
$^{93-97}$Nb and the IQPT region for $^{99-103}$Nb, which 
also incorporates strong coupling.

For the region of $^\text{93--97}$Nb the calculation is 
compared to the experimental levels in 
\cref{fig:93Nb-m,fig:95Nb-m,fig:97Nb-m}. For each 
isotope, the lowest levels with $J^\pi\eq1/2^-,3/2^-,5/2^-$ 
in each configuration are associated with the 
single-particle orbits $j\eq\pi(2p_{1/2}), \pi(2p_{3/2}), 
\pi(1f_{5/2})$.

For the region of $^\text{99--103}$Nb the calculation is 
compared to the experimental levels in 
\cref{fig:99Nb-m,fig:101Nb-m,fig:103Nb-m}. For each isotope 
the spectrum exhibits rotational bands that belong to the 
intruder B configuration, except in $^{99}$Nb where the 
normal A configuration can be identified.

\begin{figure}
\centering
\includegraphics[width=1\linewidth]{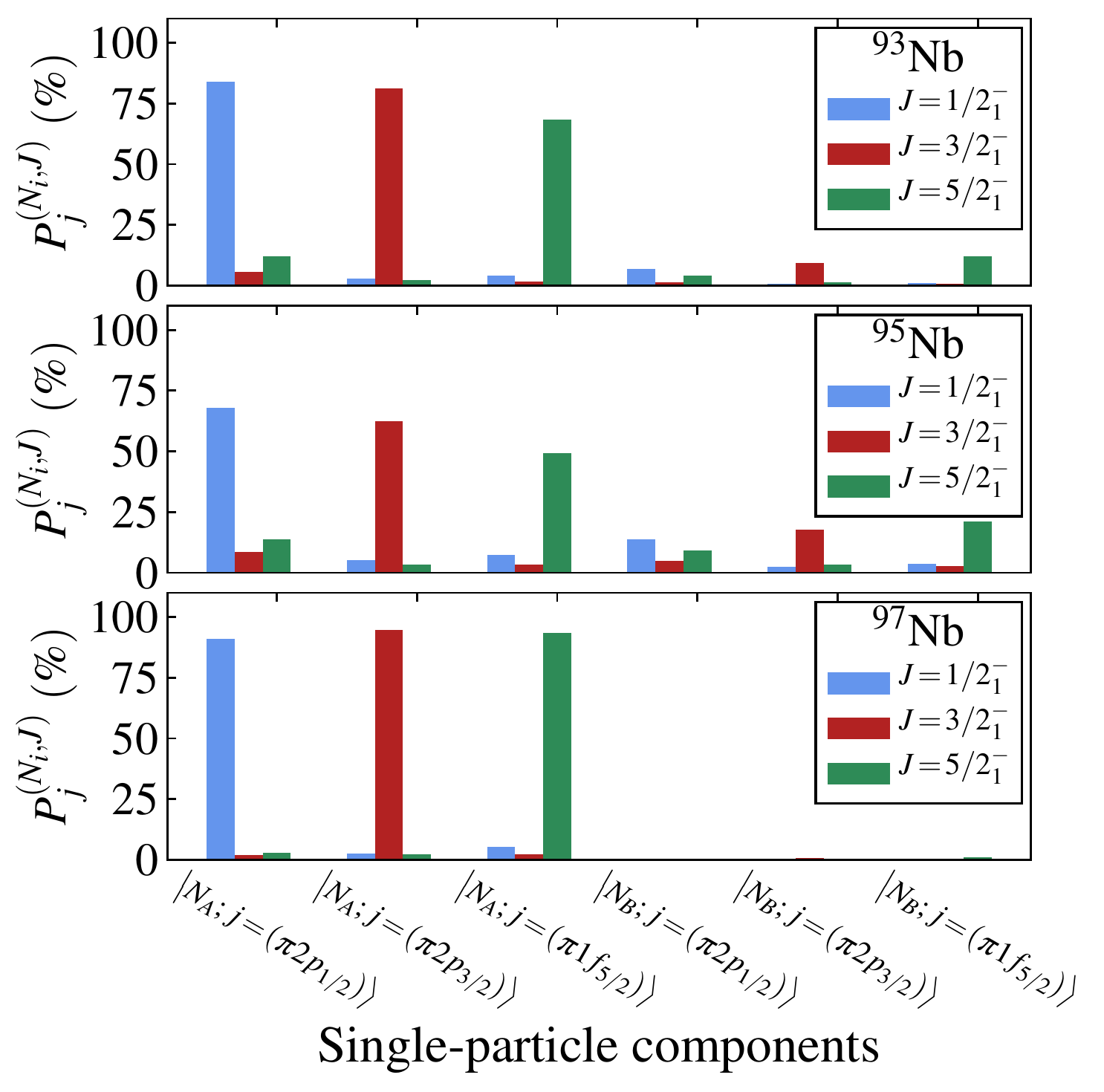}
\caption{Percentage of the single-particle components of 
the normal A and intruder B configurations, 
\cref{eq:prob_spe}, for the calculated $1/2^-_1, 3/2^-_1, 
5/2^-_1$ states of $^\text{93--97}$Nb 
isotopes.\label{fig:93-97Nb-spe}}
\end{figure}

\subsubsection{The $^\text{93--97}$Nb region: weak 
coupling} \label{sec:93-97nb_m}
As shown in \cref{fig:93Nb-m,fig:95Nb-m,fig:97Nb-m}, the 
levels with $J^\pi \eq 1/2^-_1, 3/2^-_1, 5/2^-_1$ in 
$^\text{93--97}$Nb have a quasi-particle character. They 
originate from the coupling of the $0^+_{1; \rm A}$ of the 
adjacent $^\text{92--96}$Zr isotopes with the 
$\pi(2p_{1/2})$, $\pi(2p_{3/2})$, $\pi(1f_{5/2})$ 
orbits, and are clearly identified in the calculation with 
a good agreement to the data. On top of each of them are 
other levels that have a large component $P^{(N_i,J)}_{j}$ 
of \cref{eq:prob_spe}, with the same single-particle 
$j$-orbit, where $i \eq \text{A or B}$ and $J$ is the total 
angular momentum. However, these higher lying states are 
more mixed between the different configurations.
For $^{93}$Nb in \cref{fig:93Nb-m} the single 
quasi-particle levels that are associated with 
configuration~B can be identified in the 
experimental spectrum, the $(1/2^-,3/2^-)$, $(1/2^-,3/2^-)$ 
and $5/2^-_3$ state at energy 0.97, 1.29 and 1.37~MeV, 
respectively, and are reproduced well by the calculation, 
depicted in blue in \cref{fig:93Nb-m}. 
For $^{95}$Nb, \cref{fig:95Nb-m}, there is not enough data 
to identify configuration B states, and for $^{97}$Nb some 
states could possibly belong as well to configuration~B, 
as depicted in \cref{fig:97Nb-m}.

$E2$ transitions are measured only for $^{93}$Nb and are 
reproduced more qualitatively rather than quantitatively, 
where some of them are large and at variance with the 
calculation (written in square brackets), 
$B(E2;5/2^-_2 \to 1/2^-_1) 
\eq 32^{+10}_{-9}$~W.u.~[2], $B(E2;3/2^-_2 \to 3/2^-_1) 
\eq 37^{+15}_{-30}$~W.u.~[4] and $B(E2;9/2^-_1 \to 
5/2^-_1) \eq 24(8)$~W.u.~[3]. The first value of the 
$5/2^-_2 \to 1/2^-_1$ is surprising due to the small value 
of the $B(E2; 2^+_1 \to 0^+_1) \eq 6.4(6)$~W.u. of the 
core, $^{92}$Zr, which is expected to be comparable in the 
weak coupling scenario.
They might also suggest a more unique mixing between the 
individual orbits, which is not considered in this work for 
simplicity.
\paragraph*{Wave functions.}
For $^\text{93--97}$Nb, as shown in \cref{fig:93-97Nb-spe}, 
the lowest state $1/2^-_1$ has a dominant $\pi(2p_{1/2})$ 
component of the normal A configuration, 
$P^{(N_\text{A},1/2^-_1)}_{\pi(2p_{1/2})} \!\simeq\! 80\%, 
70\%, 90\%$, with weak mixing between the 
different single-particle components of each of the 
configurations. A similar trend is observed for the 
$3/2^-_1$ and $5/2^-_1$ states, indicating these three 
states are single-quasiparticle excitations of the $\pi 
(2p_{1/2}), \pi (2p_{3/2}), \pi (1f_{5/2})$ orbits, coupled 
to the normal A configuration.

\subsubsection{The $^\text{99--103}$Nb region: 
strong coupling}\label{sec:99-103nb_m}
\begin{figure*}
\centering
\includegraphics[width=0.396\linewidth]
{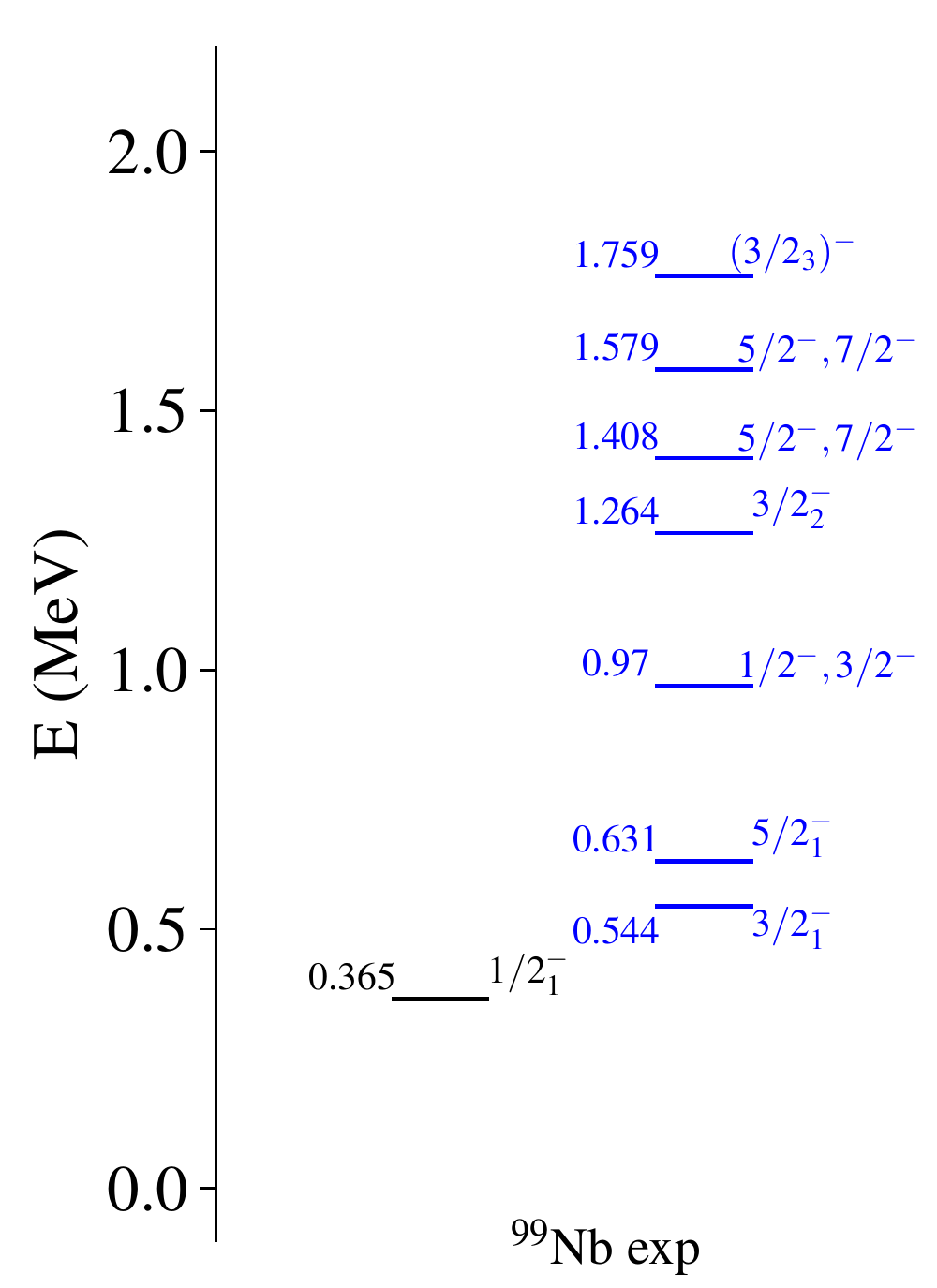}
\includegraphics[width=0.594\linewidth]
{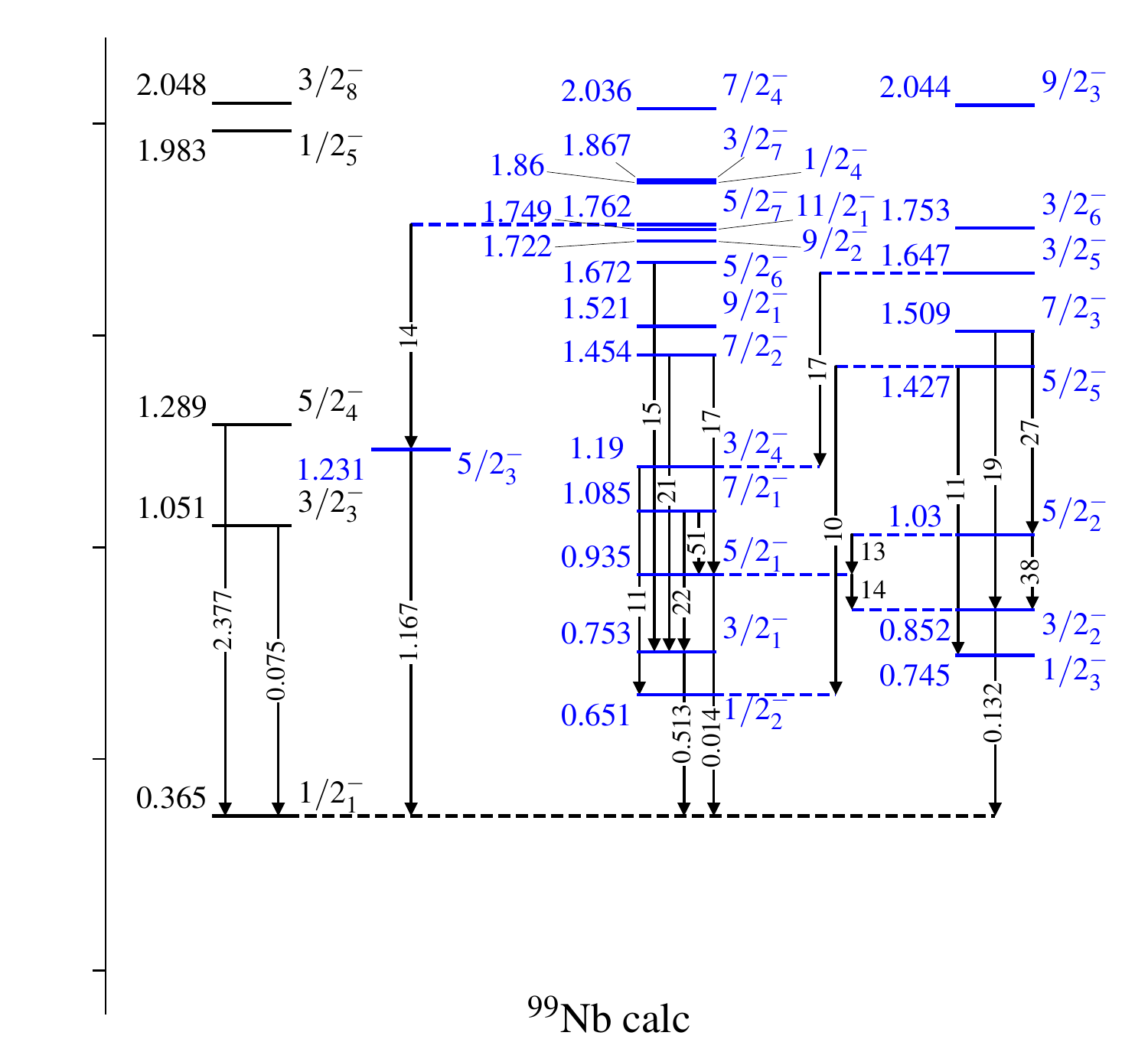}
\caption{Experimental (left) and calculated (right) energy 
levels in MeV, and $E2$ (solid arrows) transition rates in 
W.u., for $^{99}$Nb. All calculated $M1$ transitions 
(which are not shown in the figure) are smaller than 
$0.02$~W.u. Normal (intruder) states are depicted in 
black (blue). Data taken from \cite{NDS.145.25.2017}. 
\label{fig:99Nb-m}}
\end{figure*}
\begin{figure*}
\centering
\includegraphics[width=0.49\linewidth]
{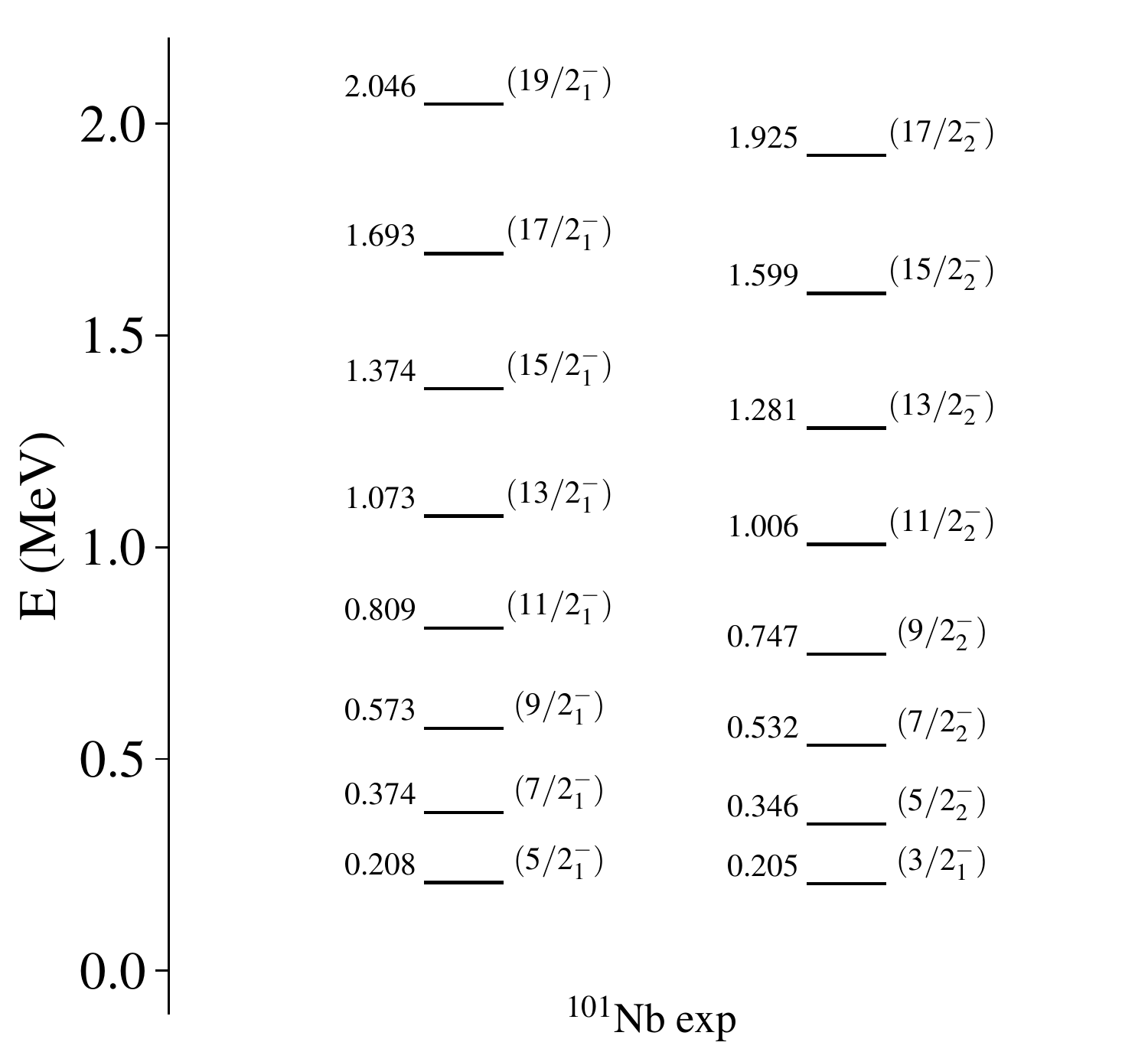}
\includegraphics[width=0.49\linewidth]
{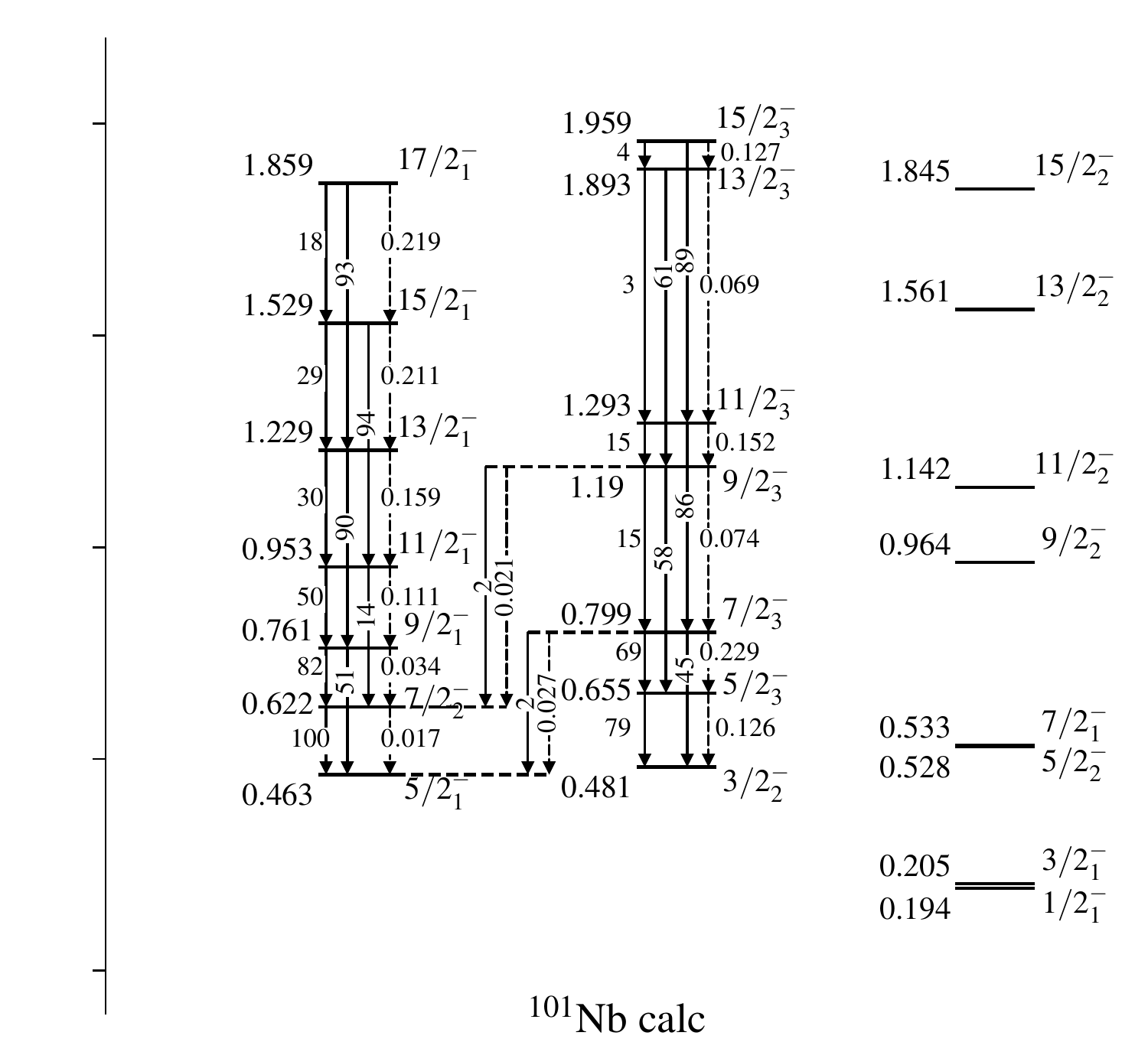}
\caption{Experimental (left) and calculated (right) energy 
levels in MeV, and $E2$ (solid arrows) and $M1$ (dashed 
arrows) transition rates in W.u., for $^{101}$Nb. Data 
taken from \cite{Luo2005}. \label{fig:101Nb-m}}
\end{figure*}
\begin{figure*}
\centering
\includegraphics[width=0.49\linewidth]
{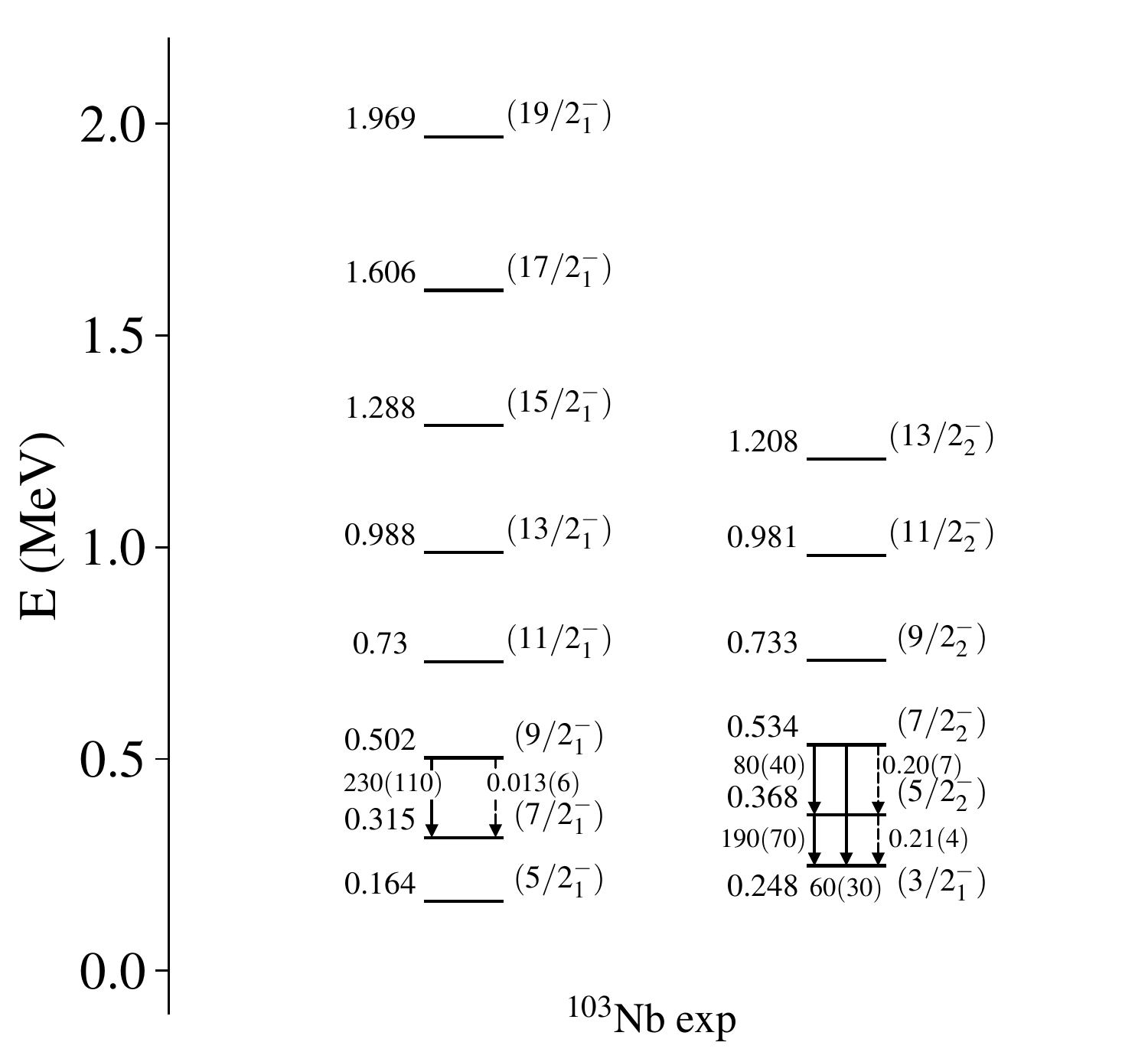}
\includegraphics[width=0.49\linewidth]
{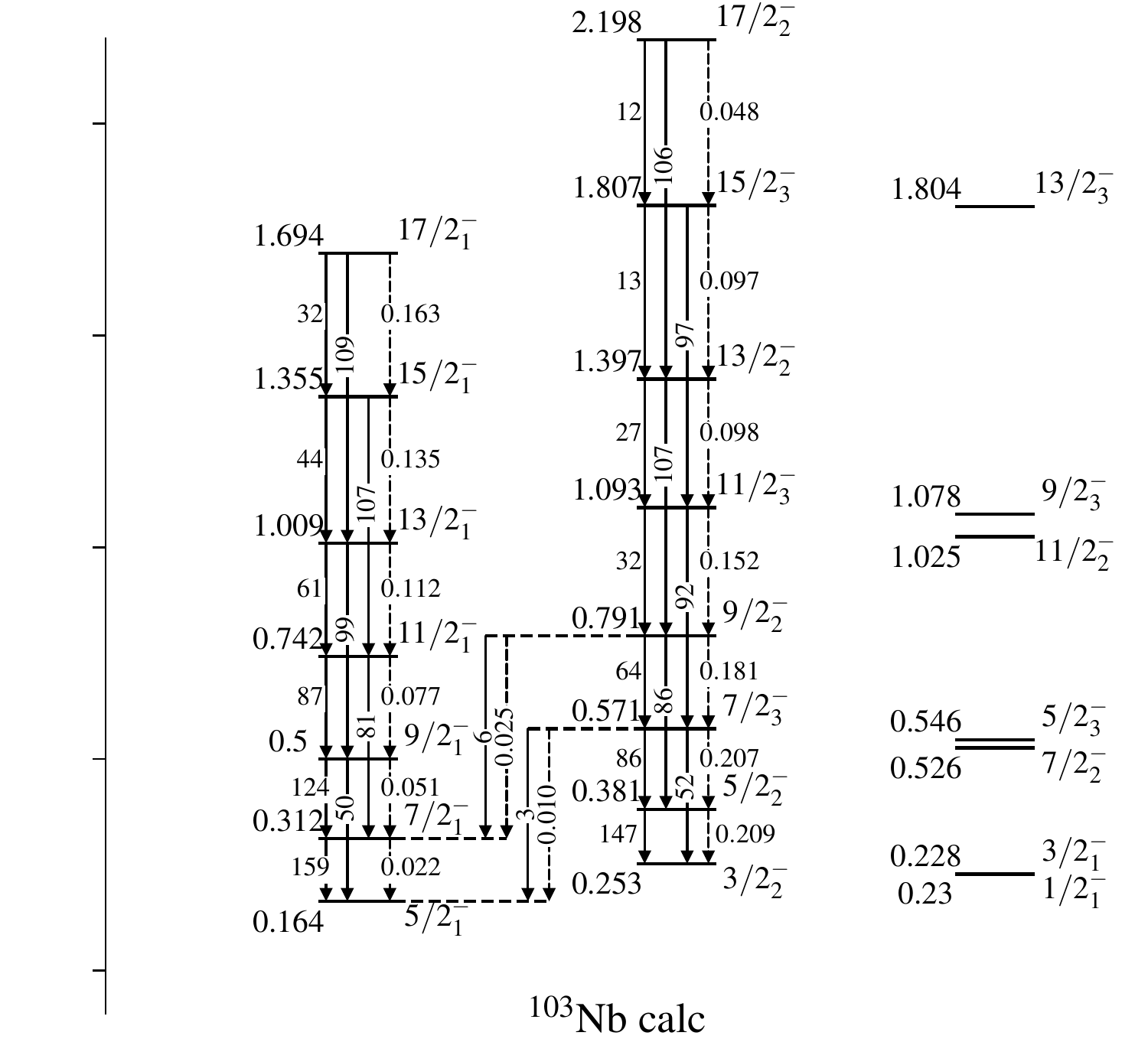}
\caption{Experimental (left) and calculated (right) energy 
levels in MeV, and $E2$ (solid arrows) and $M1$ (dashed 
arrows) transition rates in W.u., for $^{103}$Nb. Data 
taken from \cite{NDS.110.2081.2009}. \label{fig:103Nb-m}}
\end{figure*}
\begin{figure}
\centering
\begin{overpic}[width=1\linewidth]{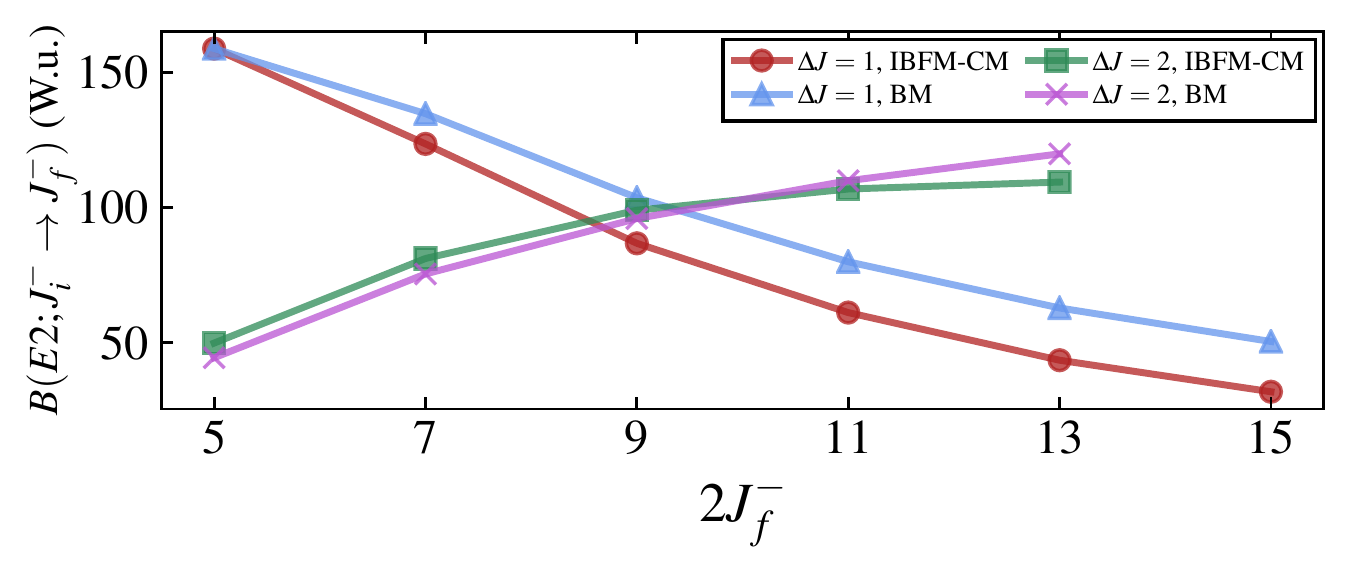}
\put (70,25) {\normalsize $K^\pi \eq 5/2^-$~(a)}
\end{overpic}\\
\begin{overpic}[width=1\linewidth]
{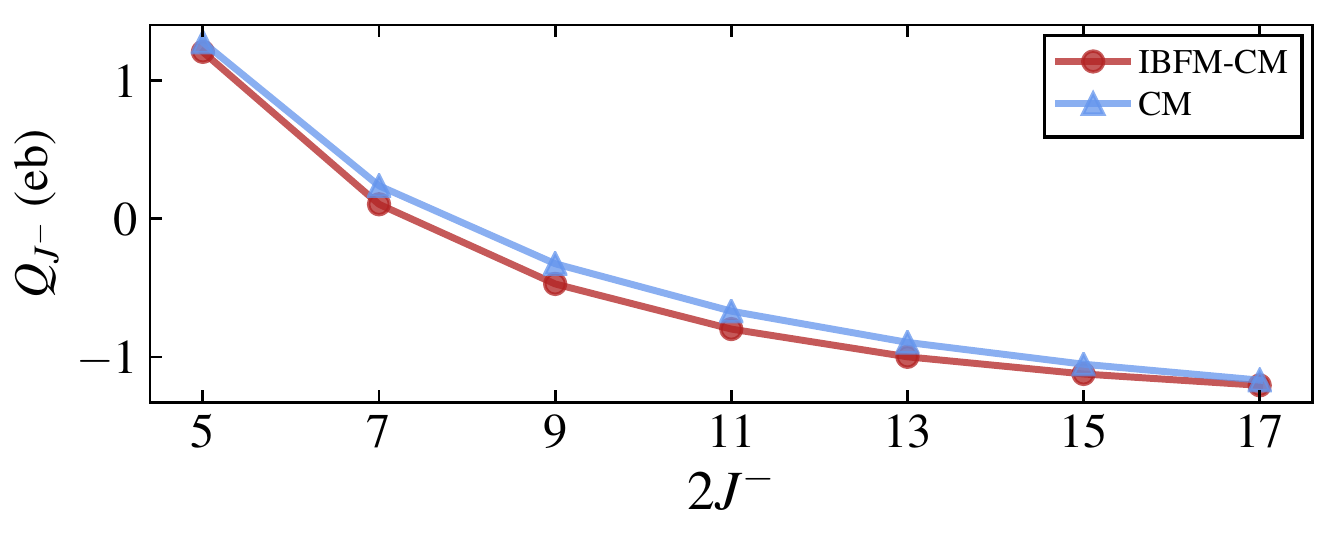}
\put (70,25) {\normalsize $K^\pi \eq 5/2^-$~(b)}
\end{overpic}\\
\caption{Comparison between the present calculation and the 
Bohr and Mottelson model (BM). (a) $E2$ transition rates in 
W.u. between members of the $K^\pi \eq 5/2^-$ band of 
$^{103}$Nb calculated in this work and using the collective 
model (BM), \cref{eq:GCM-e2}, with $\Delta J \eq J_f - 
J_i$. (b) Quadrupole moments in eb for members of the 
$K^\pi \eq 5/2^-$ band in $^{103}$Nb calculated in this 
work and using the collective model, \cref{eq:GCM-q}. 
\label{fig:BM-E2-quad-m}}
\end{figure}
\begin{figure}
\centering
\includegraphics[width=1\linewidth]{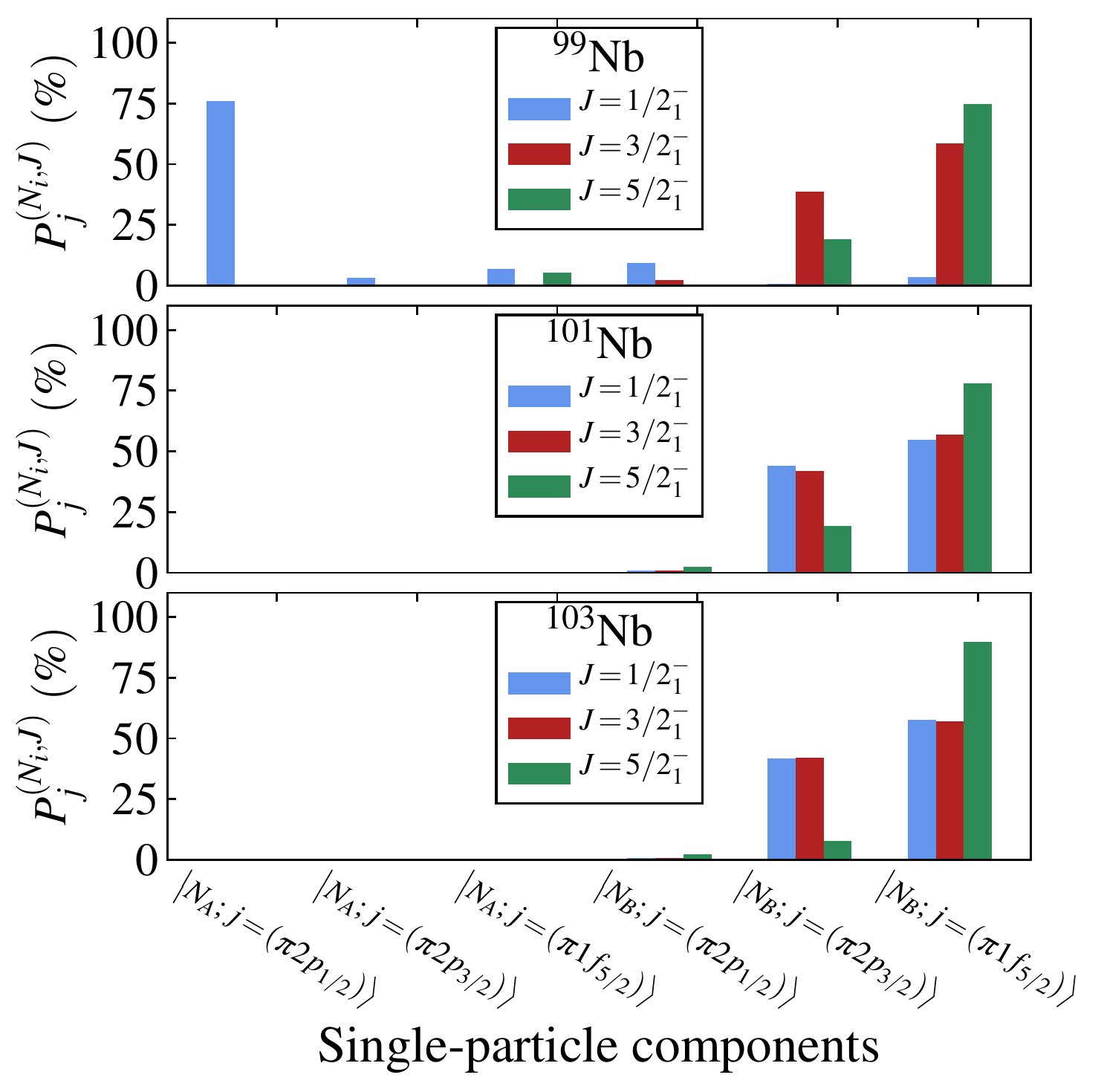}
\caption{Percentage of the single-particle components of 
the normal A and intruder B configurations, 
\cref{eq:prob_spe}, for the calculated $1/2^-_1, 3/2^-_1, 
5/2^-_1$ states in $^\text{93--97}$Nb 
isotopes.\label{fig:99-103Nb-spe}}
\end{figure}

As shown in 
\cref{fig:99Nb-m,fig:101Nb-m,fig:103Nb-m}, one can identify 
rotational bands with $K^\pi \eq 3/2^-$ and $5/2^-$ 
in $^\text{99--103}$Nb.
For $^{99}$Nb, the $1/2^-_1$ is identified as the 
configuration~A normal state that originates from the 
coupling of the $\pi(2p_{1/2})$ orbit with the $0^+_{1; \rm 
A}$ state of the adjacent $^{98}$Zr isotope. Alongside it, 
there is a rotational band with $K^\pi \eq 3/2^-$ that the 
calculation reproduces to a reasonable degree, however, a 
calculated $1/2^-_2$ appears in the spectrum, making this 
a $K^\pi \eq 1/2^-$ band. Alongside this band, the 
calculation suggests another $K^\pi \eq 1/2^-$ band 
beginning at 0.745~MeV. The $E2$ transitions within these 
bands are relatively stronger than those built upon the 
$1/2^-_1$ state, as expected from a rotational band. The 
large $E2$ transitions between the two $K^\pi \eq 
1/2^-_2,1/2^-_3$ bands indicate a strong mixing between 
them.

For $^{101,103}$Nb, all the states belong to the intruder B 
configuration and are arranged in two rotational bands with 
$K^\pi \eq 3/2^-, 5/2^-$, with a Nilsson model assignment 
$3/2^-[301],5/2^-[303]$. The calculation also suggests an 
additional $K^\pi \eq 1/2^-$ band alongside them 
with large staggering. For $^{101}$Nb, the calculated 
$K^\pi \eq 3/2^-, 5/2^-$ bands are a little 
higher in energy than experiment and the $K^\pi \eq 3/2^-$ 
band is somewhat staggered.
For $^{103}$Nb, the agreement with experiment is excellent 
with a clear particle-rotor splitting of the energy, with a 
moment of inertia \cref{eq:moment_iner} $B \eq 
0.022,0.024$~MeV for the $K^\pi \eq 5/2^-, 3/2^-$ bands, 
respectively.
The $E2$ and $M1$ transitions within the bands of both 
$^{101,103}$Nb are reproduced well. The $E2$ transitions 
trend, alongside the trend of the quadrupole moment, as a 
function of angular momentum $J$, is seen in 
\cref{fig:BM-E2-quad-m} to be very similar with that of the 
geometric collective model, \cref{eq:GCM-e2,eq:GCM-q}. As 
in the positive-parity case, \cref{sec:99-103nb_p},
the trend of the $M1$ transitions and magnetic moments is 
less similar.
The change in the experimental $K^\pi \eq 3/2^-_1$ to a 
calculated $K^\pi \eq 1/2^-_2$ band in $^{99}$Nb and the 
additional calculated $K^\pi \eq 1/2^-_1$ that do not 
appear in the spectrum might suggest the need to modify the 
proton single-particle energies used in the BCS calculation.
\paragraph*{Wave functions.}
As shown in \cref{fig:99-103Nb-spe}, for $^{99}$Nb, the 
lowest state $1/2^-_1$ has a dominant $\pi (2p_{1/2})$ 
component of the normal A configuration, 
$P^{(N_\text{A},1/2^-_1)}_{\pi (2p_{1/2})} \simeq 75\%$, 
with weak mixing between the different single-particle 
components of each of the configurations. 
This $1/2^-_1$ is the lowest configuration A state that 
resides alongside the intruder B configuration. The 
$5/2^-_1$ state has a dominant $\pi (1f_{5/2})$ B 
configuration component, while the $3/2^-_1$ is 
mixed between the intruder B configuration $\pi (2p_{3/2})$ 
and $\pi (2f_{5/2})$ components. 
For $^{101,103}$Nb, both the $1/2^-_1$ and $3/2^-_1$ states 
are mixed between the intruder B configuration 
$\pi(2p_{3/2})$ and $\pi (1f_{5/2})$ components, while the 
$5/2^-_1$ has a dominant $\pi (1f_{5/2})$ component.

\section{Results: Evolution of wave functions and 
observables along the niobium chain}\label{sec:res_evo_obs}
\subsection{Evolution of configuration and single-particle 
content}\label{sec:evo-conf}
\begin{figure}[t!]
\centering
\includegraphics[width=1\linewidth]{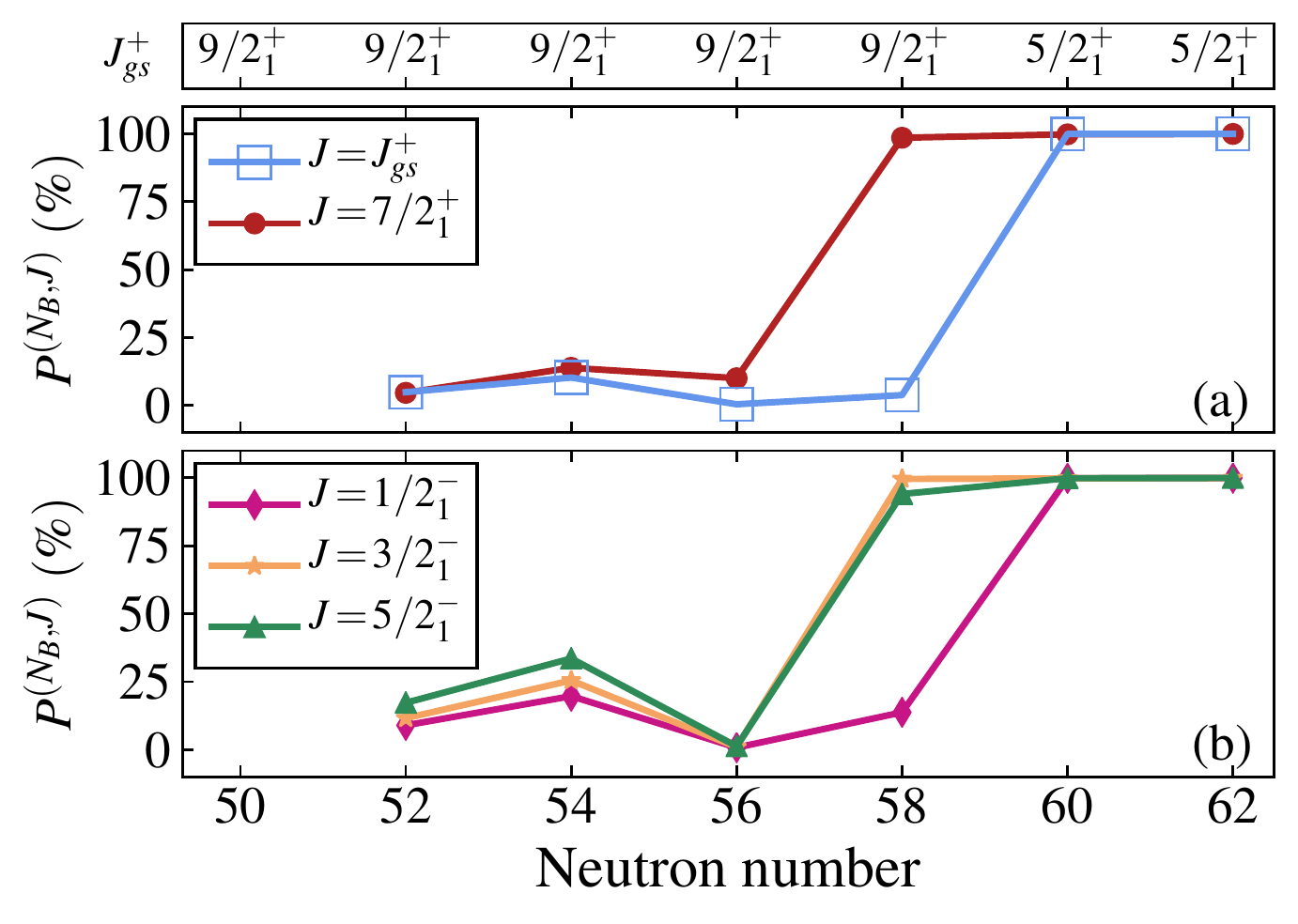} %
\caption{Percentage of the intruder (B) component
[the $P^{(N_B,J)}$ probability in \cref{eq:prob_norm_int}] 
for $^{93-103}$Nb. 
(a) The ground state ($J^+_{gs}$) and the first-excited
positive-parity state ($7/2^+_1$). (b) The $1/2^-_1, 
3/2^-_1, 5/2^-_1$ states for the negative-parity 
states. The values of $J^+_{gs}$ are 
indicated at the top panel.\label{fig:b_sqr}}
\end{figure}
A possible change in the angular momentum of the ground 
state ($J_{\rm{gs}}$) is a characteristic signature of 
QPTs in odd-mass nuclei, unlike even-even nuclei 
where the ground state remains $0^+$ after the crossing. It 
is an important measure for the quality of the 
calculations. A mean-field approach, for example, without 
configuration mixing, fails to reproduce the change between 
the $9/2^+_1$ and $5/2^+_1$ states in $J^{+}_{\rm{gs}}$ for 
the Nb isotopes~\cite{Rodriguez-Guzman2011}.
Information on configuration changes for each isotope, can 
be inferred from the evolution of the probabilities 
$P^{(N_\text{A},J)}$ or $P^{(N_\text{B},J)}$, 
\cref{eq:prob_norm_int}, of the states considered. 
\mbox{Figure \ref{fig:b_sqr}} shows the percentage of the 
wave function within the $B$ configuration, in panel (a), 
for the ground state ($J^+_\text{gs}$) and first-excited 
state ($7/2^+_1$) and in panel (b) for the $J \eq 
1/2^-_1,3/2^-_1,5/2^-_1$ states, as a function of neutron 
number across the Nb chain. 
The rapid change in structure of the $J^+_\text{gs}$ and 
$1/2^-_1$ states from the normal A~configuration (small 
$P^{(N_\text{B},J)}$ probability) for neutron number 52--58 
($^{93-99}$Nb)  to the intruder B~configuration (large 
$P^{(N_\text{B},J)}$ probability) for neutron number 60--62 
($^{101-103}$Nb)  is clearly evident, signaling a Type II 
QPT, as mentioned in \cref{sec:99-103nb_p,sec:99-103nb_m}.
The configuration change appears sooner in the $7/2^+_1$ 
and $3/2^-_1,5/2^-_1$ states, which switch to 
configuration~B already at neutron number 58 ($^{99}$Nb). 
The behavior of the $J_\text{gs}, 1/2^-_1$ and 
$7/2^+_1, 3/2^-_1,5/2^-_1$ states is inline with the 
behavior of the $0^+_1$ and $2^+_1$ states of the $_{40}$Zr 
cores with the same neutron numbers 
\cite{Gavrielov2019,Gavrielov2022}, 
which also change from configuration A to B at neutron 
number 60 and 58, respectively (see Fig.~10 of 
Ref.~\cite{Gavrielov2022}).
Outside a narrow region near neutron number 60, where the 
crossing occurs, the two configurations are weakly mixed 
and the states retain a high level of purity, except for 
the negative parity states for neutron number 54 
($^{95}$Nb), where the mixing is somewhat stronger.

\subsection{Energy levels}\label{sec:evo-energy}
\begin{figure}[t!]
\centering
\includegraphics[width=1\linewidth]{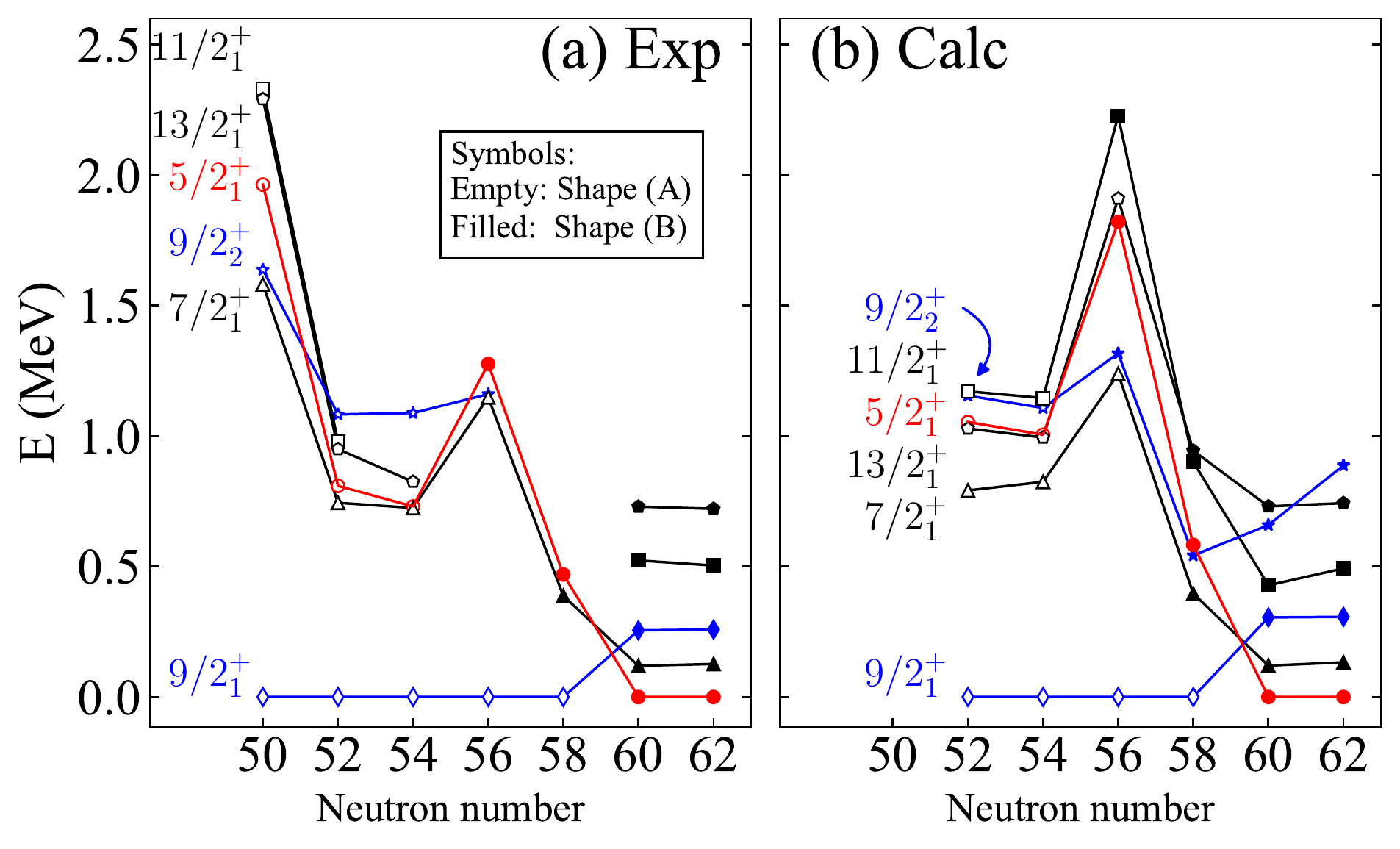} %
\caption{Comparison between
(a)~experimental~\cite{NDS.114.1293.2013,NDS.112.1163.2011,
NDS.111.2555.2010,NDS.111.525.2010,NDS.145.25.2017}
and (b)~calculated lowest-energy positive-parity levels in 
Nb isotopes. Empty (filled) symbols indicate a state 
dominated by the normal A configuration (intruder B 
configuration), with assignments based on 
\cref{eq:prob_norm_int}. In particular, the $9/2^+_1$ state 
is in the A (B) configuration for neutron number 52--58 
(60--64) and the $5/2^+_1$ state is in the A (B) 
configuration for 52--54 (56--64). Note that the calculated 
values start at 52, while the experimental values include 
the closed shell at 50.\label{fig:energies-p}}
\end{figure}
\begin{figure}[t!]
\centering
\includegraphics[width=1\linewidth]{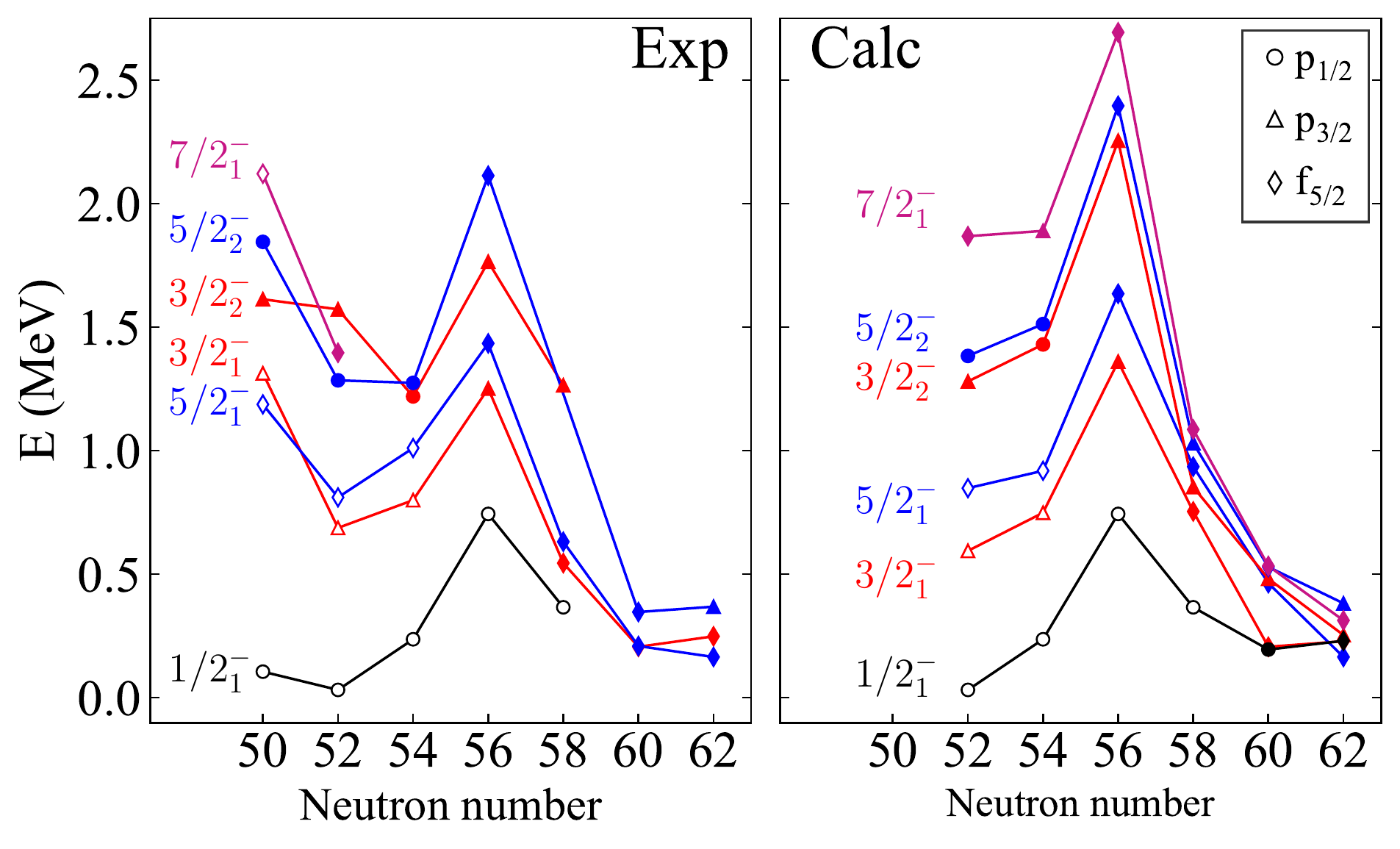} %
\caption{Comparison between 
(a)~experimental~\cite{NDS.114.1293.2013,NDS.112.1163.2011,
NDS.111.2555.2010,NDS.111.525.2010,NDS.145.25.2017}
and (b)~calculated lowest-energy negative-parity levels in 
Nb isotopes. Empty (filled) symbols indicate a state 
dominated by the normal A configuration (intruder B 
configuration), with assignments based on 
\cref{eq:prob_norm_int}. Note that the calculated values 
start at 52, while the experimental values include the 
closed shell at 50.\label{fig:energies-m}}
\end{figure}
\Cref{fig:energies-p,fig:energies-m} show the experimental 
and calculated levels of selected positive- and 
negative-parity states, respectively, along with 
assignments to configurations based on 
\cref{eq:prob_norm_int}. Open (solid) symbols indicate a 
dominantly normal (intruder) state with small (large) 
$P^{(N_\text{B},J)}$ probability. 
For the positive-parity states of \cref{fig:energies-p}, in 
the region between neutron number 50 and 56, there appear 
to be two sets of levels with a weakly deformed structure, 
associated with configurations A and B. All levels decrease 
in energy for 52--54, away from the closed shell, and rise 
again at 56 due to the $\nu(2d_{5/2})$ subshell closure. 
At neutron number 58, there is a pronounced drop in energy 
for the states of the B~configuration, due to the onset of 
deformation. At 60, the two configurations cross, 
indicating a Type~II QPT, and the ground state changes from 
$9/2^+_1$ to $5/2^+_1$, becoming the bandhead of a 
$K=5/2^+$ rotational band composed of $5/2^+_1, 
7/2^+_1,9/2^+_1, 11/2^+_1, 13/2^+_1, \ldots$ states. Beyond 
neutron number 60, the intruder B~configuration remains 
strongly deformed and the band structure persists. The 
above trend is similar to that encountered in the even-even 
$_{40}$Zr cores (see Fig.~14 of Ref.~\cite{Gavrielov2022}).

For the negative-parity states in \cref{fig:energies-m}, in 
the region between neutron number 50 and 56, there appear 
to be the $1/2^-_1$ state and two sets of levels for each 
of the $3/2^-$ and $5/2^-$ states with a weakly deformed 
structure, associated with configurations A and B. All 
levels decrease in energy for 52--54, away from the closed 
shell, and rise again at 56 due to the $\nu(2d_{5/2})$ 
subshell closure. From 58, there is a pronounced drop in 
energy for the states of the B~configuration, due to the 
onset of deformation. At 60, the two configurations cross, 
indicating a Type~II QPT. The calculated normal $1/2^-$ 
rises in energy and the $1/2^-_1$ remains the lowest 
negative-parity state in $^{101}$Nb and at $^{103}$Nb it is 
the $5/2^-_1$ that is lowest. Although not in the 
experimental data, the trend of the $1/2^-_1$ state seems 
to suggest the existence of a low-lying $1/2^-$ also in 
$^{101,103}$Nb, as suggested by the calculation. The 
$1/2^-_1$, $3/2^-_1$ and $5/2^-_1$ states become the 
bandhead of $K=1/2^-, 3/2^-, 5/2^-$ rotational bands, 
respectively.

\subsection{Two neutron separation 
energy}\label{sec:evo-s2n}
\begin{figure*}[t]
\centering
\includegraphics[width=1\linewidth]{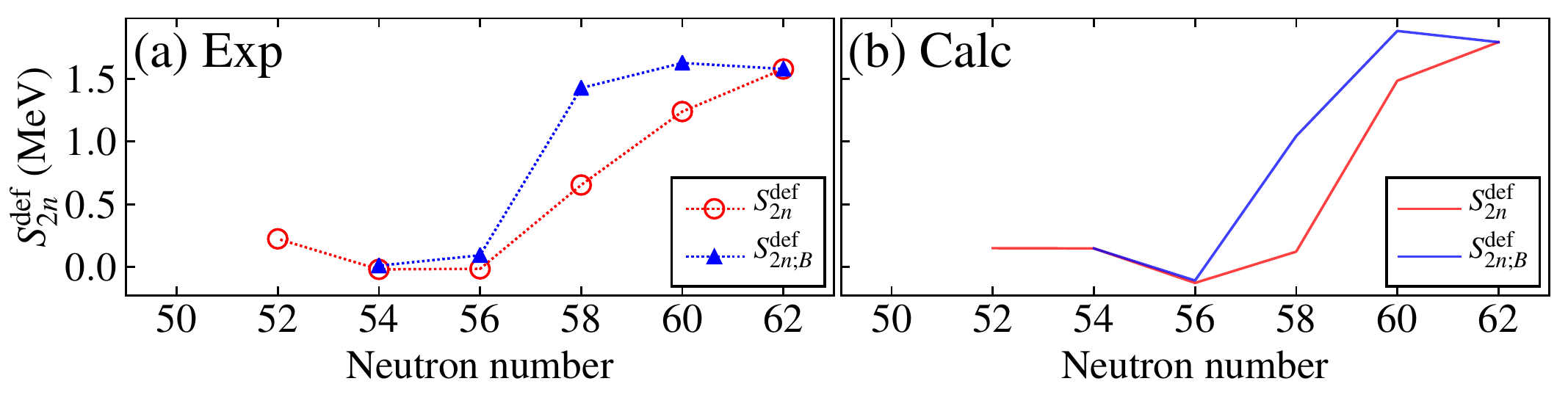} %
\caption{Comparison of the deformed part of the two neutron 
separation energies, ($S^\text{def}_{2n}$), between (a) 
experiment \cite{Huang2017} and (b) calculation. 
\label{fig:s2n-def}}
\end{figure*}
An observable that portrays both types of QPTs is two 
neutron separation energy, defined as 
\begin{equation}\label{eq:s2n_emp}
S_{2n} = 2M_n + M(N-2,Z) - M(N,Z),
\end{equation}
where $M(N,Z)$ is the mass of a nuclei with $N,Z$ 
neutrons and protons, respectively, and $M_n$ is the 
neutron mass.
It is convenient to transcribe the $S_{2n}$ as
\begin{equation}\label{eq:s2n}
S_{2n} = -\tilde A - \tilde B N_v \pm S^\text{def}_{2n} - 
\Delta_n,
\end{equation}
where $N_v$ is half the number of valence particles in the 
boson core and $S^\text{def}_{2n}$ is the contribution of 
the deformation, obtained by the expectation value of the 
Hamiltonian in the ground state. The $+$ sign applies to 
particles and the $-$ sign to holes. The $\Delta_n$ 
parameters takes into account the neutron subshell closure 
at 56, $\Delta_n=0$ for 50--56, and $\Delta_n=2$~MeV for 
58--64. 
For the Nb isotopes, the chosen values in \cref{eq:s2n} are 
$\tilde A \eq -17.25, \tilde B \eq 0.758$~MeV. The value of 
$\tilde A$ is taken to fit $^{91}$Nb, and the values of 
$\tilde B$ and $\Delta_n$ are taken from the previous 
even-even Zr calculation \cite{Gavrielov2022}. In 
\cref{fig:s2n-def}, the experimental (left) and calculated 
(right) deformed part, $S^\text{def}_{2n}$ 
\cite{Petrellis2011a, Iachello2011b}, are shown in red 
circles and lines, respectively. $S^\text{def}_{2n}$ is 
obtained by subtracting the linear part and $\Delta_n$ from 
the experimental and calculated $S_{2n}$. One can clearly 
see the onset of deformation going from neutron number 
52--56, where $S^\text{def}_{2n}$ is small, to 58--62, 
where 
it jumps and rises.

In order to denote the occurrence of both 
\mbox{Type~I} and II QPTs, in addition to \cref{eq:s2n}, 
using \cref{eq:s2n_emp} we can also estimate two neutron 
separation energies for excited states by using the mass of 
an excited state $M(N,Z) \!\equiv\! M_{exc}(N,Z) \eq 
M_{gs}(N,Z) + E_{exc}(N,Z)$, where $M_{gs}(N,Z)$ is the 
mass for the ground state and $E_{exc}(N,Z)$ is the energy 
of the excited state.
Therefore, by adding the difference $E_{exc}(N-2,Z) - 
E_{exc}(N,Z)$ to \cref{eq:s2n_emp,eq:s2n} we can obtain 
two neutron separation energy for an excited state, and for 
this we choose the lowest configuration~B state. The 
experimental and calculated results, $S^\text{def}_{2n;B}$, 
are given in blue triangles (left) and lines (right), 
respectively, in \cref{fig:s2n-def}. It is seen that 
for neutron number 54--56 $S^\text{def}_{2n;B}$ is small, 
then at 58 it jumps due to the onset of deformation at 60, 
then it flattens. 
This behavior denotes 
the \mbox{Type~I}~QPT of shape evolution from spherical to 
axially-deformed, within configuration~B. It is similar 
to the behavior of the $_{61}$Pm, $_{63}$Eu and $_{65}$Tb 
isotopes, which also undergo a QPT from spherical to 
axially-deformed shape \cite{Petrellis2011a, Iachello2011b}.
For neutron number 54--56, $S^\text{def}_{2n;B}$ 
(triangles) is close to the value of $S^\text{def}_{2n}$ 
(circles), as configuration~B is more spherical. At 58, 
there is a larger jump than $S^\text{def}_{2n}$ since 
configuration~B is more deformed than A, which 
continues at 60.
For 62, both $S^\text{def}_{2n;B}$ and $S^\text{def}_{2n}$ 
coincide since the ground state is configuration~B, which 
denotes the \mbox{Type~II} QPT.
Therefore, the deformed part of the two neutron separation 
energies in its ground and excited states serves as an 
important indicator for the occurrence of IQPTs.
\subsection{$E2$ transition rates and quadrupole 
moments}\label{sec:evo-e2-q}
\begin{figure}[t]
\centering
\includegraphics[width=1\linewidth]{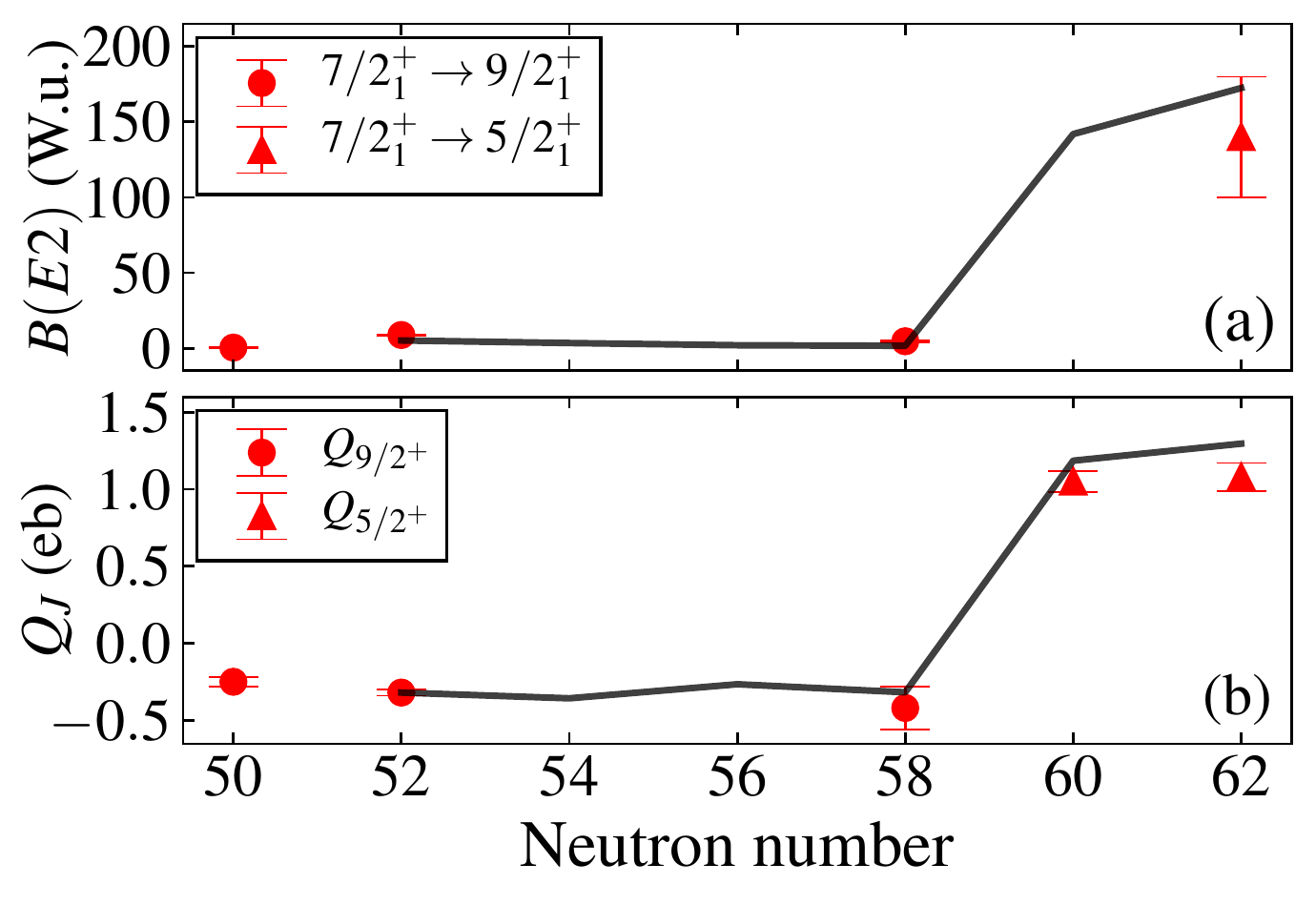} %
\caption{Evolution of (a)~$B(E2; 7/2^+_1\to J^{+}_{gs})$ in 
W.u. and (b)~Quadrupole moments of $J^+_{gs}$ in 
$eb$. Symbols (solid lines) denote experimental data 
(calculated results). Data in panels (a) and (b) are taken 
from \cite{NDS.114.1293.2013, NDS.112.1163.2011, 
NDS.110.2081.2009} and \cite{NDS.114.1293.2013, 
NDS.112.1163.2011, Cheal2009}, respectively. 
\label{fig:e2_quad}}
\end{figure}

Electromagnetic transitions and moments provide further
insight into the nature of QPTs. \Cref{fig:e2_quad} shows 
$B(E2; 7/2^+_1\to J^{+}_{gs})$ in panel (a) and quadrupole 
moment of $J^{+}_{gs}$ in panel (b). These observables are 
related to the deformation, the order parameter of the QPT.
Although the data is incomplete, one can still observe
small (large) values of these observables below (above)
neutron number 60, indicating an increase in deformation.
The calculation reproduces well this trend and
attributes it to a Type~II QPT involving a jump
between neutron number 58 and 60, from a weakly-deformed
A configuration, to a strongly-deformed B configuration.
The trend in the $E2$ transition rates is very similar to 
that of the $2^+_1\to0^+_1$ transition of the adjacent 
even-even Zr isotopes. In the Zr case, the sudden increase 
at neutron number 60 is ascribed to the IQPT, where the 
ground state configuration changes from normal to intruder, 
while the intruder configuration evolves at the same time 
from being quasi-spherical to deformed \cite{Gavrielov2019, 
Gavrielov2022}.

\subsection{$M1$ transitions and magnetic 
moments}\label{sec:evo-mag-m1}
\begin{figure}[t]
\centering
\includegraphics[width=1\linewidth]{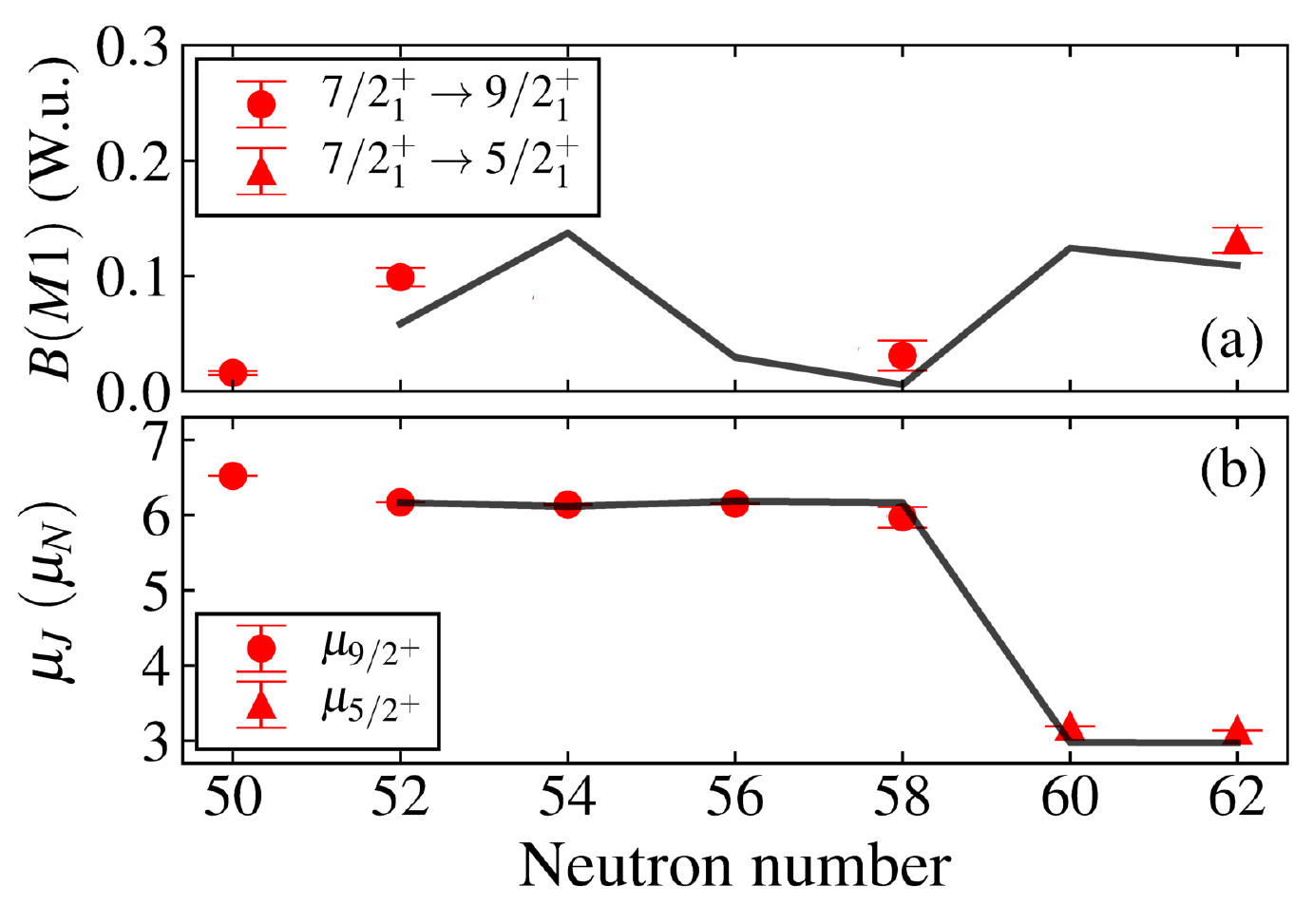} %
\caption{
Evolution of (a)~$B(M1; 7/2^+_1\to 9/2^+_1)$ and $B(M1; 
7/2^+_1\to 5/2^+_1)$ in W.u. and (b)~Magnetic moments of 
$J^+_{gs}$ in $eb$. Symbols (solid lines) denote 
experimental data (calculated results). Data in panels (a) 
and (b) are taken 
from \cite{NDS.114.1293.2013, NDS.112.1163.2011, 
NDS.145.25.2017, NDS.110.2081.2009} and \cite{Cheal2009}, 
respectively. \label{fig:m1_mag}}
\end{figure}
The trend in the experimental $B(M1;7/2^+_1 \to J^+_{\rm 
gs})$, shown in \cref{fig:m1_mag}(a), suggests a jump at 
neutron number 52 and another one at 62, which is 
reproduced by the calculation to a good degree. This 
suggests that $M1$ transition rates might be an observable 
that pronounces less the Type~II QPT, possibly due to their 
sensitivity to the single-particle degrees of freedom 
rather than the collective ones.
However, the Type~II scenario is strongly supported by 
the trend of the magnetic moments ($\mu_J$) of the ground 
state, shown in~\cref{fig:m1_mag}(b), where both the data 
and the calculation show a constant value of $\mu_J$ 
for neutron numbers 52--58, and a drop to a lower value at 
60, which persists for 60--62. This trend of approximately 
constant value for each range of neutron numbers, suggests 
a corresponding constant mixing in the ground state wave 
function, in line with the calculated weak mixing before 
and after the crossing.

\section{Conclusions and outlook}\label{sec:conc}
The general framework of the interacting boson-fermion 
model with configuration mixing (IBFM-CM) has been 
presented, allowing a quantitative description of 
shape-coexistence, configuration-mixing and related QPTs in 
odd-mass nuclei. 
A quantal analysis for the chain of the odd-even $_{41}$Nb 
isotopes involving positive- and negative-parity states was 
performed for neutron number 52--62. It examined the 
spectra and properties of individual isotopes as well as 
the evolution of energy levels and other observables 
(two-neutron separation energies, $E2$ and $M1$ transition 
rates, and magnetic and quadrupole moments) along the 
chain. Special attention has been devoted to changes in the 
configuration-content and single-particle-content of wave 
functions. In general, the calculated results, obtained by 
a fitting procedure described in \cref{app:bcs}, are found 
to be in a good agreement with the empirical data.

The results of the comprehensive analysis suggest a complex 
phase structure in these isotopes, involving two
configurations. The normal $A$ configuration remains 
spherical in all isotopes considered. The intruder $B$ 
configuration undergoes a spherical to axially deformed 
U(5)-SU(3) QPT within the boson core, with a critical point 
near $A\approx100$. In parallel to the gradual shape 
evolution within configuration $B$, the two configurations 
cross near neutron number 60, and the ground state changes 
from configuration $A$ to configuration $B$. The two 
configurations are weakly mixed and retain their purity 
before and after the crossing, thus demonstrating, IQPTs in 
odd-mass nuclei.

The new IBFM-CM framework can motivate further work in any 
medium-heavy odd-mass region with mixed configurations, 
such as $Z\!\approx\!40,50,82$ and $N\!\approx\!60, 66, 
104$, respectively, with many nuclei to be explored. The 
current results obtained for the Nb isotopes motivate 
further experiments of non-yrast spectroscopy in such 
nuclei, as well as set the path for new investigations on 
multiple QPTs and coexistence in other Bose-Fermi systems.

\begin{acknowledgements}
The author would like to acknowledge support by the Israel 
Academy of Sciences of a Postdoctoral Fellowship Program in 
Nuclear Physics. 
The author would like to thank F.~Iachello (Yale 
University) and A.~Leviatan (Hebrew University) for their 
enlightening comments, discussions and encouragement, and 
to P.~Van Isacker for providing his IBFM code, which served 
as a basis for the IBFM-CM computer program.
\end{acknowledgements}

\appendix
\section{Single-particle energies and BCS 
procedure}\label{app:bcs}
A BCS calculation is done by iterating over the equations 
for the single-quasiparticle energies ($\epsilon_j$) and 
occupation probabilities ($v^2_j$), as one varies the Fermi 
energy $\lambda_{\rm F}, $ until an equality between the 
particle number ($N_p$) and the number of valence particles 
is obtained
\begin{align}\label{eq:bcs_eqs}
\epsilon_j & = \sqrt{(E_j - \lambda_{\rm F})^2 + 
\Delta_{\rm F}}, \\
v^2_j & = \frac{1}{2}\Big(1 - \frac{E_j - 
\lambda_{\rm F}}{\epsilon_j}\Big), \\
N_p & = \sum_j (2j + 1)v^2_j.
\end{align}

In \cref{eq:bcs_eqs}, $j$ are the different shell orbits, 
$E_j$ are the experimental single-particle energies and 
$\Delta_{\rm F}$ is the pairing gap.
In this work, for $_{41}$Nb the BCS procedure is 
employed with 13 valence particle in the $Z \eq 
\text{28--50}$ shell with the $\pi(1f_{5/2}), \pi(2p_{1/2}),
\pi(2p_{3/2}), \pi(1g_{9/2})$ orbits. The same 
single-particle energies and pairing gap for both 
configurations~A and B are chosen. It might be possible to 
choose a different set of single-particle energies for the  
intruder configuration, however this is not 
done so for simplicity. The experimental single-particle 
energies are taken from Table XI of \cite{Barea2009} and 
$\Delta_{\rm F}$ is taken to be 1.5~MeV, consistent with 
the absolute values of the empirical proton pairing gaps 
(see Eq.~(2.93) of \cite{BohrMott-I}) for $^{91-97}$Nb. The 
resulting single-quasiparticle energies 
($\epsilon_j$) and occupation probabilities ($v_j^2$) are 
shown in \cref{tab:bcs} for the different orbits and 
single-particle energies ($E_j$).

\begin{table}[t!]
\begin{center}
\caption{\label{tab:bcs} Empirical single-particle energies 
($E_j$) taken from Table XI of \cite{Barea2009} and 
calculated single-quasiparticle energies ($\epsilon_j$) in 
MeV with occupation probabilities ($v^2_j$) for the 
different $j$-orbits, with a resulting Fermi energy of 
$\lambda_\text{F} \eq 2.024$~MeV.\small}
\begin{tabular}{cccc}
\hline
Orbit      & $E_j$ & $\epsilon_j$  & $v^2_j$ \\
\hline
$1g_{9/2}$ & 2.684 & 1.639 & 0.299 \\
$2p_{1/2}$ & 1.753 & 1.524 & 0.589 \\
$2p_{3/2}$ & 0.486 & 2.148 & 0.858 \\
$1f_{5/2}$ & 0.000 & 2.519 & 0.902 \\
\hline 
\end{tabular}
\end{center}
\end{table}
\begin{table}[t!]
\begin{center}
\caption{\label{tab:parameters}
\small Parameters in MeV of the boson-fermion interactions,
$\hat V^{(i)}_{\rm bf}$ of \cref{eq:V_BF_i}, obtained 
from a fit assuming $A^{(i)}_0 \eq A_0$, $\Gamma^{(i)}_0 
\eq \Gamma_0$ and $\Lambda^{(i)}_0 \eq \Lambda_0$, where 
$({i=\rm A,B})$.}
\begin{tabular}{ccccccc}
\hline
Neutron number & $52$ & $54$  & $56$ &  $58$ & $60$ & $62$\\
\hline
$A_0$       & 0   & 0	& 0   &$-0.11$&$-0.2$&$-0.2$\\
$\Gamma_0$  & 1.0 & 1.0	& 1.0 &  1.0  &  1.0 &  1.0 \\
$\Lambda_0$ & 1.0 & 1.0	& 3.0 &  3.0  &  3.8 &  3.8 \\
\hline 
\end{tabular}
\end{center}
\end{table}
Taking the derived $\epsilon_j$ and $v^2_j$, the parameters 
of the boson-fermion interaction \eqref{eq:v_bf_expl} can 
be determined from the microscopic theory of the IBFM to be
\begin{align}
A^{(i)}_j & = - \sqrt{5(2j+1)}A_0^{(i)}~,\\
\Gamma^{(i)}_{jj^\prime} & = 
\sqrt{5}\gamma_{jj^\prime}\Gamma_0^{(i)}~,\\
\Lambda^{(i)j^{\prime\prime}}_{jj^\prime} & = -2 
\sqrt{\frac{5}{2j^{\prime\prime}+1}} 
\beta_{jj^{\prime\prime}} \beta_{j^\prime 
j^{\prime\prime}} 
\Lambda_0^{(i)}~,
\end{align}
where $i=\text{A,B}$ for the different configurations and 
\begin{align}
\gamma_{jj^\prime} & = (u_j u_{j^\prime} - v_j 
v_{j^\prime})Q_{jj^\prime}~,\\
\beta_{jj^{\prime}} & = (u_j v_{j^\prime} + v_j 
u_{j^\prime})Q_{jj^\prime}~,\\
Q_{jj^\prime} & = \braket{j||Y^{(2)}||j^\prime}~,
\end{align}
where the occupation probability $u_j$ satisfy $u_j^2 \eq 1 
- v_j^2$. 

The strengths $(A_0^{(i)}, \Gamma_0^{(i)}, 
\Lambda_0^{(i)})$ are obtained by a fit, and can be 
separated to positive- and negative-parity states and to 
the different configurations.
In this work, for simplicity, we assume the same values for 
the different configurations.
They are listed in \cref{tab:parameters}, where the 
monopole term ($A_0$) vanishes for neutron number 52--56 
and corrects the quasi-particle energies at neutron number 
58--62. The quadrupole term ($\Gamma_0$) is constant for 
the entire chain. The exchange term ($\Lambda_0$) increases 
towards the neutron 
mid-shell~\cite{IachelloVanIsackerBook}. 
Altogether, the values of the parameters are either 
constant for the entire chain or segments of it and vary 
smoothly. Interestingly, these values are very similar to 
those of the $_{63}$Eu isotopes \cite{ScholtenThesis} in a 
single shell configuration and a single-$j$ calculation 
with $v^2 \eq 0.3$, which have an approximately constant 
value of $\Gamma_0 \!\simeq\! 1$ and $A_0 \eq -0.4$ and an 
increasing value of $\Lambda_0$ from $\!\approx\!1.6$ in 
the spherical region to $\!\approx\!3.8$ in the deformed 
one. For the boson-fermion mixing term, $\hat W_{\rm bf}$ 
of \cref{eq:V_BF}, the value of $\omega_j \eq 0$ is chosen, 
since for equal $\omega_j$ it coincides with the $\hat 
W_{\rm b}$ term of \cref{eq:H_b}.

For $\hat{T}_{\rm b}(M1)$ of \cref{eq:Tm1}, we use $g^{(\rm 
A)} \eq -0.21,\,-0.42\mu_N$ for neutron number 52--54 and 
zero otherwise, $g^{(\rm B)} \eq (Z/A)\mu_N$ and 
$\tilde{g}^{(\rm A)} \eq \tilde{g}^{(\rm B)} \eq 0\; 
(-0.017\mu_N)$ for 52--56 (58--62). For $\hat{T}_{\rm 
f}(M1)$ we use $g_{\ell} \eq 1\mu_N$ and a quenching of 
20.835\%, which results with a value of $g_{s} \eq 
4.4219\mu_N$.

For $\hat{T}_{\rm b}(E2)$ of \cref{eq:TsigL}, we adopt the 
same parameters $(e^{(\rm A)},e^{(\rm B)},\chi)$
used for the core Zr isotopes~\cite{Gavrielov2022},
with a slight modification of \mbox{$e^{(\rm A)} \eq 
2.45,\,1.3375~\sqrt{\text{W.u.}}$} for neutron numbers 
52--54 and \mbox{$e^{(B)} \eq 2.0325~\sqrt{\text{W.u.}}$} 
for 62. The fermion effective charge in $\hat{T}_{\rm 
f}(E2)$ is \mbox{$e_f \eq -2.361~\sqrt{\text{W.u.}}$},
determined from a fit to the ground state quadrupole moment
of $^{93}$Nb.

\bibliography{/home/noam/Desktop/Physics/Articles/paper-Nb-long.bib}

\begin{thebibliography}{85}%
\makeatletter
\providecommand \@ifxundefined [1]{%
 \@ifx{#1\undefined}
}%
\providecommand \@ifnum [1]{%
 \ifnum #1\expandafter \@firstoftwo
 \else \expandafter \@secondoftwo
 \fi
}%
\providecommand \@ifx [1]{%
 \ifx #1\expandafter \@firstoftwo
 \else \expandafter \@secondoftwo
 \fi
}%
\providecommand \natexlab [1]{#1}%
\providecommand \enquote  [1]{``#1''}%
\providecommand \bibnamefont  [1]{#1}%
\providecommand \bibfnamefont [1]{#1}%
\providecommand \citenamefont [1]{#1}%
\providecommand \href@noop [0]{\@secondoftwo}%
\providecommand \href [0]{\begingroup \@sanitize@url \@href}%
\providecommand \@href[1]{\@@startlink{#1}\@@href}%
\providecommand \@@href[1]{\endgroup#1\@@endlink}%
\providecommand \@sanitize@url [0]{\catcode `\\12\catcode `\$12\catcode
  `\&12\catcode `\#12\catcode `\^12\catcode `\_12\catcode `\%12\relax}%
\providecommand \@@startlink[1]{}%
\providecommand \@@endlink[0]{}%
\providecommand \url  [0]{\begingroup\@sanitize@url \@url }%
\providecommand \@url [1]{\endgroup\@href {#1}{\urlprefix }}%
\providecommand \urlprefix  [0]{URL }%
\providecommand \Eprint [0]{\href }%
\providecommand \doibase [0]{http://dx.doi.org/}%
\providecommand \selectlanguage [0]{\@gobble}%
\providecommand \bibinfo  [0]{\@secondoftwo}%
\providecommand \bibfield  [0]{\@secondoftwo}%
\providecommand \translation [1]{[#1]}%
\providecommand \BibitemOpen [0]{}%
\providecommand \bibitemStop [0]{}%
\providecommand \bibitemNoStop [0]{.\EOS\space}%
\providecommand \EOS [0]{\spacefactor3000\relax}%
\providecommand \BibitemShut  [1]{\csname bibitem#1\endcsname}%
\let\auto@bib@innerbib\@empty
\bibitem [{\citenamefont {Gilmore}\ and\ \citenamefont
  {Feng}(1978)}]{Gilmore1978b}%
  \BibitemOpen
  \bibfield  {author} {\bibinfo {author} {\bibfnamefont {R.}~\bibnamefont
  {Gilmore}}\ and\ \bibinfo {author} {\bibfnamefont {D.~H.}\ \bibnamefont
  {Feng}},\ }\href {\doibase http://dx.doi.org/10.1016/0370-2693(78)90090-4}
  {\bibfield  {journal} {\bibinfo  {journal} {Phys. Lett. B}\ }\textbf
  {\bibinfo {volume} {76}},\ \bibinfo {pages} {26} (\bibinfo {year}
  {1978})}\BibitemShut {NoStop}%
\bibitem [{\citenamefont {Gilmore}(1979)}]{Gilmore1979}%
  \BibitemOpen
  \bibfield  {author} {\bibinfo {author} {\bibfnamefont {R.}~\bibnamefont
  {Gilmore}},\ }\href {\doibase http://dx.doi.org/10.1063/1.524137} {\bibfield
  {journal} {\bibinfo  {journal} {J. Math. Phys.}\ }\textbf {\bibinfo {volume}
  {20}},\ \bibinfo {pages} {891} (\bibinfo {year} {1979})}\BibitemShut
  {NoStop}%
\bibitem [{\citenamefont {Cejnar}\ \emph {et~al.}(2010)\citenamefont {Cejnar},
  \citenamefont {Jolie},\ and\ \citenamefont {Casten}}]{Cejnar2010}%
  \BibitemOpen
  \bibfield  {author} {\bibinfo {author} {\bibfnamefont {P.}~\bibnamefont
  {Cejnar}}, \bibinfo {author} {\bibfnamefont {J.}~\bibnamefont {Jolie}}, \
  and\ \bibinfo {author} {\bibfnamefont {R.~F.}\ \bibnamefont {Casten}},\
  }\href {\doibase 10.1103/RevModPhys.82.2155} {\bibfield  {journal} {\bibinfo
  {journal} {Rev. Mod. Phys.}\ }\textbf {\bibinfo {volume} {82}},\ \bibinfo
  {pages} {2155} (\bibinfo {year} {2010})}\BibitemShut {NoStop}%
\bibitem [{\citenamefont {Carr}(2010)}]{carr2010QPT}%
  \BibitemOpen
  \bibinfo {editor} {\bibfnamefont {L.~D.}\ \bibnamefont {Carr}},\ ed.,\
  \href@noop {} {\emph {\bibinfo {title} {{Understanding Quantum Phase
  Transitions}}}}\ (\bibinfo  {publisher} {CRC press},\ \bibinfo {year}
  {2010})\BibitemShut {NoStop}%
\bibitem [{\citenamefont {Dieperink}\ \emph {et~al.}(1980)\citenamefont
  {Dieperink}, \citenamefont {Scholten},\ and\ \citenamefont
  {Iachello}}]{Dieperink1980}%
  \BibitemOpen
  \bibfield  {author} {\bibinfo {author} {\bibfnamefont {A.~E.~L.}\
  \bibnamefont {Dieperink}}, \bibinfo {author} {\bibfnamefont {O.}~\bibnamefont
  {Scholten}}, \ and\ \bibinfo {author} {\bibfnamefont {F.}~\bibnamefont
  {Iachello}},\ }\href {\doibase 10.1103/PhysRevLett.44.1747} {\bibfield
  {journal} {\bibinfo  {journal} {Phys. Rev. Lett.}\ }\textbf {\bibinfo
  {volume} {44}},\ \bibinfo {pages} {1747} (\bibinfo {year}
  {1980})}\BibitemShut {NoStop}%
\bibitem [{\citenamefont {Iachello}(2011)}]{Iachello2011}%
  \BibitemOpen
  \bibfield  {author} {\bibinfo {author} {\bibfnamefont {F.}~\bibnamefont
  {Iachello}},\ }\href {\doibase 10.1393/NCR/I2011-10070-7} {\bibfield
  {journal} {\bibinfo  {journal} {{Rivista} del Nuovo Cim.}\ }\textbf {\bibinfo
  {volume} {34}},\ \bibinfo {pages} {617} (\bibinfo {year} {2011})}\BibitemShut
  {NoStop}%
\bibitem [{\citenamefont {Heyde}\ and\ \citenamefont {Wood}(2011)}]{Heyde2011}%
  \BibitemOpen
  \bibfield  {author} {\bibinfo {author} {\bibfnamefont {K.}~\bibnamefont
  {Heyde}}\ and\ \bibinfo {author} {\bibfnamefont {J.~L.}\ \bibnamefont
  {Wood}},\ }\href {\doibase 10.1103/RevModPhys.83.1467} {\bibfield  {journal}
  {\bibinfo  {journal} {Rev. Mod. Phys.}\ }\textbf {\bibinfo {volume} {83}},\
  \bibinfo {pages} {1467} (\bibinfo {year} {2011})}\BibitemShut {NoStop}%
\bibitem [{\citenamefont {Frank}\ \emph {et~al.}(2006)\citenamefont {Frank},
  \citenamefont {{Van Isacker}},\ and\ \citenamefont {Iachello}}]{Frank2006}%
  \BibitemOpen
  \bibfield  {author} {\bibinfo {author} {\bibfnamefont {A.}~\bibnamefont
  {Frank}}, \bibinfo {author} {\bibfnamefont {P.}~\bibnamefont {{Van
  Isacker}}}, \ and\ \bibinfo {author} {\bibfnamefont {F.}~\bibnamefont
  {Iachello}},\ }\href {\doibase 10.1103/PhysRevC.73.061302} {\bibfield
  {journal} {\bibinfo  {journal} {Phys. Rev. C}\ }\textbf {\bibinfo {volume}
  {73}},\ \bibinfo {pages} {061302(R)} (\bibinfo {year} {2006})}\BibitemShut
  {NoStop}%
\bibitem [{\citenamefont {Federman}\ and\ \citenamefont
  {Pittel}(1979)}]{Federman1979}%
  \BibitemOpen
  \bibfield  {author} {\bibinfo {author} {\bibfnamefont {P.}~\bibnamefont
  {Federman}}\ and\ \bibinfo {author} {\bibfnamefont {S.}~\bibnamefont
  {Pittel}},\ }\href {\doibase 10.1103/PhysRevC.20.820} {\bibfield  {journal}
  {\bibinfo  {journal} {Phys. Rev. C}\ }\textbf {\bibinfo {volume} {20}},\
  \bibinfo {pages} {820} (\bibinfo {year} {1979})}\BibitemShut {NoStop}%
\bibitem [{\citenamefont {Gavrielov}\ \emph {et~al.}(2019)\citenamefont
  {Gavrielov}, \citenamefont {Leviatan},\ and\ \citenamefont
  {Iachello}}]{Gavrielov2019}%
  \BibitemOpen
  \bibfield  {author} {\bibinfo {author} {\bibfnamefont {N.}~\bibnamefont
  {Gavrielov}}, \bibinfo {author} {\bibfnamefont {A.}~\bibnamefont {Leviatan}},
  \ and\ \bibinfo {author} {\bibfnamefont {F.}~\bibnamefont {Iachello}},\
  }\href {\doibase 10.1103/PhysRevC.99.064324} {\bibfield  {journal} {\bibinfo
  {journal} {Phys. Rev. C}\ }\textbf {\bibinfo {volume} {99}},\ \bibinfo
  {pages} {064324} (\bibinfo {year} {2019})}\BibitemShut {NoStop}%
\bibitem [{\citenamefont {Gavrielov}\ \emph {et~al.}(2020)\citenamefont
  {Gavrielov}, \citenamefont {Leviatan},\ and\ \citenamefont
  {Iachello}}]{Gavrielov2020}%
  \BibitemOpen
  \bibfield  {author} {\bibinfo {author} {\bibfnamefont {N.}~\bibnamefont
  {Gavrielov}}, \bibinfo {author} {\bibfnamefont {A.}~\bibnamefont {Leviatan}},
  \ and\ \bibinfo {author} {\bibfnamefont {F.}~\bibnamefont {Iachello}},\
  }\href {\doibase 10.1088/1402-4896/ab456b} {\bibfield  {journal} {\bibinfo
  {journal} {Phys. Scr.}\ }\textbf {\bibinfo {volume} {95}},\ \bibinfo {pages}
  {024001} (\bibinfo {year} {2020})}\BibitemShut {NoStop}%
\bibitem [{\citenamefont {Gavrielov}\ \emph
  {et~al.}(2022{\natexlab{a}})\citenamefont {Gavrielov}, \citenamefont
  {Leviatan},\ and\ \citenamefont {Iachello}}]{Gavrielov2022}%
  \BibitemOpen
  \bibfield  {author} {\bibinfo {author} {\bibfnamefont {N.}~\bibnamefont
  {Gavrielov}}, \bibinfo {author} {\bibfnamefont {A.}~\bibnamefont {Leviatan}},
  \ and\ \bibinfo {author} {\bibfnamefont {F.}~\bibnamefont {Iachello}},\
  }\href {\doibase 10.1103/PhysRevC.105.014305} {\bibfield  {journal} {\bibinfo
   {journal} {Phys. Rev. C}\ }\textbf {\bibinfo {volume} {105}},\ \bibinfo
  {pages} {014305} (\bibinfo {year} {2022}{\natexlab{a}})}\BibitemShut
  {NoStop}%
\bibitem [{\citenamefont {Casten}(2009)}]{Casten2009}%
  \BibitemOpen
  \bibfield  {author} {\bibinfo {author} {\bibfnamefont {R.~F.}\ \bibnamefont
  {Casten}},\ }\href {\doibase http://dx.doi.org/10.1016/j.ppnp.2008.06.002}
  {\bibfield  {journal} {\bibinfo  {journal} {Prog. Part. Nucl. Phys.}\
  }\textbf {\bibinfo {volume} {62}},\ \bibinfo {pages} {183} (\bibinfo {year}
  {2009})}\BibitemShut {NoStop}%
\bibitem [{\citenamefont {Fortunato}(2021)}]{Fortunato2021}%
  \BibitemOpen
  \bibfield  {author} {\bibinfo {author} {\bibfnamefont {L.}~\bibnamefont
  {Fortunato}},\ }\href {\doibase 10.1016/J.PPNP.2021.103891} {\bibfield
  {journal} {\bibinfo  {journal} {Prog. Part. Nucl. Phys.}\ }\textbf {\bibinfo
  {volume} {121}},\ \bibinfo {pages} {103891} (\bibinfo {year}
  {2021})}\BibitemShut {NoStop}%
\bibitem [{\citenamefont {Caurier}\ \emph {et~al.}(2005)\citenamefont
  {Caurier}, \citenamefont {Mart{\'{i}}nez-Pinedo}, \citenamefont {Nowacki},
  \citenamefont {Poves},\ and\ \citenamefont {Zuker}}]{Caurier2005}%
  \BibitemOpen
  \bibfield  {author} {\bibinfo {author} {\bibfnamefont {E.}~\bibnamefont
  {Caurier}}, \bibinfo {author} {\bibfnamefont {G.}~\bibnamefont
  {Mart{\'{i}}nez-Pinedo}}, \bibinfo {author} {\bibfnamefont {F.}~\bibnamefont
  {Nowacki}}, \bibinfo {author} {\bibfnamefont {A.}~\bibnamefont {Poves}}, \
  and\ \bibinfo {author} {\bibfnamefont {A.~P.}\ \bibnamefont {Zuker}},\ }\href
  {\doibase 10.1103/RevModPhys.77.427} {\bibfield  {journal} {\bibinfo
  {journal} {Rev. Mod. Phys.}\ }\textbf {\bibinfo {volume} {77}},\ \bibinfo
  {pages} {427} (\bibinfo {year} {2005})},\ \Eprint
  {http://arxiv.org/abs/0402046v1} {arXiv:0402046v1 [arXiv:nucl-th]}
  \BibitemShut {NoStop}%
\bibitem [{\citenamefont {Bally}\ \emph {et~al.}(2014)\citenamefont {Bally},
  \citenamefont {Avez}, \citenamefont {Bender},\ and\ \citenamefont
  {Heenen}}]{Bally2014}%
  \BibitemOpen
  \bibfield  {author} {\bibinfo {author} {\bibfnamefont {B.}~\bibnamefont
  {Bally}}, \bibinfo {author} {\bibfnamefont {B.}~\bibnamefont {Avez}},
  \bibinfo {author} {\bibfnamefont {M.}~\bibnamefont {Bender}}, \ and\ \bibinfo
  {author} {\bibfnamefont {P.-H.}\ \bibnamefont {Heenen}},\ }\href {\doibase
  10.1103/PhysRevLett.113.162501} {\bibfield  {journal} {\bibinfo  {journal}
  {Phys. Rev. Lett.}\ }\textbf {\bibinfo {volume} {113}},\ \bibinfo {pages}
  {162501} (\bibinfo {year} {2014})}\BibitemShut {NoStop}%
\bibitem [{\citenamefont {Scholten}\ and\ \citenamefont
  {Blasi}(1982)}]{Scholten1982}%
  \BibitemOpen
  \bibfield  {author} {\bibinfo {author} {\bibfnamefont {O.}~\bibnamefont
  {Scholten}}\ and\ \bibinfo {author} {\bibfnamefont {N.}~\bibnamefont
  {Blasi}},\ }\href {\doibase 10.1016/0375-9474(82)90575-9} {\bibfield
  {journal} {\bibinfo  {journal} {Nucl. Phys. A}\ }\textbf {\bibinfo {volume}
  {380}},\ \bibinfo {pages} {509} (\bibinfo {year} {1982})}\BibitemShut
  {NoStop}%
\bibitem [{\citenamefont {Iachello}\ and\ \citenamefont {{Van
  Isacker}}(1991)}]{IachelloVanIsackerBook}%
  \BibitemOpen
  \bibfield  {author} {\bibinfo {author} {\bibfnamefont {F.}~\bibnamefont
  {Iachello}}\ and\ \bibinfo {author} {\bibfnamefont {P.}~\bibnamefont {{Van
  Isacker}}},\ }\href@noop {} {\emph {\bibinfo {title} {{The Interacting
  Boson-Fermion Model}}}}\ (\bibinfo  {publisher} {Cambridge University
  Press},\ \bibinfo {year} {1991})\BibitemShut {NoStop}%
\bibitem [{\citenamefont {Jolie}\ \emph {et~al.}(2004)\citenamefont {Jolie},
  \citenamefont {Heinze}, \citenamefont {{Van Isacker}},\ and\ \citenamefont
  {Casten}}]{Jolie2004}%
  \BibitemOpen
  \bibfield  {author} {\bibinfo {author} {\bibfnamefont {J.}~\bibnamefont
  {Jolie}}, \bibinfo {author} {\bibfnamefont {S.}~\bibnamefont {Heinze}},
  \bibinfo {author} {\bibfnamefont {P.}~\bibnamefont {{Van Isacker}}}, \ and\
  \bibinfo {author} {\bibfnamefont {R.~F.}\ \bibnamefont {Casten}},\ }\href
  {\doibase 10.1103/PhysRevC.70.011305} {\bibfield  {journal} {\bibinfo
  {journal} {Phys. Rev. C}\ }\textbf {\bibinfo {volume} {70}},\ \bibinfo
  {pages} {011305(R)} (\bibinfo {year} {2004})}\BibitemShut {NoStop}%
\bibitem [{\citenamefont {Alonso}\ \emph {et~al.}(2005)\citenamefont {Alonso},
  \citenamefont {Arias}, \citenamefont {Fortunato},\ and\ \citenamefont
  {Vitturi}}]{Alonso2005}%
  \BibitemOpen
  \bibfield  {author} {\bibinfo {author} {\bibfnamefont {C.~E.}\ \bibnamefont
  {Alonso}}, \bibinfo {author} {\bibfnamefont {J.~M.}\ \bibnamefont {Arias}},
  \bibinfo {author} {\bibfnamefont {L.}~\bibnamefont {Fortunato}}, \ and\
  \bibinfo {author} {\bibfnamefont {A.}~\bibnamefont {Vitturi}},\ }\href
  {\doibase 10.1103/PhysRevC.72.061302} {\bibfield  {journal} {\bibinfo
  {journal} {Phys. Rev. C}\ }\textbf {\bibinfo {volume} {72}},\ \bibinfo
  {pages} {061302(R)} (\bibinfo {year} {2005})}\BibitemShut {NoStop}%
\bibitem [{\citenamefont {Alonso}\ \emph {et~al.}(2007)\citenamefont {Alonso},
  \citenamefont {Arias},\ and\ \citenamefont {Vitturi}}]{Alonso2007}%
  \BibitemOpen
  \bibfield  {author} {\bibinfo {author} {\bibfnamefont {C.~E.}\ \bibnamefont
  {Alonso}}, \bibinfo {author} {\bibfnamefont {J.~M.}\ \bibnamefont {Arias}}, \
  and\ \bibinfo {author} {\bibfnamefont {A.}~\bibnamefont {Vitturi}},\ }\href
  {\doibase 10.1103/PhysRevC.75.064316} {\bibfield  {journal} {\bibinfo
  {journal} {Phys. Rev. C}\ }\textbf {\bibinfo {volume} {75}},\ \bibinfo
  {pages} {064316} (\bibinfo {year} {2007})}\BibitemShut {NoStop}%
\bibitem [{\citenamefont {Alonso}\ \emph {et~al.}(2009)\citenamefont {Alonso},
  \citenamefont {Arias}, \citenamefont {Fortunato},\ and\ \citenamefont
  {Vitturi}}]{Alonso2009}%
  \BibitemOpen
  \bibfield  {author} {\bibinfo {author} {\bibfnamefont {C.~E.}\ \bibnamefont
  {Alonso}}, \bibinfo {author} {\bibfnamefont {J.~M.}\ \bibnamefont {Arias}},
  \bibinfo {author} {\bibfnamefont {L.}~\bibnamefont {Fortunato}}, \ and\
  \bibinfo {author} {\bibfnamefont {A.}~\bibnamefont {Vitturi}},\ }\href
  {\doibase 10.1103/PhysRevC.79.014306} {\bibfield  {journal} {\bibinfo
  {journal} {Phys. Rev. C}\ }\textbf {\bibinfo {volume} {79}},\ \bibinfo
  {pages} {014306} (\bibinfo {year} {2009})}\BibitemShut {NoStop}%
\bibitem [{\citenamefont {B\"{o}y\"{u}kata}\ \emph {et~al.}(2010)\citenamefont
  {B\"{o}y\"{u}kata}, \citenamefont {{Van Isacker}},\ and\ \citenamefont
  {Uluer}}]{Boyukata2010}%
  \BibitemOpen
  \bibfield  {author} {\bibinfo {author} {\bibfnamefont {M.}~\bibnamefont
  {B\"{o}y\"{u}kata}}, \bibinfo {author} {\bibfnamefont {P.}~\bibnamefont {{Van
  Isacker}}}, \ and\ \bibinfo {author} {\bibfnamefont {I.}~\bibnamefont
  {Uluer}},\ }\href {\doibase 10.1088/0954-3899/37/10/105102} {\bibfield
  {journal} {\bibinfo  {journal} {J. Phys. G Nucl. Part. Phys.}\ }\textbf
  {\bibinfo {volume} {37}},\ \bibinfo {pages} {105102} (\bibinfo {year}
  {2010})}\BibitemShut {NoStop}%
\bibitem [{\citenamefont {Petrellis}\ \emph {et~al.}(2011)\citenamefont
  {Petrellis}, \citenamefont {Leviatan},\ and\ \citenamefont
  {Iachello}}]{Petrellis2011a}%
  \BibitemOpen
  \bibfield  {author} {\bibinfo {author} {\bibfnamefont {D.}~\bibnamefont
  {Petrellis}}, \bibinfo {author} {\bibfnamefont {A.}~\bibnamefont {Leviatan}},
  \ and\ \bibinfo {author} {\bibfnamefont {F.}~\bibnamefont {Iachello}},\
  }\href {\doibase http://dx.doi.org/10.1016/j.aop.2010.12.001} {\bibfield
  {journal} {\bibinfo  {journal} {Ann. Phys.}\ }\textbf {\bibinfo {volume}
  {326}},\ \bibinfo {pages} {926} (\bibinfo {year} {2011})}\BibitemShut
  {NoStop}%
\bibitem [{\citenamefont {Iachello}\ \emph {et~al.}(2011)\citenamefont
  {Iachello}, \citenamefont {Leviatan},\ and\ \citenamefont
  {Petrellis}}]{Iachello2011b}%
  \BibitemOpen
  \bibfield  {author} {\bibinfo {author} {\bibfnamefont {F.}~\bibnamefont
  {Iachello}}, \bibinfo {author} {\bibfnamefont {A.}~\bibnamefont {Leviatan}},
  \ and\ \bibinfo {author} {\bibfnamefont {D.}~\bibnamefont {Petrellis}},\
  }\href {\doibase 10.1016/j.physletb.2011.10.024} {\bibfield  {journal}
  {\bibinfo  {journal} {Phys. Lett. B}\ }\textbf {\bibinfo {volume} {705}},\
  \bibinfo {pages} {379} (\bibinfo {year} {2011})}\BibitemShut {NoStop}%
\bibitem [{\citenamefont {B\"{o}y\"{u}kata}\ \emph {et~al.}(2021)\citenamefont
  {B\"{o}y\"{u}kata}, \citenamefont {Alonso}, \citenamefont {Arias},
  \citenamefont {Fortunato},\ and\ \citenamefont {Vitturi}}]{Boyukata2021}%
  \BibitemOpen
  \bibfield  {author} {\bibinfo {author} {\bibfnamefont {M.}~\bibnamefont
  {B\"{o}y\"{u}kata}}, \bibinfo {author} {\bibfnamefont {C.~E.}\ \bibnamefont
  {Alonso}}, \bibinfo {author} {\bibfnamefont {J.~M.}\ \bibnamefont {Arias}},
  \bibinfo {author} {\bibfnamefont {L.}~\bibnamefont {Fortunato}}, \ and\
  \bibinfo {author} {\bibfnamefont {A.}~\bibnamefont {Vitturi}},\ }\href
  {\doibase 10.3390/sym13020215} {\bibfield  {journal} {\bibinfo  {journal}
  {Symmetry (Basel).}\ }\textbf {\bibinfo {volume} {13}},\ \bibinfo {pages}
  {215} (\bibinfo {year} {2021})}\BibitemShut {NoStop}%
\bibitem [{\citenamefont {Nomura}\ \emph
  {et~al.}(2016{\natexlab{a}})\citenamefont {Nomura}, \citenamefont {Otsuka},\
  and\ \citenamefont {{Van Isacker}}}]{Nomura2016a}%
  \BibitemOpen
  \bibfield  {author} {\bibinfo {author} {\bibfnamefont {K.}~\bibnamefont
  {Nomura}}, \bibinfo {author} {\bibfnamefont {T.}~\bibnamefont {Otsuka}}, \
  and\ \bibinfo {author} {\bibfnamefont {P.}~\bibnamefont {{Van Isacker}}},\
  }\href {\doibase 10.1088/0954-3899/43/2/024008} {\bibfield  {journal}
  {\bibinfo  {journal} {J. Phys. G}\ }\textbf {\bibinfo {volume} {43}},\
  \bibinfo {pages} {024008} (\bibinfo {year} {2016}{\natexlab{a}})}\BibitemShut
  {NoStop}%
\bibitem [{\citenamefont {Nomura}\ \emph
  {et~al.}(2016{\natexlab{b}})\citenamefont {Nomura}, \citenamefont
  {Nik{\v{s}}i{\'{c}}},\ and\ \citenamefont {Vretenar}}]{Nomura2016b}%
  \BibitemOpen
  \bibfield  {author} {\bibinfo {author} {\bibfnamefont {K.}~\bibnamefont
  {Nomura}}, \bibinfo {author} {\bibfnamefont {T.}~\bibnamefont
  {Nik{\v{s}}i{\'{c}}}}, \ and\ \bibinfo {author} {\bibfnamefont
  {D.}~\bibnamefont {Vretenar}},\ }\href {\doibase 10.1103/PhysRevC.93.054305}
  {\bibfield  {journal} {\bibinfo  {journal} {Phys. Rev. C}\ }\textbf {\bibinfo
  {volume} {93}},\ \bibinfo {pages} {054305} (\bibinfo {year}
  {2016}{\natexlab{b}})}\BibitemShut {NoStop}%
\bibitem [{\citenamefont {Nomura}\ \emph {et~al.}(2020)\citenamefont {Nomura},
  \citenamefont {Nik{\v{s}}i{\'{c}}},\ and\ \citenamefont
  {Vretenar}}]{Nomura2020}%
  \BibitemOpen
  \bibfield  {author} {\bibinfo {author} {\bibfnamefont {K.}~\bibnamefont
  {Nomura}}, \bibinfo {author} {\bibfnamefont {T.}~\bibnamefont
  {Nik{\v{s}}i{\'{c}}}}, \ and\ \bibinfo {author} {\bibfnamefont
  {D.}~\bibnamefont {Vretenar}},\ }\href {\doibase 10.1103/PhysRevC.102.034315}
  {\bibfield  {journal} {\bibinfo  {journal} {Phys. Rev. C}\ }\textbf {\bibinfo
  {volume} {102}},\ \bibinfo {pages} {034315} (\bibinfo {year} {2020})},\
  \Eprint {http://arxiv.org/abs/2006.16662} {arXiv:2006.16662} \BibitemShut
  {NoStop}%
\bibitem [{\citenamefont {Quan}\ \emph {et~al.}(2018)\citenamefont {Quan},
  \citenamefont {Li}, \citenamefont {Vretenar},\ and\ \citenamefont
  {Meng}}]{Quan2018}%
  \BibitemOpen
  \bibfield  {author} {\bibinfo {author} {\bibfnamefont {S.}~\bibnamefont
  {Quan}}, \bibinfo {author} {\bibfnamefont {Z.~P.}\ \bibnamefont {Li}},
  \bibinfo {author} {\bibfnamefont {D.}~\bibnamefont {Vretenar}}, \ and\
  \bibinfo {author} {\bibfnamefont {J.}~\bibnamefont {Meng}},\ }\href {\doibase
  10.1103/PhysRevC.97.031301} {\bibfield  {journal} {\bibinfo  {journal} {Phys.
  Rev. C}\ }\textbf {\bibinfo {volume} {97}},\ \bibinfo {pages} {031301(R)}
  (\bibinfo {year} {2018})}\BibitemShut {NoStop}%
\bibitem [{\citenamefont {Brant}\ \emph {et~al.}(1998)\citenamefont {Brant},
  \citenamefont {Paar},\ and\ \citenamefont {Wolf}}]{Brant1998}%
  \BibitemOpen
  \bibfield  {author} {\bibinfo {author} {\bibfnamefont {S.}~\bibnamefont
  {Brant}}, \bibinfo {author} {\bibfnamefont {V.}~\bibnamefont {Paar}}, \ and\
  \bibinfo {author} {\bibfnamefont {A.}~\bibnamefont {Wolf}},\ }\href {\doibase
  10.1103/PhysRevC.58.1349} {\bibfield  {journal} {\bibinfo  {journal} {Phys.
  Rev. C}\ }\textbf {\bibinfo {volume} {58}},\ \bibinfo {pages} {1349}
  (\bibinfo {year} {1998})}\BibitemShut {NoStop}%
\bibitem [{\citenamefont {Rodriguez-Guzman}\ \emph {et~al.}(2011)\citenamefont
  {Rodriguez-Guzman}, \citenamefont {Sarriguren},\ and\ \citenamefont
  {Robledo}}]{Rodriguez-Guzman2011}%
  \BibitemOpen
  \bibfield  {author} {\bibinfo {author} {\bibfnamefont {R.}~\bibnamefont
  {Rodriguez-Guzman}}, \bibinfo {author} {\bibfnamefont {P.}~\bibnamefont
  {Sarriguren}}, \ and\ \bibinfo {author} {\bibfnamefont {L.~M.}\ \bibnamefont
  {Robledo}},\ }\href {\doibase 10.1103/PhysRevC.83.044307} {\bibfield
  {journal} {\bibinfo  {journal} {Phys. Rev. C}\ }\textbf {\bibinfo {volume}
  {83}},\ \bibinfo {pages} {044307} (\bibinfo {year} {2011})}\BibitemShut
  {NoStop}%
\bibitem [{\citenamefont {Spagnoletti}\ \emph {et~al.}(2019)\citenamefont
  {Spagnoletti}, \citenamefont {Simpson}, \citenamefont {Kisyov}, \citenamefont
  {Bucurescu}, \citenamefont {R\'{e}gis}, \citenamefont {Saed-Samii},
  \citenamefont {Blanc}, \citenamefont {Jentschel}, \citenamefont {K\"{o}ster},
  \citenamefont {Mutti}, \citenamefont {Soldner}, \citenamefont {de~France},
  \citenamefont {Ur}, \citenamefont {Urban}, \citenamefont {Bruce},
  \citenamefont {Bernards}, \citenamefont {Drouet}, \citenamefont {Fraile},
  \citenamefont {Gaffney}, \citenamefont {Ghit\v{a}}, \citenamefont {Ilieva},
  \citenamefont {Jolie}, \citenamefont {Korten}, \citenamefont {Kr\"{o}ll},
  \citenamefont {Lalkovski}, \citenamefont {Larijarni}, \citenamefont
  {Lic\v{a}}, \citenamefont {Mach}, \citenamefont {M\v{a}rginean},
  \citenamefont {Paziy}, \citenamefont {Podoly\`{a}k}, \citenamefont {Regan},
  \citenamefont {Scheck}, \citenamefont {Smith}, \citenamefont {Thiamova},
  \citenamefont {Townsley}, \citenamefont {Vancraeyenest}, \citenamefont
  {Vedia}, \citenamefont {Warr}, \citenamefont {Werner},\ and\ \citenamefont
  {Zieli\`{n}ska}}]{Spagnoletti2019a}%
  \BibitemOpen
  \bibfield  {author} {\bibinfo {author} {\bibfnamefont {P.}~\bibnamefont
  {Spagnoletti}}, \bibinfo {author} {\bibfnamefont {G.}~\bibnamefont
  {Simpson}}, \bibinfo {author} {\bibfnamefont {S.}~\bibnamefont {Kisyov}},
  \bibinfo {author} {\bibfnamefont {D.}~\bibnamefont {Bucurescu}}, \bibinfo
  {author} {\bibfnamefont {J.-M.}\ \bibnamefont {R\'{e}gis}}, \bibinfo {author}
  {\bibfnamefont {N.}~\bibnamefont {Saed-Samii}}, \bibinfo {author}
  {\bibfnamefont {A.}~\bibnamefont {Blanc}}, \bibinfo {author} {\bibfnamefont
  {M.}~\bibnamefont {Jentschel}}, \bibinfo {author} {\bibfnamefont
  {U.}~\bibnamefont {K\"{o}ster}}, \bibinfo {author} {\bibfnamefont
  {P.}~\bibnamefont {Mutti}}, \bibinfo {author} {\bibfnamefont
  {T.}~\bibnamefont {Soldner}}, \bibinfo {author} {\bibfnamefont
  {G.}~\bibnamefont {de~France}}, \bibinfo {author} {\bibfnamefont {C.~A.}\
  \bibnamefont {Ur}}, \bibinfo {author} {\bibfnamefont {W.}~\bibnamefont
  {Urban}}, \bibinfo {author} {\bibfnamefont {A.~M.}\ \bibnamefont {Bruce}},
  \bibinfo {author} {\bibfnamefont {C.}~\bibnamefont {Bernards}}, \bibinfo
  {author} {\bibfnamefont {F.}~\bibnamefont {Drouet}}, \bibinfo {author}
  {\bibfnamefont {L.~M.}\ \bibnamefont {Fraile}}, \bibinfo {author}
  {\bibfnamefont {L.~P.}\ \bibnamefont {Gaffney}}, \bibinfo {author}
  {\bibfnamefont {D.~G.}\ \bibnamefont {Ghit\v{a}}}, \bibinfo {author}
  {\bibfnamefont {S.}~\bibnamefont {Ilieva}}, \bibinfo {author} {\bibfnamefont
  {J.}~\bibnamefont {Jolie}}, \bibinfo {author} {\bibfnamefont
  {W.}~\bibnamefont {Korten}}, \bibinfo {author} {\bibfnamefont
  {T.}~\bibnamefont {Kr\"{o}ll}}, \bibinfo {author} {\bibfnamefont
  {S.}~\bibnamefont {Lalkovski}}, \bibinfo {author} {\bibfnamefont
  {C.}~\bibnamefont {Larijarni}}, \bibinfo {author} {\bibfnamefont
  {R.}~\bibnamefont {Lic\v{a}}}, \bibinfo {author} {\bibfnamefont
  {H.}~\bibnamefont {Mach}}, \bibinfo {author} {\bibfnamefont {N.}~\bibnamefont
  {M\v{a}rginean}}, \bibinfo {author} {\bibfnamefont {V.}~\bibnamefont
  {Paziy}}, \bibinfo {author} {\bibfnamefont {Z.}~\bibnamefont {Podoly\`{a}k}},
  \bibinfo {author} {\bibfnamefont {P.~H.}\ \bibnamefont {Regan}}, \bibinfo
  {author} {\bibfnamefont {M.}~\bibnamefont {Scheck}}, \bibinfo {author}
  {\bibfnamefont {J.~F.}\ \bibnamefont {Smith}}, \bibinfo {author}
  {\bibfnamefont {G.}~\bibnamefont {Thiamova}}, \bibinfo {author}
  {\bibfnamefont {C.}~\bibnamefont {Townsley}}, \bibinfo {author}
  {\bibfnamefont {A.}~\bibnamefont {Vancraeyenest}}, \bibinfo {author}
  {\bibfnamefont {V.}~\bibnamefont {Vedia}}, \bibinfo {author} {\bibfnamefont
  {N.}~\bibnamefont {Warr}}, \bibinfo {author} {\bibfnamefont {V.}~\bibnamefont
  {Werner}}, \ and\ \bibinfo {author} {\bibfnamefont {M.}~\bibnamefont
  {Zieli\`{n}ska}},\ }\href {\doibase 10.1103/PhysRevC.100.014311} {\bibfield
  {journal} {\bibinfo  {journal} {Phys. Rev. C}\ }\textbf {\bibinfo {volume}
  {100}},\ \bibinfo {pages} {014311} (\bibinfo {year} {2019})}\BibitemShut
  {NoStop}%
\bibitem [{\citenamefont {Garrett}(2021)}]{Garrett2021}%
  \BibitemOpen
  \bibfield  {author} {\bibinfo {author} {\bibfnamefont {P.~E.}\ \bibnamefont
  {Garrett}},\ }\href {\doibase 10.1103/PhysRevLett.127.169201} {\bibfield
  {journal} {\bibinfo  {journal} {Phys. Rev. Lett.}\ }\textbf {\bibinfo
  {volume} {127}},\ \bibinfo {pages} {169201} (\bibinfo {year}
  {2021})}\BibitemShut {NoStop}%
\bibitem [{\citenamefont {Cheifetz}\ \emph {et~al.}(1970)\citenamefont
  {Cheifetz}, \citenamefont {Jared}, \citenamefont {Thompson},\ and\
  \citenamefont {Wilhelmy}}]{Cheifetz1970}%
  \BibitemOpen
  \bibfield  {author} {\bibinfo {author} {\bibfnamefont {E.}~\bibnamefont
  {Cheifetz}}, \bibinfo {author} {\bibfnamefont {R.~C.}\ \bibnamefont {Jared}},
  \bibinfo {author} {\bibfnamefont {S.~G.}\ \bibnamefont {Thompson}}, \ and\
  \bibinfo {author} {\bibfnamefont {J.~B.}\ \bibnamefont {Wilhelmy}},\ }\href
  {\doibase 10.1103/PhysRevLett.25.38} {\bibfield  {journal} {\bibinfo
  {journal} {Phys. Rev. Lett.}\ }\textbf {\bibinfo {volume} {25}},\ \bibinfo
  {pages} {38} (\bibinfo {year} {1970})}\BibitemShut {NoStop}%
\bibitem [{\citenamefont {Heyde}\ \emph {et~al.}(1985)\citenamefont {Heyde},
  \citenamefont {{Van Isacker}}, \citenamefont {Casten},\ and\ \citenamefont
  {Wood}}]{Heyde1985}%
  \BibitemOpen
  \bibfield  {author} {\bibinfo {author} {\bibfnamefont {K.}~\bibnamefont
  {Heyde}}, \bibinfo {author} {\bibfnamefont {P.}~\bibnamefont {{Van
  Isacker}}}, \bibinfo {author} {\bibfnamefont {R.~F.}\ \bibnamefont {Casten}},
  \ and\ \bibinfo {author} {\bibfnamefont {J.~L.}\ \bibnamefont {Wood}},\
  }\href {\doibase http://dx.doi.org/10.1016/0370-2693(85)91575-8} {\bibfield
  {journal} {\bibinfo  {journal} {Phys. Lett. B}\ }\textbf {\bibinfo {volume}
  {155}},\ \bibinfo {pages} {303} (\bibinfo {year} {1985})}\BibitemShut
  {NoStop}%
\bibitem [{\citenamefont {Heyde}\ \emph {et~al.}(1987)\citenamefont {Heyde},
  \citenamefont {Jolie}, \citenamefont {Moreau}, \citenamefont {Ryckebusch},
  \citenamefont {Waroquier}, \citenamefont {Duppen}, \citenamefont {Huyse},\
  and\ \citenamefont {Wood}}]{Heyde1987}%
  \BibitemOpen
  \bibfield  {author} {\bibinfo {author} {\bibfnamefont {K.}~\bibnamefont
  {Heyde}}, \bibinfo {author} {\bibfnamefont {J.}~\bibnamefont {Jolie}},
  \bibinfo {author} {\bibfnamefont {J.}~\bibnamefont {Moreau}}, \bibinfo
  {author} {\bibfnamefont {J.}~\bibnamefont {Ryckebusch}}, \bibinfo {author}
  {\bibfnamefont {M.}~\bibnamefont {Waroquier}}, \bibinfo {author}
  {\bibfnamefont {P.~V.}\ \bibnamefont {Duppen}}, \bibinfo {author}
  {\bibfnamefont {M.}~\bibnamefont {Huyse}}, \ and\ \bibinfo {author}
  {\bibfnamefont {J.~L.}\ \bibnamefont {Wood}},\ }\href {\doibase
  http://dx.doi.org/10.1016/0375-9474(87)90439-8} {\bibfield  {journal}
  {\bibinfo  {journal} {Nucl. Phys. A}\ }\textbf {\bibinfo {volume} {466}},\
  \bibinfo {pages} {189} (\bibinfo {year} {1987})}\BibitemShut {NoStop}%
\bibitem [{\citenamefont {Federman}\ and\ \citenamefont
  {Pittel}(1977)}]{Federman1977}%
  \BibitemOpen
  \bibfield  {author} {\bibinfo {author} {\bibfnamefont {P.}~\bibnamefont
  {Federman}}\ and\ \bibinfo {author} {\bibfnamefont {S.}~\bibnamefont
  {Pittel}},\ }\href {\doibase http://dx.doi.org/10.1016/0370-2693(77)90825-5}
  {\bibfield  {journal} {\bibinfo  {journal} {Phys. Lett. B}\ }\textbf
  {\bibinfo {volume} {69}},\ \bibinfo {pages} {385} (\bibinfo {year}
  {1977})}\BibitemShut {NoStop}%
\bibitem [{\citenamefont {Federman}\ \emph {et~al.}(1979)\citenamefont
  {Federman}, \citenamefont {Pittel},\ and\ \citenamefont
  {Campos}}]{Federman1979b}%
  \BibitemOpen
  \bibfield  {author} {\bibinfo {author} {\bibfnamefont {P.}~\bibnamefont
  {Federman}}, \bibinfo {author} {\bibfnamefont {S.}~\bibnamefont {Pittel}}, \
  and\ \bibinfo {author} {\bibfnamefont {R.}~\bibnamefont {Campos}},\ }\href
  {\doibase 10.1016/0370-2693(79)90412-X} {\bibfield  {journal} {\bibinfo
  {journal} {Phys. Lett. B}\ }\textbf {\bibinfo {volume} {82}},\ \bibinfo
  {pages} {9} (\bibinfo {year} {1979})}\BibitemShut {NoStop}%
\bibitem [{\citenamefont {Federman}\ \emph {et~al.}(1984)\citenamefont
  {Federman}, \citenamefont {Pittel},\ and\ \citenamefont
  {Etchegoyen}}]{Federman1984}%
  \BibitemOpen
  \bibfield  {author} {\bibinfo {author} {\bibfnamefont {P.}~\bibnamefont
  {Federman}}, \bibinfo {author} {\bibfnamefont {S.}~\bibnamefont {Pittel}}, \
  and\ \bibinfo {author} {\bibfnamefont {A.}~\bibnamefont {Etchegoyen}},\
  }\href {\doibase 10.1016/0370-2693(84)90750-0} {\bibfield  {journal}
  {\bibinfo  {journal} {Phys. Lett. B}\ }\textbf {\bibinfo {volume} {140}},\
  \bibinfo {pages} {269} (\bibinfo {year} {1984})}\BibitemShut {NoStop}%
\bibitem [{\citenamefont {Mach}\ \emph {et~al.}(1990)\citenamefont {Mach},
  \citenamefont {Warburton}, \citenamefont {Krips}, \citenamefont {Gill},\ and\
  \citenamefont {Moszy{\'{n}}ski}}]{Mach1990}%
  \BibitemOpen
  \bibfield  {author} {\bibinfo {author} {\bibfnamefont {H.}~\bibnamefont
  {Mach}}, \bibinfo {author} {\bibfnamefont {E.~K.}\ \bibnamefont {Warburton}},
  \bibinfo {author} {\bibfnamefont {W.}~\bibnamefont {Krips}}, \bibinfo
  {author} {\bibfnamefont {R.~L.}\ \bibnamefont {Gill}}, \ and\ \bibinfo
  {author} {\bibfnamefont {M.}~\bibnamefont {Moszy{\'{n}}ski}},\ }\href
  {\doibase 10.1103/PhysRevC.42.568} {\bibfield  {journal} {\bibinfo  {journal}
  {Phys. Rev. C}\ }\textbf {\bibinfo {volume} {42}},\ \bibinfo {pages} {568}
  (\bibinfo {year} {1990})}\BibitemShut {NoStop}%
\bibitem [{\citenamefont {Togashi}\ \emph {et~al.}(2016)\citenamefont
  {Togashi}, \citenamefont {Tsunoda}, \citenamefont {Otsuka},\ and\
  \citenamefont {Shimizu}}]{Togashi2016}%
  \BibitemOpen
  \bibfield  {author} {\bibinfo {author} {\bibfnamefont {T.}~\bibnamefont
  {Togashi}}, \bibinfo {author} {\bibfnamefont {Y.}~\bibnamefont {Tsunoda}},
  \bibinfo {author} {\bibfnamefont {T.}~\bibnamefont {Otsuka}}, \ and\ \bibinfo
  {author} {\bibfnamefont {N.}~\bibnamefont {Shimizu}},\ }\href {\doibase
  10.1103/PhysRevLett.117.172502} {\bibfield  {journal} {\bibinfo  {journal}
  {Phys. Rev. Lett.}\ }\textbf {\bibinfo {volume} {117}},\ \bibinfo {pages}
  {172502} (\bibinfo {year} {2016})}\BibitemShut {NoStop}%
\bibitem [{\citenamefont {Garrett}\ \emph {et~al.}(2022)\citenamefont
  {Garrett}, \citenamefont {Zieli{\'{n}}ska},\ and\ \citenamefont
  {Cl{\'{e}}ment}}]{Garrett2022}%
  \BibitemOpen
  \bibfield  {author} {\bibinfo {author} {\bibfnamefont {P.~E.}\ \bibnamefont
  {Garrett}}, \bibinfo {author} {\bibfnamefont {M.}~\bibnamefont
  {Zieli{\'{n}}ska}}, \ and\ \bibinfo {author} {\bibfnamefont {E.}~\bibnamefont
  {Cl{\'{e}}ment}},\ }\href {\doibase 10.1016/J.PPNP.2021.103931} {\bibfield
  {journal} {\bibinfo  {journal} {Prog. Part. Nucl. Phys.}\ }\textbf {\bibinfo
  {volume} {124}},\ \bibinfo {pages} {103931} (\bibinfo {year}
  {2022})}\BibitemShut {NoStop}%
\bibitem [{\citenamefont {Lhersonneau}\ \emph {et~al.}(1995)\citenamefont
  {Lhersonneau}, \citenamefont {Gabelmann}, \citenamefont {Liang},
  \citenamefont {Pfeiffer}, \citenamefont {Kratz}, \citenamefont {Ohm},\ and\
  \citenamefont {ISOLDE}}]{Lhersonneau1995}%
  \BibitemOpen
  \bibfield  {author} {\bibinfo {author} {\bibfnamefont {G.}~\bibnamefont
  {Lhersonneau}}, \bibinfo {author} {\bibfnamefont {H.}~\bibnamefont
  {Gabelmann}}, \bibinfo {author} {\bibfnamefont {M.}~\bibnamefont {Liang}},
  \bibinfo {author} {\bibfnamefont {B.}~\bibnamefont {Pfeiffer}}, \bibinfo
  {author} {\bibfnamefont {K.-L.}\ \bibnamefont {Kratz}}, \bibinfo {author}
  {\bibfnamefont {H.}~\bibnamefont {Ohm}}, \ and\ \bibinfo {author}
  {\bibfnamefont {C.}~\bibnamefont {ISOLDE}},\ }\href {\doibase
  10.1103/PhysRevC.51.1211} {\bibfield  {journal} {\bibinfo  {journal} {Phys.
  Rev. C}\ }\textbf {\bibinfo {volume} {51}},\ \bibinfo {pages} {1211}
  (\bibinfo {year} {1995})}\BibitemShut {NoStop}%
\bibitem [{\citenamefont {Lhersonneau}\ \emph {et~al.}(1997)\citenamefont
  {Lhersonneau}, \citenamefont {Pfeiffer}, \citenamefont {Persson},
  \citenamefont {Suhonen}, \citenamefont {Toivanen}, \citenamefont {Campbell},
  \citenamefont {Dendooven}, \citenamefont {Honkanen}, \citenamefont {Huhta},
  \citenamefont {Jones}, \citenamefont {Julin}, \citenamefont {Juutinen},
  \citenamefont {Oinonen}, \citenamefont {Penttil{\"{a}}}, \citenamefont
  {Per{\"{a}}j{\"{a}}rvi}, \citenamefont {Savelius}, \citenamefont {Jicheng},
  \citenamefont {Wang},\ and\ \citenamefont
  {{\"{A}}yst{\"{o}}}}]{Lhersonneau1997a}%
  \BibitemOpen
  \bibfield  {author} {\bibinfo {author} {\bibfnamefont {G.}~\bibnamefont
  {Lhersonneau}}, \bibinfo {author} {\bibfnamefont {B.}~\bibnamefont
  {Pfeiffer}}, \bibinfo {author} {\bibfnamefont {J.~R.}\ \bibnamefont
  {Persson}}, \bibinfo {author} {\bibfnamefont {J.}~\bibnamefont {Suhonen}},
  \bibinfo {author} {\bibfnamefont {J.}~\bibnamefont {Toivanen}}, \bibinfo
  {author} {\bibfnamefont {P.}~\bibnamefont {Campbell}}, \bibinfo {author}
  {\bibfnamefont {P.}~\bibnamefont {Dendooven}}, \bibinfo {author}
  {\bibfnamefont {A.}~\bibnamefont {Honkanen}}, \bibinfo {author}
  {\bibfnamefont {M.}~\bibnamefont {Huhta}}, \bibinfo {author} {\bibfnamefont
  {P.~M.}\ \bibnamefont {Jones}}, \bibinfo {author} {\bibfnamefont
  {R.}~\bibnamefont {Julin}}, \bibinfo {author} {\bibfnamefont
  {S.}~\bibnamefont {Juutinen}}, \bibinfo {author} {\bibfnamefont
  {M.}~\bibnamefont {Oinonen}}, \bibinfo {author} {\bibfnamefont
  {H.}~\bibnamefont {Penttil{\"{a}}}}, \bibinfo {author} {\bibfnamefont
  {K.}~\bibnamefont {Per{\"{a}}j{\"{a}}rvi}}, \bibinfo {author} {\bibfnamefont
  {A.}~\bibnamefont {Savelius}}, \bibinfo {author} {\bibfnamefont
  {W.}~\bibnamefont {Jicheng}}, \bibinfo {author} {\bibfnamefont {J.~C.}\
  \bibnamefont {Wang}}, \ and\ \bibinfo {author} {\bibfnamefont
  {J.}~\bibnamefont {{\"{A}}yst{\"{o}}}},\ }\href {\doibase
  10.1007/s002180050335} {\bibfield  {journal} {\bibinfo  {journal}
  {Zeitschrift f{\"{u}}r Phys. A Hadron. Nucl.}\ }\textbf {\bibinfo {volume}
  {358}},\ \bibinfo {pages} {317} (\bibinfo {year} {1997})}\BibitemShut
  {NoStop}%
\bibitem [{\citenamefont {Esmaylzadeh}\ \emph {et~al.}(2019)\citenamefont
  {Esmaylzadeh}, \citenamefont {R{\'{e}}gis}, \citenamefont {Kim},
  \citenamefont {K{\"{o}}ster}, \citenamefont {Jolie}, \citenamefont
  {Karayonchev}, \citenamefont {Knafla}, \citenamefont {Nomura}, \citenamefont
  {Robledo},\ and\ \citenamefont {Rodriguez-Guzman}}]{Esmaylzadeh2019}%
  \BibitemOpen
  \bibfield  {author} {\bibinfo {author} {\bibfnamefont {A.}~\bibnamefont
  {Esmaylzadeh}}, \bibinfo {author} {\bibfnamefont {J.-M.}\ \bibnamefont
  {R{\'{e}}gis}}, \bibinfo {author} {\bibfnamefont {Y.~H.}\ \bibnamefont
  {Kim}}, \bibinfo {author} {\bibfnamefont {U.}~\bibnamefont {K{\"{o}}ster}},
  \bibinfo {author} {\bibfnamefont {J.}~\bibnamefont {Jolie}}, \bibinfo
  {author} {\bibfnamefont {V.}~\bibnamefont {Karayonchev}}, \bibinfo {author}
  {\bibfnamefont {L.}~\bibnamefont {Knafla}}, \bibinfo {author} {\bibfnamefont
  {K.}~\bibnamefont {Nomura}}, \bibinfo {author} {\bibfnamefont {L.~M.}\
  \bibnamefont {Robledo}}, \ and\ \bibinfo {author} {\bibfnamefont
  {R.}~\bibnamefont {Rodriguez-Guzman}},\ }\href {\doibase
  10.1103/PhysRevC.100.064309} {\bibfield  {journal} {\bibinfo  {journal}
  {Phys. Rev. C}\ }\textbf {\bibinfo {volume} {100}},\ \bibinfo {pages}
  {064309} (\bibinfo {year} {2019})}\BibitemShut {NoStop}%
\bibitem [{\citenamefont {Gloeckner}(1975)}]{Gloeckner1975}%
  \BibitemOpen
  \bibfield  {author} {\bibinfo {author} {\bibfnamefont {D.}~\bibnamefont
  {Gloeckner}},\ }\href {\doibase 10.1016/0375-9474(75)90484-4} {\bibfield
  {journal} {\bibinfo  {journal} {Nucl. Phys. A}\ }\textbf {\bibinfo {volume}
  {253}},\ \bibinfo {pages} {301} (\bibinfo {year} {1975})}\BibitemShut
  {NoStop}%
\bibitem [{\citenamefont {Bucurescu}\ \emph {et~al.}(2005)\citenamefont
  {Bucurescu}, \citenamefont {Podoly\`{a}k}, \citenamefont {Rusu},
  \citenamefont {de~Angelis}, \citenamefont {Zhang}, \citenamefont
  {C\v{a}ta-Danil}, \citenamefont {C\v{a}ta-Danil}, \citenamefont {Ivaşcu},
  \citenamefont {Mărginean}, \citenamefont {Mărginean}, \citenamefont
  {Mihăilescu}, \citenamefont {Suliman}, \citenamefont {Regan}, \citenamefont
  {Gelletly}, \citenamefont {Langdown}, \citenamefont {Valiente-Dob\'{o}n},
  \citenamefont {Bazzacco}, \citenamefont {Lunardi}, \citenamefont {Ur},
  \citenamefont {Axiotis}, \citenamefont {Gadea}, \citenamefont {Farnea},
  \citenamefont {Ionescu-Bujor}, \citenamefont {Iordăchescu}, \citenamefont
  {Kr{\"{o}}ll}, \citenamefont {Martinez}, \citenamefont {Bizzetti},
  \citenamefont {Broda}, \citenamefont {Medina}, \citenamefont {Quintana},\
  and\ \citenamefont {Rubio}}]{Bucurescu2005}%
  \BibitemOpen
  \bibfield  {author} {\bibinfo {author} {\bibfnamefont {D.}~\bibnamefont
  {Bucurescu}}, \bibinfo {author} {\bibfnamefont {Z.}~\bibnamefont
  {Podoly\`{a}k}}, \bibinfo {author} {\bibfnamefont {C.}~\bibnamefont {Rusu}},
  \bibinfo {author} {\bibfnamefont {G.}~\bibnamefont {de~Angelis}}, \bibinfo
  {author} {\bibfnamefont {Y.~H.}\ \bibnamefont {Zhang}}, \bibinfo {author}
  {\bibfnamefont {G.}~\bibnamefont {C\v{a}ta-Danil}}, \bibinfo {author}
  {\bibfnamefont {I.}~\bibnamefont {C\v{a}ta-Danil}}, \bibinfo {author}
  {\bibfnamefont {M.}~\bibnamefont {Ivaşcu}}, \bibinfo {author} {\bibfnamefont
  {N.}~\bibnamefont {Mărginean}}, \bibinfo {author} {\bibfnamefont
  {R.}~\bibnamefont {Mărginean}}, \bibinfo {author} {\bibfnamefont {L.~C.}\
  \bibnamefont {Mihăilescu}}, \bibinfo {author} {\bibfnamefont {G.~A.}\
  \bibnamefont {Suliman}}, \bibinfo {author} {\bibfnamefont {P.~H.}\
  \bibnamefont {Regan}}, \bibinfo {author} {\bibfnamefont {W.}~\bibnamefont
  {Gelletly}}, \bibinfo {author} {\bibfnamefont {S.~D.}\ \bibnamefont
  {Langdown}}, \bibinfo {author} {\bibfnamefont {J.~J.}\ \bibnamefont
  {Valiente-Dob\'{o}n}}, \bibinfo {author} {\bibfnamefont {D.}~\bibnamefont
  {Bazzacco}}, \bibinfo {author} {\bibfnamefont {S.}~\bibnamefont {Lunardi}},
  \bibinfo {author} {\bibfnamefont {C.~A.}\ \bibnamefont {Ur}}, \bibinfo
  {author} {\bibfnamefont {M.}~\bibnamefont {Axiotis}}, \bibinfo {author}
  {\bibfnamefont {A.}~\bibnamefont {Gadea}}, \bibinfo {author} {\bibfnamefont
  {E.}~\bibnamefont {Farnea}}, \bibinfo {author} {\bibfnamefont
  {M.}~\bibnamefont {Ionescu-Bujor}}, \bibinfo {author} {\bibfnamefont
  {A.}~\bibnamefont {Iordăchescu}}, \bibinfo {author} {\bibfnamefont
  {T.}~\bibnamefont {Kr{\"{o}}ll}}, \bibinfo {author} {\bibfnamefont
  {T.}~\bibnamefont {Martinez}}, \bibinfo {author} {\bibfnamefont {P.~G.}\
  \bibnamefont {Bizzetti}}, \bibinfo {author} {\bibfnamefont {R.}~\bibnamefont
  {Broda}}, \bibinfo {author} {\bibfnamefont {N.~H.}\ \bibnamefont {Medina}},
  \bibinfo {author} {\bibfnamefont {B.}~\bibnamefont {Quintana}}, \ and\
  \bibinfo {author} {\bibfnamefont {B.}~\bibnamefont {Rubio}},\ }\href
  {\doibase 10.1103/PhysRevC.71.034315} {\bibfield  {journal} {\bibinfo
  {journal} {Phys. Rev. C}\ }\textbf {\bibinfo {volume} {71}},\ \bibinfo
  {pages} {034315} (\bibinfo {year} {2005})}\BibitemShut {NoStop}%
\bibitem [{\citenamefont {Orce}\ \emph {et~al.}(2006)\citenamefont {Orce},
  \citenamefont {Holt}, \citenamefont {Linnemann}, \citenamefont {McKay},
  \citenamefont {Lesher}, \citenamefont {Fransen}, \citenamefont {Holt},
  \citenamefont {Kumar}, \citenamefont {Warr}, \citenamefont {Werner},
  \citenamefont {Jolie}, \citenamefont {Kuo}, \citenamefont {McEllistrem},
  \citenamefont {Pietralla},\ and\ \citenamefont {Yates}}]{Orce2006}%
  \BibitemOpen
  \bibfield  {author} {\bibinfo {author} {\bibfnamefont {J.~N.}\ \bibnamefont
  {Orce}}, \bibinfo {author} {\bibfnamefont {J.~D.}\ \bibnamefont {Holt}},
  \bibinfo {author} {\bibfnamefont {A.}~\bibnamefont {Linnemann}}, \bibinfo
  {author} {\bibfnamefont {C.~J.}\ \bibnamefont {McKay}}, \bibinfo {author}
  {\bibfnamefont {S.~R.}\ \bibnamefont {Lesher}}, \bibinfo {author}
  {\bibfnamefont {C.}~\bibnamefont {Fransen}}, \bibinfo {author} {\bibfnamefont
  {J.~W.}\ \bibnamefont {Holt}}, \bibinfo {author} {\bibfnamefont
  {A.}~\bibnamefont {Kumar}}, \bibinfo {author} {\bibfnamefont
  {N.}~\bibnamefont {Warr}}, \bibinfo {author} {\bibfnamefont {V.}~\bibnamefont
  {Werner}}, \bibinfo {author} {\bibfnamefont {J.}~\bibnamefont {Jolie}},
  \bibinfo {author} {\bibfnamefont {T.~T.~S.}\ \bibnamefont {Kuo}}, \bibinfo
  {author} {\bibfnamefont {M.~T.}\ \bibnamefont {McEllistrem}}, \bibinfo
  {author} {\bibfnamefont {N.}~\bibnamefont {Pietralla}}, \ and\ \bibinfo
  {author} {\bibfnamefont {S.~W.}\ \bibnamefont {Yates}},\ }\href {\doibase
  10.1103/PhysRevLett.97.062504} {\bibfield  {journal} {\bibinfo  {journal}
  {Phys. Rev. Lett.}\ }\textbf {\bibinfo {volume} {97}},\ \bibinfo {pages}
  {062504} (\bibinfo {year} {2006})}\BibitemShut {NoStop}%
\bibitem [{\citenamefont {Orce}\ \emph {et~al.}(2010)\citenamefont {Orce},
  \citenamefont {Holt}, \citenamefont {Linnemann}, \citenamefont {McKay},
  \citenamefont {Fransen}, \citenamefont {Jolie}, \citenamefont {Kuo},
  \citenamefont {Lesher}, \citenamefont {McEllistrem}, \citenamefont
  {Pietralla}, \citenamefont {Warr}, \citenamefont {Werner},\ and\
  \citenamefont {Yates}}]{Orce2010}%
  \BibitemOpen
  \bibfield  {author} {\bibinfo {author} {\bibfnamefont {J.~N.}\ \bibnamefont
  {Orce}}, \bibinfo {author} {\bibfnamefont {J.~D.}\ \bibnamefont {Holt}},
  \bibinfo {author} {\bibfnamefont {A.}~\bibnamefont {Linnemann}}, \bibinfo
  {author} {\bibfnamefont {C.~J.}\ \bibnamefont {McKay}}, \bibinfo {author}
  {\bibfnamefont {C.}~\bibnamefont {Fransen}}, \bibinfo {author} {\bibfnamefont
  {J.}~\bibnamefont {Jolie}}, \bibinfo {author} {\bibfnamefont {T.~T.~S.}\
  \bibnamefont {Kuo}}, \bibinfo {author} {\bibfnamefont {S.~R.}\ \bibnamefont
  {Lesher}}, \bibinfo {author} {\bibfnamefont {M.~T.}\ \bibnamefont
  {McEllistrem}}, \bibinfo {author} {\bibfnamefont {N.}~\bibnamefont
  {Pietralla}}, \bibinfo {author} {\bibfnamefont {N.}~\bibnamefont {Warr}},
  \bibinfo {author} {\bibfnamefont {V.}~\bibnamefont {Werner}}, \ and\ \bibinfo
  {author} {\bibfnamefont {S.~W.}\ \bibnamefont {Yates}},\ }\href {\doibase
  10.1103/PhysRevC.82.044317} {\bibfield  {journal} {\bibinfo  {journal} {Phys.
  Rev. C}\ }\textbf {\bibinfo {volume} {82}},\ \bibinfo {pages} {044317}
  (\bibinfo {year} {2010})}\BibitemShut {NoStop}%
\bibitem [{\citenamefont {Brant}\ \emph {et~al.}(1988)\citenamefont {Brant},
  \citenamefont {Sistemich}, \citenamefont {Paar},\ and\ \citenamefont
  {Lhersonneau}}]{Brant1988}%
  \BibitemOpen
  \bibfield  {author} {\bibinfo {author} {\bibfnamefont {S.}~\bibnamefont
  {Brant}}, \bibinfo {author} {\bibfnamefont {K.}~\bibnamefont {Sistemich}},
  \bibinfo {author} {\bibfnamefont {V.}~\bibnamefont {Paar}}, \ and\ \bibinfo
  {author} {\bibfnamefont {G.}~\bibnamefont {Lhersonneau}},\ }\href {\doibase
  10.1007/BF01290121} {\bibfield  {journal} {\bibinfo  {journal} {Zeitschrift
  f{{\"{u}}}r Phys. A At. Nucl.}\ }\textbf {\bibinfo {volume} {330}},\ \bibinfo
  {pages} {365} (\bibinfo {year} {1988})}\BibitemShut {NoStop}%
\bibitem [{\citenamefont {Lhersonneau}\ \emph {et~al.}(1990)\citenamefont
  {Lhersonneau}, \citenamefont {Pfeiffer}, \citenamefont {Kratz}, \citenamefont
  {Ohm}, \citenamefont {Sistemich}, \citenamefont {Brant},\ and\ \citenamefont
  {Paar}}]{Lhersonneau1990}%
  \BibitemOpen
  \bibfield  {author} {\bibinfo {author} {\bibfnamefont {G.}~\bibnamefont
  {Lhersonneau}}, \bibinfo {author} {\bibfnamefont {B.}~\bibnamefont
  {Pfeiffer}}, \bibinfo {author} {\bibfnamefont {K.~L.}\ \bibnamefont {Kratz}},
  \bibinfo {author} {\bibfnamefont {H.}~\bibnamefont {Ohm}}, \bibinfo {author}
  {\bibfnamefont {K.}~\bibnamefont {Sistemich}}, \bibinfo {author}
  {\bibfnamefont {S.}~\bibnamefont {Brant}}, \ and\ \bibinfo {author}
  {\bibfnamefont {V.}~\bibnamefont {Paar}},\ }\href {\doibase
  10.1007/BF01294286} {\bibfield  {journal} {\bibinfo  {journal} {Zeitschrift
  f{{\"{u}}}r Phys. A At. Nucl.}\ }\textbf {\bibinfo {volume} {337}},\ \bibinfo
  {pages} {149} (\bibinfo {year} {1990})}\BibitemShut {NoStop}%
\bibitem [{\citenamefont {Lhersonneau}\ \emph {et~al.}(1998)\citenamefont
  {Lhersonneau}, \citenamefont {Suhonen}, \citenamefont {Dendooven},
  \citenamefont {Honkanen}, \citenamefont {Huhta}, \citenamefont {Jones},
  \citenamefont {Julin}, \citenamefont {Juutinen}, \citenamefont {Oinonen},
  \citenamefont {Penttil{\"{a}}}, \citenamefont {Persson}, \citenamefont
  {Per\"{a}j\"{a}rvi}, \citenamefont {Savelius}, \citenamefont {Wang},
  \citenamefont {{\"{A}}yst{\"{o}}}, \citenamefont {Brant}, \citenamefont
  {Paar},\ and\ \citenamefont {Vretenar}}]{Lhersonneau1998}%
  \BibitemOpen
  \bibfield  {author} {\bibinfo {author} {\bibfnamefont {G.}~\bibnamefont
  {Lhersonneau}}, \bibinfo {author} {\bibfnamefont {J.}~\bibnamefont
  {Suhonen}}, \bibinfo {author} {\bibfnamefont {P.}~\bibnamefont {Dendooven}},
  \bibinfo {author} {\bibfnamefont {A.}~\bibnamefont {Honkanen}}, \bibinfo
  {author} {\bibfnamefont {M.}~\bibnamefont {Huhta}}, \bibinfo {author}
  {\bibfnamefont {P.}~\bibnamefont {Jones}}, \bibinfo {author} {\bibfnamefont
  {R.}~\bibnamefont {Julin}}, \bibinfo {author} {\bibfnamefont
  {S.}~\bibnamefont {Juutinen}}, \bibinfo {author} {\bibfnamefont
  {M.}~\bibnamefont {Oinonen}}, \bibinfo {author} {\bibfnamefont
  {H.}~\bibnamefont {Penttil{\"{a}}}}, \bibinfo {author} {\bibfnamefont
  {J.~R.}\ \bibnamefont {Persson}}, \bibinfo {author} {\bibfnamefont
  {K.}~\bibnamefont {Per\"{a}j\"{a}rvi}}, \bibinfo {author} {\bibfnamefont
  {A.}~\bibnamefont {Savelius}}, \bibinfo {author} {\bibfnamefont {J.~C.}\
  \bibnamefont {Wang}}, \bibinfo {author} {\bibfnamefont {J.}~\bibnamefont
  {{\"{A}}yst{\"{o}}}}, \bibinfo {author} {\bibfnamefont {S.}~\bibnamefont
  {Brant}}, \bibinfo {author} {\bibfnamefont {V.}~\bibnamefont {Paar}}, \ and\
  \bibinfo {author} {\bibfnamefont {D.}~\bibnamefont {Vretenar}},\ }\href
  {\doibase 10.1103/PhysRevC.57.2974} {\bibfield  {journal} {\bibinfo
  {journal} {Phys. Rev. C}\ }\textbf {\bibinfo {volume} {57}},\ \bibinfo
  {pages} {2974} (\bibinfo {year} {1998})}\BibitemShut {NoStop}%
\bibitem [{\citenamefont {Boulay}\ \emph {et~al.}(2020)\citenamefont {Boulay},
  \citenamefont {Simpson}, \citenamefont {Ichikawa}, \citenamefont {Kisyov},
  \citenamefont {Bucurescu}, \citenamefont {Takamine}, \citenamefont {Ahn},
  \citenamefont {Asahi}, \citenamefont {Baba}, \citenamefont {Balabanski},
  \citenamefont {Egami}, \citenamefont {Fujita}, \citenamefont {Fukuda},
  \citenamefont {Funayama}, \citenamefont {Furukawa}, \citenamefont {Georgiev},
  \citenamefont {Gladkov}, \citenamefont {Hass}, \citenamefont {Imamura},
  \citenamefont {Inabe}, \citenamefont {Ishibashi}, \citenamefont {Kawaguchi},
  \citenamefont {Kawamura}, \citenamefont {Kim}, \citenamefont {Kobayashi},
  \citenamefont {Kojima}, \citenamefont {Kusoglu}, \citenamefont {Lozeva},
  \citenamefont {Momiyama}, \citenamefont {Mukul}, \citenamefont {Niikura},
  \citenamefont {Nishibata}, \citenamefont {Nishizaka}, \citenamefont
  {Odahara}, \citenamefont {Ohtomo}, \citenamefont {Ralet}, \citenamefont
  {Sato}, \citenamefont {Shimizu}, \citenamefont {Sumikama}, \citenamefont
  {Suzuki}, \citenamefont {Takeda}, \citenamefont {Tao}, \citenamefont
  {Togano}, \citenamefont {Tominaga}, \citenamefont {Ueno}, \citenamefont
  {Yamazaki}, \citenamefont {Yang},\ and\ \citenamefont {Daugas}}]{Boulay2020}%
  \BibitemOpen
  \bibfield  {author} {\bibinfo {author} {\bibfnamefont {F.}~\bibnamefont
  {Boulay}}, \bibinfo {author} {\bibfnamefont {G.~S.}\ \bibnamefont {Simpson}},
  \bibinfo {author} {\bibfnamefont {Y.}~\bibnamefont {Ichikawa}}, \bibinfo
  {author} {\bibfnamefont {S.}~\bibnamefont {Kisyov}}, \bibinfo {author}
  {\bibfnamefont {D.}~\bibnamefont {Bucurescu}}, \bibinfo {author}
  {\bibfnamefont {A.}~\bibnamefont {Takamine}}, \bibinfo {author}
  {\bibfnamefont {D.~S.}\ \bibnamefont {Ahn}}, \bibinfo {author} {\bibfnamefont
  {K.}~\bibnamefont {Asahi}}, \bibinfo {author} {\bibfnamefont
  {H.}~\bibnamefont {Baba}}, \bibinfo {author} {\bibfnamefont {D.~L.}\
  \bibnamefont {Balabanski}}, \bibinfo {author} {\bibfnamefont
  {T.}~\bibnamefont {Egami}}, \bibinfo {author} {\bibfnamefont
  {T.}~\bibnamefont {Fujita}}, \bibinfo {author} {\bibfnamefont
  {N.}~\bibnamefont {Fukuda}}, \bibinfo {author} {\bibfnamefont
  {C.}~\bibnamefont {Funayama}}, \bibinfo {author} {\bibfnamefont
  {T.}~\bibnamefont {Furukawa}}, \bibinfo {author} {\bibfnamefont
  {G.}~\bibnamefont {Georgiev}}, \bibinfo {author} {\bibfnamefont
  {A.}~\bibnamefont {Gladkov}}, \bibinfo {author} {\bibfnamefont
  {M.}~\bibnamefont {Hass}}, \bibinfo {author} {\bibfnamefont {K.}~\bibnamefont
  {Imamura}}, \bibinfo {author} {\bibfnamefont {N.}~\bibnamefont {Inabe}},
  \bibinfo {author} {\bibfnamefont {Y.}~\bibnamefont {Ishibashi}}, \bibinfo
  {author} {\bibfnamefont {T.}~\bibnamefont {Kawaguchi}}, \bibinfo {author}
  {\bibfnamefont {T.}~\bibnamefont {Kawamura}}, \bibinfo {author}
  {\bibfnamefont {W.}~\bibnamefont {Kim}}, \bibinfo {author} {\bibfnamefont
  {Y.}~\bibnamefont {Kobayashi}}, \bibinfo {author} {\bibfnamefont
  {S.}~\bibnamefont {Kojima}}, \bibinfo {author} {\bibfnamefont
  {A.}~\bibnamefont {Kusoglu}}, \bibinfo {author} {\bibfnamefont
  {R.}~\bibnamefont {Lozeva}}, \bibinfo {author} {\bibfnamefont
  {S.}~\bibnamefont {Momiyama}}, \bibinfo {author} {\bibfnamefont
  {I.}~\bibnamefont {Mukul}}, \bibinfo {author} {\bibfnamefont
  {M.}~\bibnamefont {Niikura}}, \bibinfo {author} {\bibfnamefont
  {H.}~\bibnamefont {Nishibata}}, \bibinfo {author} {\bibfnamefont
  {T.}~\bibnamefont {Nishizaka}}, \bibinfo {author} {\bibfnamefont
  {A.}~\bibnamefont {Odahara}}, \bibinfo {author} {\bibfnamefont
  {Y.}~\bibnamefont {Ohtomo}}, \bibinfo {author} {\bibfnamefont
  {D.}~\bibnamefont {Ralet}}, \bibinfo {author} {\bibfnamefont
  {T.}~\bibnamefont {Sato}}, \bibinfo {author} {\bibfnamefont {Y.}~\bibnamefont
  {Shimizu}}, \bibinfo {author} {\bibfnamefont {T.}~\bibnamefont {Sumikama}},
  \bibinfo {author} {\bibfnamefont {H.}~\bibnamefont {Suzuki}}, \bibinfo
  {author} {\bibfnamefont {H.}~\bibnamefont {Takeda}}, \bibinfo {author}
  {\bibfnamefont {L.~C.}\ \bibnamefont {Tao}}, \bibinfo {author} {\bibfnamefont
  {Y.}~\bibnamefont {Togano}}, \bibinfo {author} {\bibfnamefont
  {D.}~\bibnamefont {Tominaga}}, \bibinfo {author} {\bibfnamefont
  {H.}~\bibnamefont {Ueno}}, \bibinfo {author} {\bibfnamefont {H.}~\bibnamefont
  {Yamazaki}}, \bibinfo {author} {\bibfnamefont {X.~F.}\ \bibnamefont {Yang}},
  \ and\ \bibinfo {author} {\bibfnamefont {J.~M.}\ \bibnamefont {Daugas}},\
  }\href {\doibase 10.1103/PhysRevLett.124.112501} {\bibfield  {journal}
  {\bibinfo  {journal} {Phys. Rev. Lett.}\ }\textbf {\bibinfo {volume} {124}},\
  \bibinfo {pages} {112501} (\bibinfo {year} {2020})}\BibitemShut {NoStop}%
\bibitem [{\citenamefont {Sieja}\ \emph {et~al.}(2009)\citenamefont {Sieja},
  \citenamefont {Nowacki}, \citenamefont {Langanke},\ and\ \citenamefont
  {Mart{\'{i}}nez-Pinedo}}]{Sieja2009}%
  \BibitemOpen
  \bibfield  {author} {\bibinfo {author} {\bibfnamefont {K.}~\bibnamefont
  {Sieja}}, \bibinfo {author} {\bibfnamefont {F.}~\bibnamefont {Nowacki}},
  \bibinfo {author} {\bibfnamefont {K.}~\bibnamefont {Langanke}}, \ and\
  \bibinfo {author} {\bibfnamefont {G.}~\bibnamefont {Mart{\'{i}}nez-Pinedo}},\
  }\href {\doibase 10.1103/PhysRevC.79.064310} {\bibfield  {journal} {\bibinfo
  {journal} {Phys. Rev. C}\ }\textbf {\bibinfo {volume} {79}},\ \bibinfo
  {pages} {064310} (\bibinfo {year} {2009})}\BibitemShut {NoStop}%
\bibitem [{\citenamefont {Sieja}(2021)}]{Sieja2021}%
  \BibitemOpen
  \bibfield  {author} {\bibinfo {author} {\bibfnamefont {K.}~\bibnamefont
  {Sieja}},\ }\href {\doibase 10.3390/universe8010023} {\bibfield  {journal}
  {\bibinfo  {journal} {Universe}\ }\textbf {\bibinfo {volume} {8}},\ \bibinfo
  {pages} {23} (\bibinfo {year} {2021})}\BibitemShut {NoStop}%
\bibitem [{\citenamefont {Gavrielov}\ \emph
  {et~al.}(2022{\natexlab{b}})\citenamefont {Gavrielov}, \citenamefont
  {Leviatan},\ and\ \citenamefont {Iachello}}]{Gavrielov2022c}%
  \BibitemOpen
  \bibfield  {author} {\bibinfo {author} {\bibfnamefont {N.}~\bibnamefont
  {Gavrielov}}, \bibinfo {author} {\bibfnamefont {A.}~\bibnamefont {Leviatan}},
  \ and\ \bibinfo {author} {\bibfnamefont {F.}~\bibnamefont {Iachello}},\
  }\href {\doibase 10.1103/PhysRevC.106.L051304} {\bibfield  {journal}
  {\bibinfo  {journal} {Phys. Rev. C}\ }\textbf {\bibinfo {volume} {106}},\
  \bibinfo {pages} {L051304} (\bibinfo {year}
  {2022}{\natexlab{b}})}\BibitemShut {NoStop}%
\bibitem [{\citenamefont {Iachello}\ and\ \citenamefont
  {Talmi}(1987)}]{IachelloTalmi1987}%
  \BibitemOpen
  \bibfield  {author} {\bibinfo {author} {\bibfnamefont {F.}~\bibnamefont
  {Iachello}}\ and\ \bibinfo {author} {\bibfnamefont {I.}~\bibnamefont
  {Talmi}},\ }\href {\doibase 10.1103/RevModPhys.59.339} {\bibfield  {journal}
  {\bibinfo  {journal} {Rev. Mod. Phys.}\ }\textbf {\bibinfo {volume} {59}},\
  \bibinfo {pages} {339} (\bibinfo {year} {1987})}\BibitemShut {NoStop}%
\bibitem [{\citenamefont {Duval}\ and\ \citenamefont
  {Barrett}(1981)}]{Duval1981}%
  \BibitemOpen
  \bibfield  {author} {\bibinfo {author} {\bibfnamefont {P.~D.}\ \bibnamefont
  {Duval}}\ and\ \bibinfo {author} {\bibfnamefont {B.~R.}\ \bibnamefont
  {Barrett}},\ }\href {\doibase 10.1016/0370-2693(81)90321-X} {\bibfield
  {journal} {\bibinfo  {journal} {Phys. Lett. B}\ }\textbf {\bibinfo {volume}
  {100}},\ \bibinfo {pages} {223} (\bibinfo {year} {1981})}\BibitemShut
  {NoStop}%
\bibitem [{\citenamefont {Duval}\ and\ \citenamefont
  {Barrett}(1982)}]{Duval1982}%
  \BibitemOpen
  \bibfield  {author} {\bibinfo {author} {\bibfnamefont {P.~D.}\ \bibnamefont
  {Duval}}\ and\ \bibinfo {author} {\bibfnamefont {B.~R.}\ \bibnamefont
  {Barrett}},\ }\href {\doibase 10.1016/0375-9474(82)90061-6} {\bibfield
  {journal} {\bibinfo  {journal} {Nucl. Phys. A}\ }\textbf {\bibinfo {volume}
  {376}},\ \bibinfo {pages} {213} (\bibinfo {year} {1982})}\BibitemShut
  {NoStop}%
\bibitem [{\citenamefont {Sambataro}\ and\ \citenamefont
  {Moln{\'{a}}r}(1982)}]{Sambataro1982}%
  \BibitemOpen
  \bibfield  {author} {\bibinfo {author} {\bibfnamefont {M.}~\bibnamefont
  {Sambataro}}\ and\ \bibinfo {author} {\bibfnamefont {G.}~\bibnamefont
  {Moln{\'{a}}r}},\ }\href {\doibase 10.1016/0375-9474(82)90060-4} {\bibfield
  {journal} {\bibinfo  {journal} {Nucl. Phys. A}\ }\textbf {\bibinfo {volume}
  {376}},\ \bibinfo {pages} {201} (\bibinfo {year} {1982})}\BibitemShut
  {NoStop}%
\bibitem [{\citenamefont {Garc\'ia-Ramos}\ and\ \citenamefont
  {Heyde}(2014)}]{GarciaRamos2014a}%
  \BibitemOpen
  \bibfield  {author} {\bibinfo {author} {\bibfnamefont {J.~E.}\ \bibnamefont
  {Garc\'ia-Ramos}}\ and\ \bibinfo {author} {\bibfnamefont {K.}~\bibnamefont
  {Heyde}},\ }\href {\doibase 10.1103/PhysRevC.89.014306} {\bibfield  {journal}
  {\bibinfo  {journal} {Phys. Rev. C}\ }\textbf {\bibinfo {volume} {89}},\
  \bibinfo {pages} {014306} (\bibinfo {year} {2014})}\BibitemShut {NoStop}%
\bibitem [{\citenamefont {Garc\'ia-Ramos}\ \emph {et~al.}(2014)\citenamefont
  {Garc\'ia-Ramos}, \citenamefont {Heyde}, \citenamefont {Robledo},\ and\
  \citenamefont {Rodriguez-Guzman}}]{GarciaRamos2014b}%
  \BibitemOpen
  \bibfield  {author} {\bibinfo {author} {\bibfnamefont {J.~E.}\ \bibnamefont
  {Garc\'ia-Ramos}}, \bibinfo {author} {\bibfnamefont {K.}~\bibnamefont
  {Heyde}}, \bibinfo {author} {\bibfnamefont {L.~M.}\ \bibnamefont {Robledo}},
  \ and\ \bibinfo {author} {\bibfnamefont {R.}~\bibnamefont
  {Rodriguez-Guzman}},\ }\href {\doibase 10.1103/PhysRevC.89.034313} {\bibfield
   {journal} {\bibinfo  {journal} {Phys. Rev. C}\ }\textbf {\bibinfo {volume}
  {89}},\ \bibinfo {pages} {034313} (\bibinfo {year} {2014})}\BibitemShut
  {NoStop}%
\bibitem [{\citenamefont {Garc\'ia-Ramos}\ and\ \citenamefont
  {Heyde}(2015)}]{GarciaRamos2015a}%
  \BibitemOpen
  \bibfield  {author} {\bibinfo {author} {\bibfnamefont {J.~E.}\ \bibnamefont
  {Garc\'ia-Ramos}}\ and\ \bibinfo {author} {\bibfnamefont {K.}~\bibnamefont
  {Heyde}},\ }\href {\doibase 10.1103/PhysRevC.92.034309} {\bibfield  {journal}
  {\bibinfo  {journal} {Phys. Rev. C}\ }\textbf {\bibinfo {volume} {92}},\
  \bibinfo {pages} {034309} (\bibinfo {year} {2015})},\ \Eprint
  {http://arxiv.org/abs/1507.08035} {arXiv:1507.08035} \BibitemShut {NoStop}%
\bibitem [{\citenamefont {Nomura}\ \emph
  {et~al.}(2016{\natexlab{c}})\citenamefont {Nomura}, \citenamefont
  {Rodriguez-Guzman},\ and\ \citenamefont {Robledo}}]{Nomura2016c}%
  \BibitemOpen
  \bibfield  {author} {\bibinfo {author} {\bibfnamefont {K.}~\bibnamefont
  {Nomura}}, \bibinfo {author} {\bibfnamefont {R.}~\bibnamefont
  {Rodriguez-Guzman}}, \ and\ \bibinfo {author} {\bibfnamefont {L.~M.}\
  \bibnamefont {Robledo}},\ }\href {\doibase 10.1103/PhysRevC.94.044314}
  {\bibfield  {journal} {\bibinfo  {journal} {Phys. Rev. C}\ }\textbf {\bibinfo
  {volume} {94}},\ \bibinfo {pages} {044314} (\bibinfo {year}
  {2016}{\natexlab{c}})}\BibitemShut {NoStop}%
\bibitem [{\citenamefont {Leviatan}\ \emph {et~al.}(2018)\citenamefont
  {Leviatan}, \citenamefont {Gavrielov}, \citenamefont {Garc\'ia-Ramos},\ and\
  \citenamefont {{Van Isacker}}}]{Leviatan2018a}%
  \BibitemOpen
  \bibfield  {author} {\bibinfo {author} {\bibfnamefont {A.}~\bibnamefont
  {Leviatan}}, \bibinfo {author} {\bibfnamefont {N.}~\bibnamefont {Gavrielov}},
  \bibinfo {author} {\bibfnamefont {J.~E.}\ \bibnamefont {Garc\'ia-Ramos}}, \
  and\ \bibinfo {author} {\bibfnamefont {P.}~\bibnamefont {{Van Isacker}}},\
  }\href {\doibase 10.1103/PhysRevC.98.031302} {\bibfield  {journal} {\bibinfo
  {journal} {Phys. Rev. C}\ }\textbf {\bibinfo {volume} {98}},\ \bibinfo
  {pages} {031302(R)} (\bibinfo {year} {2018})}\BibitemShut {NoStop}%
\bibitem [{\citenamefont {Maya-Barbecho}\ and\ \citenamefont
  {Garc\'ia-Ramos}(2022)}]{MayaBarbecho2022}%
  \BibitemOpen
  \bibfield  {author} {\bibinfo {author} {\bibfnamefont {E.}~\bibnamefont
  {Maya-Barbecho}}\ and\ \bibinfo {author} {\bibfnamefont {J.~E.}\ \bibnamefont
  {Garc\'ia-Ramos}},\ }\href {\doibase 10.1103/PhysRevC.105.034341} {\bibfield
  {journal} {\bibinfo  {journal} {Phys. Rev. C}\ }\textbf {\bibinfo {volume}
  {105}},\ \bibinfo {pages} {034341} (\bibinfo {year} {2022})}\BibitemShut
  {NoStop}%
\bibitem [{\citenamefont {Lawson}\ and\ \citenamefont
  {Uretsky}(1957)}]{Lawson1957}%
  \BibitemOpen
  \bibfield  {author} {\bibinfo {author} {\bibfnamefont {R.~D.}\ \bibnamefont
  {Lawson}}\ and\ \bibinfo {author} {\bibfnamefont {J.~L.}\ \bibnamefont
  {Uretsky}},\ }\href {\doibase 10.1103/PhysRev.108.1300} {\bibfield  {journal}
  {\bibinfo  {journal} {Phys. Rev.}\ }\textbf {\bibinfo {volume} {108}},\
  \bibinfo {pages} {1300} (\bibinfo {year} {1957})}\BibitemShut {NoStop}%
\bibitem [{\citenamefont {Bohr}\ and\ \citenamefont
  {Mottelson}(1998{\natexlab{a}})}]{BohrMott-II}%
  \BibitemOpen
  \bibfield  {author} {\bibinfo {author} {\bibfnamefont {A.}~\bibnamefont
  {Bohr}}\ and\ \bibinfo {author} {\bibfnamefont {B.~R.}\ \bibnamefont
  {Mottelson}},\ }\href@noop {} {\emph {\bibinfo {title} {{Nuclear
  structure}}}},\ Vol.~\bibinfo {volume} {2}\ (\bibinfo  {publisher} {World
  Scientific},\ \bibinfo {year} {1998})\BibitemShut {NoStop}%
\bibitem [{\citenamefont {Baglin}(2011)}]{NDS.112.1163.2011}%
  \BibitemOpen
  \bibfield  {author} {\bibinfo {author} {\bibfnamefont {C.~M.}\ \bibnamefont
  {Baglin}},\ }\href {\doibase 10.1016/J.NDS.2011.04.001} {\bibfield  {journal}
  {\bibinfo  {journal} {Nucl. Data Sheets}\ }\textbf {\bibinfo {volume}
  {112}},\ \bibinfo {pages} {1163} (\bibinfo {year} {2011})}\BibitemShut
  {NoStop}%
\bibitem [{\citenamefont {Basu}\ \emph {et~al.}(2010)\citenamefont {Basu},
  \citenamefont {Mukherjee},\ and\ \citenamefont
  {Sonzogni}}]{NDS.111.2555.2010}%
  \BibitemOpen
  \bibfield  {author} {\bibinfo {author} {\bibfnamefont {S.}~\bibnamefont
  {Basu}}, \bibinfo {author} {\bibfnamefont {G.}~\bibnamefont {Mukherjee}}, \
  and\ \bibinfo {author} {\bibfnamefont {A.}~\bibnamefont {Sonzogni}},\ }\href
  {\doibase 10.1016/J.NDS.2010.10.001} {\bibfield  {journal} {\bibinfo
  {journal} {Nucl. Data Sheets}\ }\textbf {\bibinfo {volume} {111}},\ \bibinfo
  {pages} {2555} (\bibinfo {year} {2010})}\BibitemShut {NoStop}%
\bibitem [{\citenamefont {Nica}(2010)}]{NDS.111.525.2010}%
  \BibitemOpen
  \bibfield  {author} {\bibinfo {author} {\bibfnamefont {N.}~\bibnamefont
  {Nica}},\ }\href {\doibase 10.1016/J.NDS.2010.03.001} {\bibfield  {journal}
  {\bibinfo  {journal} {Nucl. Data Sheets}\ }\textbf {\bibinfo {volume}
  {111}},\ \bibinfo {pages} {525} (\bibinfo {year} {2010})}\BibitemShut
  {NoStop}%
\bibitem [{\citenamefont {Browne}\ and\ \citenamefont
  {Tuli}(2017)}]{NDS.145.25.2017}%
  \BibitemOpen
  \bibfield  {author} {\bibinfo {author} {\bibfnamefont {E.}~\bibnamefont
  {Browne}}\ and\ \bibinfo {author} {\bibfnamefont {J.}~\bibnamefont {Tuli}},\
  }\href {\doibase 10.1016/J.NDS.2017.09.002} {\bibfield  {journal} {\bibinfo
  {journal} {Nucl. Data Sheets}\ }\textbf {\bibinfo {volume} {145}},\ \bibinfo
  {pages} {25} (\bibinfo {year} {2017})}\BibitemShut {NoStop}%
\bibitem [{\citenamefont {{Van Heerden}}\ \emph {et~al.}(1973)\citenamefont
  {{Van Heerden}}, \citenamefont {McMurray},\ and\ \citenamefont
  {Saayman}}]{VanHeerden1973}%
  \BibitemOpen
  \bibfield  {author} {\bibinfo {author} {\bibfnamefont {I.}~\bibnamefont {{Van
  Heerden}}}, \bibinfo {author} {\bibfnamefont {W.~R.}\ \bibnamefont
  {McMurray}}, \ and\ \bibinfo {author} {\bibfnamefont {R.}~\bibnamefont
  {Saayman}},\ }\href {\doibase https://doi.org/10.1007/BF01398067} {\bibfield
  {journal} {\bibinfo  {journal} {Zeitschrift f{{\"{u}}}r Phys. A Hadron.
  Nucl.}\ }\textbf {\bibinfo {volume} {260}},\ \bibinfo {pages} {9} (\bibinfo
  {year} {1973})}\BibitemShut {NoStop}%
\bibitem [{\citenamefont {{Evaluated Nuclear Structure Data File
  (ENSDF)}}()}]{ensdf}%
  \BibitemOpen
  \bibfield  {author} {\bibinfo {author} {\bibnamefont {{Evaluated Nuclear
  Structure Data File (ENSDF)}}},\ }\href {https://www.nndc.bnl.gov/ensdf/}
  {\enquote {\bibinfo {title} {https://www.nndc.bnl.gov/ensdf},}\ }\BibitemShut
  {NoStop}%
\bibitem [{\citenamefont {Hagen}\ \emph {et~al.}(2017)\citenamefont {Hagen},
  \citenamefont {G{\"{o}}rgen}, \citenamefont {Korten}, \citenamefont {Grente},
  \citenamefont {Salsac}, \citenamefont {Farget}, \citenamefont {Ragnarsson},
  \citenamefont {Braunroth}, \citenamefont {Bruyneel}, \citenamefont
  {Celikovic}, \citenamefont {Cl{\'{e}}ment}, \citenamefont {de~France},
  \citenamefont {Delaune}, \citenamefont {Dewald}, \citenamefont {Dijon},
  \citenamefont {Hackstein}, \citenamefont {Jacquot}, \citenamefont
  {Litzinger}, \citenamefont {Ljungvall}, \citenamefont {Louchart},
  \citenamefont {Michelagnoli}, \citenamefont {Napoli}, \citenamefont
  {Recchia}, \citenamefont {Rother}, \citenamefont {Sahin}, \citenamefont
  {Siem}, \citenamefont {Sulignano}, \citenamefont {Theisen},\ and\
  \citenamefont {Valiente-Dob\'{o}n}}]{Hagen2017}%
  \BibitemOpen
  \bibfield  {author} {\bibinfo {author} {\bibfnamefont {T.~W.}\ \bibnamefont
  {Hagen}}, \bibinfo {author} {\bibfnamefont {A.}~\bibnamefont {G{\"{o}}rgen}},
  \bibinfo {author} {\bibfnamefont {W.}~\bibnamefont {Korten}}, \bibinfo
  {author} {\bibfnamefont {L.}~\bibnamefont {Grente}}, \bibinfo {author}
  {\bibfnamefont {M.-D.}\ \bibnamefont {Salsac}}, \bibinfo {author}
  {\bibfnamefont {F.}~\bibnamefont {Farget}}, \bibinfo {author} {\bibfnamefont
  {I.}~\bibnamefont {Ragnarsson}}, \bibinfo {author} {\bibfnamefont
  {T.}~\bibnamefont {Braunroth}}, \bibinfo {author} {\bibfnamefont
  {B.}~\bibnamefont {Bruyneel}}, \bibinfo {author} {\bibfnamefont
  {I.}~\bibnamefont {Celikovic}}, \bibinfo {author} {\bibfnamefont
  {E.}~\bibnamefont {Cl{\'{e}}ment}}, \bibinfo {author} {\bibfnamefont
  {G.}~\bibnamefont {de~France}}, \bibinfo {author} {\bibfnamefont
  {O.}~\bibnamefont {Delaune}}, \bibinfo {author} {\bibfnamefont
  {A.}~\bibnamefont {Dewald}}, \bibinfo {author} {\bibfnamefont
  {A.}~\bibnamefont {Dijon}}, \bibinfo {author} {\bibfnamefont
  {M.}~\bibnamefont {Hackstein}}, \bibinfo {author} {\bibfnamefont
  {B.}~\bibnamefont {Jacquot}}, \bibinfo {author} {\bibfnamefont
  {J.}~\bibnamefont {Litzinger}}, \bibinfo {author} {\bibfnamefont
  {J.}~\bibnamefont {Ljungvall}}, \bibinfo {author} {\bibfnamefont
  {C.}~\bibnamefont {Louchart}}, \bibinfo {author} {\bibfnamefont
  {C.}~\bibnamefont {Michelagnoli}}, \bibinfo {author} {\bibfnamefont {D.~R.}\
  \bibnamefont {Napoli}}, \bibinfo {author} {\bibfnamefont {F.}~\bibnamefont
  {Recchia}}, \bibinfo {author} {\bibfnamefont {W.}~\bibnamefont {Rother}},
  \bibinfo {author} {\bibfnamefont {E.}~\bibnamefont {Sahin}}, \bibinfo
  {author} {\bibfnamefont {S.}~\bibnamefont {Siem}}, \bibinfo {author}
  {\bibfnamefont {B.}~\bibnamefont {Sulignano}}, \bibinfo {author}
  {\bibfnamefont {C.}~\bibnamefont {Theisen}}, \ and\ \bibinfo {author}
  {\bibfnamefont {J.~J.}\ \bibnamefont {Valiente-Dob\'{o}n}},\ }\href {\doibase
  10.1103/PhysRevC.95.034302} {\bibfield  {journal} {\bibinfo  {journal} {Phys.
  Rev. C}\ }\textbf {\bibinfo {volume} {95}},\ \bibinfo {pages} {034302}
  (\bibinfo {year} {2017})}\BibitemShut {NoStop}%
\bibitem [{\citenamefont {{De Frenne}}(2009)}]{NDS.110.2081.2009}%
  \BibitemOpen
  \bibfield  {author} {\bibinfo {author} {\bibfnamefont {D.}~\bibnamefont {{De
  Frenne}}},\ }\href {\doibase 10.1016/J.NDS.2009.08.002} {\bibfield  {journal}
  {\bibinfo  {journal} {Nucl. Data Sheets}\ }\textbf {\bibinfo {volume}
  {110}},\ \bibinfo {pages} {2081} (\bibinfo {year} {2009})}\BibitemShut
  {NoStop}%
\bibitem [{\citenamefont {Hotchkis}\ \emph {et~al.}(1991)\citenamefont
  {Hotchkis}, \citenamefont {Durell}, \citenamefont {Fitzgerald}, \citenamefont
  {Mowbray}, \citenamefont {Phillips}, \citenamefont {Ahmad}, \citenamefont
  {Carpenter}, \citenamefont {Janssens}, \citenamefont {Khoo}, \citenamefont
  {Moore}, \citenamefont {Morss}, \citenamefont {Benet},\ and\ \citenamefont
  {Ye}}]{Hotchkis1991}%
  \BibitemOpen
  \bibfield  {author} {\bibinfo {author} {\bibfnamefont {M.}~\bibnamefont
  {Hotchkis}}, \bibinfo {author} {\bibfnamefont {J.}~\bibnamefont {Durell}},
  \bibinfo {author} {\bibfnamefont {J.}~\bibnamefont {Fitzgerald}}, \bibinfo
  {author} {\bibfnamefont {A.}~\bibnamefont {Mowbray}}, \bibinfo {author}
  {\bibfnamefont {W.}~\bibnamefont {Phillips}}, \bibinfo {author}
  {\bibfnamefont {I.}~\bibnamefont {Ahmad}}, \bibinfo {author} {\bibfnamefont
  {M.~P.}\ \bibnamefont {Carpenter}}, \bibinfo {author} {\bibfnamefont
  {R.~V.~F.}\ \bibnamefont {Janssens}}, \bibinfo {author} {\bibfnamefont
  {T.}~\bibnamefont {Khoo}}, \bibinfo {author} {\bibfnamefont {E.}~\bibnamefont
  {Moore}}, \bibinfo {author} {\bibfnamefont {L.}~\bibnamefont {Morss}},
  \bibinfo {author} {\bibfnamefont {P.}~\bibnamefont {Benet}}, \ and\ \bibinfo
  {author} {\bibfnamefont {D.}~\bibnamefont {Ye}},\ }\href {\doibase
  10.1016/0375-9474(91)90758-X} {\bibfield  {journal} {\bibinfo  {journal}
  {Nucl. Phys. A}\ }\textbf {\bibinfo {volume} {530}},\ \bibinfo {pages} {111}
  (\bibinfo {year} {1991})}\BibitemShut {NoStop}%
\bibitem [{\citenamefont {Luo}\ \emph {et~al.}(2005)\citenamefont {Luo},
  \citenamefont {Rasmussen}, \citenamefont {Stefanescu}, \citenamefont
  {Gelberg}, \citenamefont {Hamilton}, \citenamefont {Ramayya}, \citenamefont
  {Hwang}, \citenamefont {Zhu}, \citenamefont {Gore}, \citenamefont {Fong},
  \citenamefont {Jones}, \citenamefont {Wu}, \citenamefont {Lee}, \citenamefont
  {Ginter}, \citenamefont {Ma}, \citenamefont {Ter-Akopian}, \citenamefont
  {Daniel}, \citenamefont {Stoyer},\ and\ \citenamefont {Donangelo}}]{Luo2005}%
  \BibitemOpen
  \bibfield  {author} {\bibinfo {author} {\bibfnamefont {Y.-A.}\ \bibnamefont
  {Luo}}, \bibinfo {author} {\bibfnamefont {J.~O.}\ \bibnamefont {Rasmussen}},
  \bibinfo {author} {\bibfnamefont {I.}~\bibnamefont {Stefanescu}}, \bibinfo
  {author} {\bibfnamefont {A.}~\bibnamefont {Gelberg}}, \bibinfo {author}
  {\bibfnamefont {J.~H.}\ \bibnamefont {Hamilton}}, \bibinfo {author}
  {\bibfnamefont {A.~V.}\ \bibnamefont {Ramayya}}, \bibinfo {author}
  {\bibfnamefont {J.~K.}\ \bibnamefont {Hwang}}, \bibinfo {author}
  {\bibfnamefont {S.~J.}\ \bibnamefont {Zhu}}, \bibinfo {author} {\bibfnamefont
  {P.~M.}\ \bibnamefont {Gore}}, \bibinfo {author} {\bibfnamefont
  {D.}~\bibnamefont {Fong}}, \bibinfo {author} {\bibfnamefont {E.~F.}\
  \bibnamefont {Jones}}, \bibinfo {author} {\bibfnamefont {S.~C.}\ \bibnamefont
  {Wu}}, \bibinfo {author} {\bibfnamefont {I.~Y.}\ \bibnamefont {Lee}},
  \bibinfo {author} {\bibfnamefont {T.~N.}\ \bibnamefont {Ginter}}, \bibinfo
  {author} {\bibfnamefont {W.~C.}\ \bibnamefont {Ma}}, \bibinfo {author}
  {\bibfnamefont {G.~M.}\ \bibnamefont {Ter-Akopian}}, \bibinfo {author}
  {\bibfnamefont {A.~V.}\ \bibnamefont {Daniel}}, \bibinfo {author}
  {\bibfnamefont {M.~A.}\ \bibnamefont {Stoyer}}, \ and\ \bibinfo {author}
  {\bibfnamefont {R.}~\bibnamefont {Donangelo}},\ }\href {\doibase
  10.1088/0954-3899/31/11/013} {\bibfield  {journal} {\bibinfo  {journal} {J.
  Phys. G Nucl. Part. Phys.}\ }\textbf {\bibinfo {volume} {31}},\ \bibinfo
  {pages} {1303} (\bibinfo {year} {2005})}\BibitemShut {NoStop}%
\bibitem [{\citenamefont {Baglin}(2013)}]{NDS.114.1293.2013}%
  \BibitemOpen
  \bibfield  {author} {\bibinfo {author} {\bibfnamefont {C.~M.}\ \bibnamefont
  {Baglin}},\ }\href {\doibase 10.1016/J.NDS.2013.10.002} {\bibfield  {journal}
  {\bibinfo  {journal} {Nucl. Data Sheets}\ }\textbf {\bibinfo {volume}
  {114}},\ \bibinfo {pages} {1293} (\bibinfo {year} {2013})}\BibitemShut
  {NoStop}%
\bibitem [{\citenamefont {Huang}\ \emph {et~al.}(2017)\citenamefont {Huang},
  \citenamefont {Audi}, \citenamefont {Wang}, \citenamefont {Kondev},
  \citenamefont {Naimi},\ and\ \citenamefont {Xu}}]{Huang2017}%
  \BibitemOpen
  \bibfield  {author} {\bibinfo {author} {\bibfnamefont {W.}~\bibnamefont
  {Huang}}, \bibinfo {author} {\bibfnamefont {G.}~\bibnamefont {Audi}},
  \bibinfo {author} {\bibfnamefont {M.}~\bibnamefont {Wang}}, \bibinfo {author}
  {\bibfnamefont {F.~G.}\ \bibnamefont {Kondev}}, \bibinfo {author}
  {\bibfnamefont {S.}~\bibnamefont {Naimi}}, \ and\ \bibinfo {author}
  {\bibfnamefont {X.}~\bibnamefont {Xu}},\ }\href {\doibase
  10.1088/1674-1137/41/3/030002} {\bibfield  {journal} {\bibinfo  {journal}
  {Chinese Phys. C}\ }\textbf {\bibinfo {volume} {41}},\ \bibinfo {pages}
  {030002} (\bibinfo {year} {2017})}\BibitemShut {NoStop}%
\bibitem [{\citenamefont {Cheal}\ \emph {et~al.}(2009)\citenamefont {Cheal},
  \citenamefont {Baczynska}, \citenamefont {Billowes}, \citenamefont
  {Campbell}, \citenamefont {Charlwood}, \citenamefont {Eronen}, \citenamefont
  {Forest}, \citenamefont {Jokinen}, \citenamefont {Kessler}, \citenamefont
  {Moore}, \citenamefont {Reponen}, \citenamefont {Rothe}, \citenamefont
  {R{\"{u}}ffer}, \citenamefont {Saastamoinen}, \citenamefont {Tungate},\ and\
  \citenamefont {{\"{A}}yst{\"{o}}}}]{Cheal2009}%
  \BibitemOpen
  \bibfield  {author} {\bibinfo {author} {\bibfnamefont {B.}~\bibnamefont
  {Cheal}}, \bibinfo {author} {\bibfnamefont {K.}~\bibnamefont {Baczynska}},
  \bibinfo {author} {\bibfnamefont {J.}~\bibnamefont {Billowes}}, \bibinfo
  {author} {\bibfnamefont {P.}~\bibnamefont {Campbell}}, \bibinfo {author}
  {\bibfnamefont {F.~C.}\ \bibnamefont {Charlwood}}, \bibinfo {author}
  {\bibfnamefont {T.}~\bibnamefont {Eronen}}, \bibinfo {author} {\bibfnamefont
  {D.~H.}\ \bibnamefont {Forest}}, \bibinfo {author} {\bibfnamefont
  {A.}~\bibnamefont {Jokinen}}, \bibinfo {author} {\bibfnamefont
  {T.}~\bibnamefont {Kessler}}, \bibinfo {author} {\bibfnamefont {I.~D.}\
  \bibnamefont {Moore}}, \bibinfo {author} {\bibfnamefont {M.}~\bibnamefont
  {Reponen}}, \bibinfo {author} {\bibfnamefont {S.}~\bibnamefont {Rothe}},
  \bibinfo {author} {\bibfnamefont {M.}~\bibnamefont {R{\"{u}}ffer}}, \bibinfo
  {author} {\bibfnamefont {A.}~\bibnamefont {Saastamoinen}}, \bibinfo {author}
  {\bibfnamefont {G.}~\bibnamefont {Tungate}}, \ and\ \bibinfo {author}
  {\bibfnamefont {J.}~\bibnamefont {{\"{A}}yst{\"{o}}}},\ }\href {\doibase
  10.1103/PhysRevLett.102.222501} {\bibfield  {journal} {\bibinfo  {journal}
  {Phys. Rev. Lett.}\ }\textbf {\bibinfo {volume} {102}},\ \bibinfo {pages}
  {222501} (\bibinfo {year} {2009})}\BibitemShut {NoStop}%
\bibitem [{\citenamefont {Barea}\ and\ \citenamefont
  {Iachello}(2009)}]{Barea2009}%
  \BibitemOpen
  \bibfield  {author} {\bibinfo {author} {\bibfnamefont {J.}~\bibnamefont
  {Barea}}\ and\ \bibinfo {author} {\bibfnamefont {F.}~\bibnamefont
  {Iachello}},\ }\href {\doibase 10.1103/PhysRevC.79.044301} {\bibfield
  {journal} {\bibinfo  {journal} {Phys. Rev. C}\ }\textbf {\bibinfo {volume}
  {79}},\ \bibinfo {pages} {044301} (\bibinfo {year} {2009})}\BibitemShut
  {NoStop}%
\bibitem [{\citenamefont {Bohr}\ and\ \citenamefont
  {Mottelson}(1998{\natexlab{b}})}]{BohrMott-I}%
  \BibitemOpen
  \bibfield  {author} {\bibinfo {author} {\bibfnamefont {A.}~\bibnamefont
  {Bohr}}\ and\ \bibinfo {author} {\bibfnamefont {B.~R.}\ \bibnamefont
  {Mottelson}},\ }\href@noop {} {\emph {\bibinfo {title} {{Nuclear
  structure}}}},\ Vol.~\bibinfo {volume} {1}\ (\bibinfo  {publisher} {World
  Scientific},\ \bibinfo {year} {1998})\BibitemShut {NoStop}%
\bibitem [{\citenamefont {Scholten}(1980)}]{ScholtenThesis}%
  \BibitemOpen
  \bibfield  {author} {\bibinfo {author} {\bibfnamefont {O.}~\bibnamefont
  {Scholten}},\ }\emph {\bibinfo {title} {{The Interacting Boson Approximation
  Model and Applications}}},\ \href@noop {} {Ph.D. thesis},\ \bibinfo  {school}
  {Groningen} (\bibinfo {year} {1980})\BibitemShut {NoStop}%
\end{thebibliography}%
\end{document}